\documentclass[a4paper,10pt]{article}

\usepackage{jheppub} 

\usepackage[T1]{fontenc} 

    \title{\boldmath A general method for the resummation of
  event-shape distributions in $e^+e^-$ annihilation}

\author[a]{Andrea Banfi,}
\author[a]{Heather McAslan,}
\author[b]{Pier Francesco Monni,}
\author[b,c]{Giulia Zanderighi}

\affiliation[a]{Department of Physics and Astronomy, University of Sussex,\\Sussex House, Brighton, BN1 9RH, UK}
\affiliation[b]{Rudolf Peierls Centre for Theoretical
  Physics,University of Oxford,\\1 Keble Road, Oxford OX1 3NP, UK}
  \affiliation[c]{CERN, Theory Division, \\
    CH-1211 Geneva 23, Switzerland}

\emailAdd{a.banfi@sussex.ac.uk}
\emailAdd{H.Mcaslan@sussex.ac.uk}
\emailAdd{pier.monni@physics.ox.ac.uk}
\emailAdd{giulia.zanderighi@cern.ch}

\abstract{We present a novel method for resummation of event shapes to
  next-to-next-to-leading-logarithmic (NNLL) accuracy. We discuss the
  technique and describe its implementation in a numerical program in
  the case of $\ee$ collisions where the resummed prediction is
  matched to NNLO.  We reproduce all the existing predictions and
  present new results for oblateness and thrust major.}

\preprint{OUTP-14-18P}

\newcommand{\as}{\alpha_s}

\newcommand{\cO}[1]{{\cal O}\left(#1\right)}

\newcommand{\ee}{e^+e^-}

\newcommand{\fullF}{\mathcal{F}}
\newcommand{\FNLL}{\mathcal{F}_{\rm NLL}}
\newcommand{\FNNLL}{\mathcal{F}_{\rm NNLL}}
\newcommand{\ie}{i.e.~}

\newcommand{\Vsc}{V_{\rm sc}}
\newcommand{\Vwa}{V_{\rm wa}}
\newcommand{\Vfull}{V}
\newcommand{\Vhc}{V_{\rm hc}}

\newcommand{\dZ}{d{\cal Z}[\{R'_{\mathrm{NLL}, \ell_i}, k_i\}]}
\newcommand{\RpNLL}{R'_{\mathrm{NLL}}}

\begin{document} 
\maketitle
\flushbottom

\section{Introduction}

Event-shape variables in $\ee$ annihilation are among the most studied
QCD observables. Since they are very sensitive to the pattern of QCD
radiation, they have been widely used in the past to measure the QCD
coupling constant, and to test non-perturbative hadronization models
(see e.g. ref.~\cite{Dasgupta:2003iq} and references therein).
The study of event shapes also led to important advances in the
understanding of all-order properties of QCD radiation, for instance
through the ``discovery'' of non-global
logarithms~\cite{NG1,NG2,Banfi:2002hw}.
Fixed order predictions for observables involving up to three jets in
$e^+e^-$ collisions have been available up to next-to-next-to-leading
order
(NNLO)~\cite{GehrmannDeRidder:2007hr,GehrmannDeRidder:2008ug,Weinzierl:2008iv,Weinzierl:2009ms}
for some years.
While fixed order calculations provide a good approximation of hard
radiation, which contributes to the region where event shapes have
rather large values, resummed calculations are required where the bulk
of data lies, i.e. in the region dominated by multiple soft-collinear
emissions.
Next-to-leading logarithmic (NLL) resummations, that include all terms
${\cal O}(\alpha_s^nL^n)$ in the exponent of integrated
distributions are available for specific
observables~\cite{CSS,CTTW,Bonciani:2003nt,thr_res,mh_ee,cpar_res,CTWbroad,DLMSBroadPT,y3-kt_ee,CatDokWeb,JetsDisSchmell,BKS03,Larkoski:2014uqa}. In
ref.~\cite{Banfi:2001bz} a semi-numerical approach was presented to
compute the NLL resummation for all event shapes and jet rates that
are recursive infrared and collinear (rIRC) safe and are continuously
global. Most NLL resummations have been performed for observables that
satisfy these minimal requirements. Some recent works address the
problem of resumming ratios of angularities which happen to be not IR
safe, but still resummable~\cite{Larkoski:2013paa}. Their resummations
rely on factorisation theorems for double differential distributions
of angularities~\cite{Larkoski:2014tva,Procura:2014cba}.
The method of ref.~\cite{Banfi:2001bz} was subsequently extended and
implemented in the computer program {\tt CAESAR}~\cite{Banfi:2004yd},
that also verifies whether a given observable satisfies these
properties.
This led to a first systematic study of event shapes in hadronic dijet
production at NLL accuracy matched to next-to-leading order (NLO)
results at hadron colliders~\cite{Banfi:2010xy,Banfi:2004nk}.

More recently, some observables have been resummed beyond NLL
accuracy.
These resummations have been so far obtained through
observable-dependent factorisation theorems which lead to a full
decomposition of the cross section in the infrared limit in terms of
different kinematical subprocesses (i.e. soft, collinear, hard) which
are then resummed individually through evolution equations.  Despite
being systematically extendable to all orders, this approach is
strictly observable-dependent and requires that the observable can be
factorised in some conjugate space.
In particular, full next-to-next-to-leading logarithmic (NNLL)
predictions are available for a number of event shapes at lepton
colliders like thrust $1-T$~\cite{Becher:2008cf,Monni:2011gb}, heavy
jet mass $\rho_H$~\cite{Chien:2010kc}, jet broadenings
$B_T$,~$B_W$~\cite{Becher:2012qc}, $C$-parameter~\cite{Alioli:2012fc}
and energy-energy-correlation~\cite{deFlorian:2004mp}.\footnote{Note
  that the NNLL $A^{(3)}$ coefficient in ref.~\cite{deFlorian:2004mp}
  is incomplete. The correct coefficient has been derived in
  ref.~\cite{Becher:2010tm}.} For $1-T$ and $\rho_H$ all N$^3$LL
corrections but the four-loop cusp anomalous dimension are also known.
Similar observables have been resummed at the same accuracy also in
deep inelastic
scattering~\cite{Kang:2013nha,Kang:2013wca,Kang:2013lga}.  For
hadronic collisions, full NNLL resummations are available for
processes where a colour singlet is produced at Born level,
specifically for the boson's transverse
momentum~\cite{Bozzi:2005wk,Becher:2010tm} and
$\phi^*$~\cite{Banfi:2011dx}, the beam
thrust~\cite{Stewart:2010pd,Berger:2010xi} and the leading jet's
transverse
momentum~\cite{Becher:2012qa,Becher:2013xia,Banfi:2012jm,Stewart:2013faa},
and for heavy quark pair's transverse
momentum~\cite{Zhu:2012ts,Catani:2014qha}. For an arbitrary number of
legs, a NNLL accurate resummation is available for the $N$-jettiness
variable~\cite{Stewart:2010tn,Jouttenus:2011wh}.

Currently most of the phenomenological interest is devoted to
hadron-hadron collisions. However, in view of a possible future
$e^+e^-$ machine (see e.g.\ \cite{Fan:2014vta,Brock:2014tja}), it is
desirable to improve our description of {\it generic} $e^+e^-$ event
shapes to next-to-next-to-leading logarithmic (NNLL) level, matched to
exact next-to-next-to-leading order (NNLO) results.
Furthermore, $\ee$ observables provide a simpler laboratory in which
to develop new methods, compared to jet production in hadronic
collisions.  Therefore, in this work we focus on $\ee$ collisions,
with the aim to extend the method suggested here to hadron colliders
in a future publication.

In this article we derive a general and systematic method to compute
NNLL corrections to event shape distributions in $\ee$ collisions. The
method is flexible and can handle any rIRC safe observable which is
continuously global, without any additional requirement on
factorisability of the observable into kinematic subprocesses. The
method relies on a semi-numerical approach in which all real
corrections can be expressed in terms of four-dimensional phase space
integrals to all orders, and can be efficiently implemented using
Monte Carlo techniques. The remaining analytic ingredient is a Sudakov
form factor, i.e. the exponential of the so-called ``radiator''.  In
the present paper we do not derive a general expression for the NNLL
radiator, but we show that the only unknown contribution is universal
for classes of observables which scale in the same fashion for a
single soft-collinear emission. The latter property allows us to resum
a number of observables by using the radiator of those for which a
NNLL resummation was previously known. We derive the method and
describe its numerical implementation in the program {\tt ARES}
(Automated Resummation for Event Shapes).  In the present article we
limit ourselves to the resummation of NNLL terms, nevertheless the
technique described here can be extended systematically to higher
logarithmic orders.

The paper is structured as follows. In Section~\ref{sec:nll} we recall
the NLL method of ref.~\cite{Banfi:2001bz} in detail, revisiting all
the approximations that lead to the derivation of the master
resummation formula.  In Section~\ref{sec:nnll} we describe all NNLL
corrections showing how to derive them systematically. We then apply
the resummation method to the following seven event-shape observables:
the thrust $1-T$, the $C$ parameter, the heavy-jet mass $\rho_H$, the
total and wide-jet broadenings $B_T$, $B_W$, the thrust major $T_M$,
and the oblateness $O$, for which data from LEP are available. In
Section~\ref{sec:checkandmatch} we test the resummation program by
expanding the resummed cross section to fixed order in the strong
coupling. For observables for which an analytic NNLL resummation was
previously available in the literature (i.e. thrust, heavy jet mass
and jet broadenings), we check our results against the analytic ones
up to (and including) ${\cal O}(\alpha_s^3)$. For the remaining
observables, for which a NNLL analytic result was not available so far
(i.e. $C$ parameter, thrust major $T_M$ and oblateness $O$) we check
the expansion of the resummed result against the NLO generator {\tt
  Event2}~\cite{Catani:1996vz}. In the second part of
Section~\ref{sec:checkandmatch} we perform a matching to the NNLO
distributions obtained with the event generator {\tt
  EERAD3}~\cite{Ridder:2014wza}.  Our conclusions are reported in
Section~\ref{sec:conclusions}.  A definition of the observables
studied here can be found in Appendix~\ref{sec:obsdef}.  All analytic
ingredients used in this article are reported in
Appendix~\ref{sec:radiator}.  In Appendix~\ref{sec:additive} we show
that for a class of additive observables (e.g. $1-T$, $C$ and
$\rho_H$), all the necessary NNLL corrections can be computed
analytically, and we give explicit analytic results.  The numerical
implementation of our method in a Monte Carlo code is discussed in
Appendix~\ref{sec:monte-carlo-determ}.

\section{Review of NLL resummation}
\label{sec:nll}
We consider the resummation of a generic continuously global,
recursive infrared and collinear (rIRC) safe event-shape observable
$V$, a function of all final-state momenta, in $e^+e^-$ annihilation.
We review here the next-to-leading logarithmic (NLL) resummation for
these observables.  This section is largely inspired by Sec.~2 of
ref.~\cite{Banfi:2004yd}, which contains a detailed derivation of the
NLL resummation for generic event-shapes within the {\tt CAESAR}
approach.

At Born level, the final state consists of a quark $\tilde p_{1}$ and
an antiquark $\tilde p_{2}$, which are back-to-back. All event shapes
we consider vanish in the Born limit, \ie $\Vfull(\{\tilde
p_{1},\tilde p_{2}\})=0$.\footnote{In the case of the thrust, the
  resummation is actually performed for $\tau\equiv 1-T$.}
Beyond Born level, further radiation (of gluons or gluons splitting
into quarks) is present and the final state consists in general of $n$
secondary emissions, $k_1,\dots,k_n$, and of the primary quark and
antiquark which recoil against these additional emissions.
We denote the value of an event shape by $\Vfull(\{\tilde
 p\},k_1,\dots,k_n)$, with $\{\tilde p\}= \{\tilde p_1,\tilde p_2\}$.

 For any final state event, it is possible to use the thrust axis
 $\vec n_T$ to define two like-light vectors, $p_1$ and $p_2$ as
\begin{equation}
\label{eq:com-frame}
p_1 = \frac{Q}{2}(1,\vec n_T)\,,\qquad p_2 = \frac{Q}{2}(1,-\vec n_T)\,,
\end{equation}
where $Q$ denotes the total centre of mass energy of the collision.
At Born level clearly $\tilde p_1$ and $\tilde p_2$ coincide with
$p_1$ and $p_2$.

In order to compute the resummed distribution for an observable $V$,
it is useful to parametrise each emission $k_i$ and its phase-space in
terms of Sudakov variables:
\begin{equation}
  \label{eq:Sudakov}
  k_i = z_i^{(1)}p_1 +  z_i^{(2)}p_2 + \kappa_{t,i}\,,
\end{equation}
where $\kappa_{t,i}$ is a space-like four-vector, orthogonal to $p_1$
and $p_2$. In the reference frame in which $p_1$ and $p_2$ are given
by eq.~(\ref{eq:com-frame}), each $\kappa_{t,i}$ has no timelike
component and can be written as $\kappa_{t,i}=(0,\vec k_{t,i})$, such
that $\kappa_{t,i}^2 = - k_{t,i}^2$.  Notice that since $k_i$ is
massless
\[
k_{t,i}^2 = \frac{2(p_1 k_i) 2(p_2 k_i)}{2(p_1 p_2)}\,.
\]
We recall that the thrust axis divides each event in two hemispheres
$\mathcal{H}^{(1)}$ and $\mathcal{H}^{(2)}$. If all emissions are soft
and/or collinear, $\tilde p_1$ and $\tilde p_2$ belong to different
hemispheres. We denote by $\mathcal{H}^{(i)}$ the hemisphere
containing $\tilde p_i$.
Finally, we introduce the emission's rapidity $\eta_i$ with respect to
the thrust axis, which is given by
\begin{equation}
  \label{eq:rapidity}
  \eta_i = \frac{1}{2}\ln \frac{z_i^{(1)}}{z_i^{(2)}}\,,\quad {\rm with}\quad 
  |\eta_i| < \ln\frac{Q}{k_{t,i}}\,,
\end{equation}
where the boundary for $\eta_i$ is obtained by imposing
$z_i^{(\ell)}<1$ for any leg $\ell=1,2$.

We consider observables $\Vfull$ that obey the following general
parametrisation\footnote{All event shapes for which a NLL resummation
  is known obey this form.}  for a single soft emission $k$ collinear
to leg $\ell$ (\ie parton $\tilde p_\ell$):
\begin{equation}
\label{eq:v-scaling}
 \Vsc(\{\tilde{p}\},k) = d_\ell \,g_\ell(\phi) \left(\frac{k_t}{Q}\right)^a
 e^{-b_\ell \eta^{(\ell)}}\,,
\end{equation}
where $\eta^{(1)}=\eta$ and $\eta^{(2)}=-\eta$, and $\phi$ is the
angle that the transverse momentum $\vec k_t$ forms with a fixed reference
vector $\vec n$ orthogonal to the thrust axis. Collinear and infrared
safety imposes that $a > 0$ and $b_\ell> -a$.

In order to build the NLL resummed cumulative distribution $\Sigma(v)$
\begin{equation}
  \label{eq:Sigma}
\Sigma(v) =\frac{1}{\sigma}\int_0^v dv' \frac{d \sigma(v')}{d v'},
\end{equation}
it is enough to consider an ensemble of soft-collinear partons,
emitted independently off the hard legs, together with the
corresponding virtual corrections, as follows:
\begin{equation}
  \label{eq:Sigma-start-1}
  \Sigma(v) = {\cal H}(Q^2) \sum_{n=0}^{\infty} \frac{1}{n!} 
  \int\prod_i [dk_i] M^2(k_i)\,\Theta\left(v-V(\{\tilde p\},k_1,\dots,k_n)\right)\,.
\end{equation}
Here ${\cal H}(Q^2)$ represents virtual corrections to the Born
process, normalised to the total cross section $\sigma$, and $[dk]
M^2(k)$ is the one-gluon emission probability
\begin{equation}
  \label{eq:M2}
  [dk] M^2(k)= dz^{(1)}dz^{(2)}\frac{d\phi}{2\pi}
  \frac{dk_t^2}{k_t^2} \delta\left(z^{(1)} z^{(2)}-\frac{k_t^2}{Q^2}\right)\frac{\alpha^{\rm CMW}_s(k_t) C_F}{4\pi} \frac{z^{(1)} p_{gq}(z^{(1)})}{C_F}
 \frac{z^{(2)} p_{gq}(z^{(2)})}{C_F}\,,
\end{equation}
with\footnote{The azimuthal dependence of the squared amplitude can be
  ignored in the quark-initiated branching. In hadron-hadron and
  hadron-lepton collisions the primary branching $g\to gg$ may occur,
  and the corresponding azimuth-unaveraged splitting functions must be
  used for a NNLL resummation~\cite{Catani:2010pd}. However, in
  special configurations like colour-singlet production, this
  azimuthal dependence contributes at most at N$^3$LL.}
\begin{equation}
  \label{eq:pgq}
  p_{gq}(z) =C_F \frac{1+(1-z)^2}{z}\,.
\end{equation}
Notice that $\alpha^{\rm CMW}_s(k_t)$ is the QCD coupling in the CMW
scheme~\cite{Catani:1990rr}. In this scheme the QCD coupling is
defined as the strength of the soft radiation, inclusive in its
branchings, and is related to the coupling in the $\overline{\rm MS }$
scheme ($\alpha_s=\alpha_s^{\overline{\rm MS}}$) by
\begin{equation}
\label{eq:CMW-scheme}
\alpha^{\rm CMW}_s(k_t) = \alpha_s(k_t) \left (1 +  \frac{\alpha_s(k_t)}{2\pi} K\right)\, +\cO{\alpha^3_s(k_t)},
\quad
K = \left(\frac{67}{18}-\frac{\pi^2}{6}\right)C_A - \frac{5}{9} n_f\,.
\end{equation}
The constant $K$ is a remainder of the cancellation of infrared and
collinear singularities between unresolved real
emissions\footnote{For a definition of resolved and unresolved
  emissions see text after Eq.~\eqref{eq:Sigma-start}.} and virtual
corrections. This term gives NLL contributions starting at order
$\alpha_s^2L^2$, which are universal for all rIRC safe observables,
and proportional to the two-loop cusp anomalous
dimension~\cite{Banfi:2004yd}. The CMW scheme is an effective way of
incorporating such corrections into a redefinition of the coupling.

The soft-collinear limit of eq.~(\ref{eq:M2}) is obtained by taking
the limit $z^{(1)},z^{(2)}\to 0$, giving
\begin{equation}
  \label{eq:matrix-element}
[dk]M^2_{\rm sc}(k)= \sum_{\ell=1,2} 2 C_\ell \frac{\alpha^{\rm CMW}_s(k_{t})}{\pi}\frac{dk_{t}}{k_{t}} 
 d\eta^{(\ell)}\, \Theta\left(\ln\left(\frac{Q}{k_{t}}\right)-\eta^{(\ell)}\right) \Theta(\eta^{(\ell)})
\frac{d\phi}{2\pi}\,,
\end{equation}
where $C_\ell$ is the Casimir relative to leg $\ell$ ($C_F$ in the
present case) and $\eta^{(\ell)}$ is the rapidity with respect to leg
$\ell$, as defined after eq.~\eqref{eq:v-scaling}.


%
Notice that all integrals in eq.~(\ref{eq:Sigma-start-1}), as well as
the function ${\cal H}(Q^2)$, are to be considered as suitably
regulated, for instance using dimensional regularisation.
At NLL the observable is well
approximated by its soft-collinear scaling~\eqref{eq:v-scaling}. We
thus decide to rewrite eq.~\eqref{eq:Sigma-start-1} as
\begin{align}
  \label{eq:Sigma-start}
  \Sigma(v) &= {\cal H}(Q^2) \sum_{n=0}^{\infty} \frac{1}{n!} 
  \int\prod_i [dk_i] M^2(k_i)\,\{\Theta\left(v-\Vsc(\{\tilde
    p\},k_1,\dots,k_n)\right) \notag\\
& + \left[ \Theta\left(v-V(\{\tilde
    p\},k_1,\dots,k_n)\right) -\Theta\left(v-\Vsc(\{\tilde
    p\},k_1,\dots,k_n)\right) \right]\}\,,
\end{align}
where $\Vsc(\{\tilde p\},k_1,\dots,k_n) $ denotes the observable with
all emissions treated as if they were soft and collinear.  We decide
to divide the integrals in the real term into a contribution due to
emissions with $\Vsc(\{\tilde p\},k)> \epsilon v$ (that we refer to as
{\it resolved}), and one due to emissions with $\Vsc(\{\tilde p\},k)<
\epsilon v$ (that we refer to as {\it unresolved}). Here $\epsilon$ is
a small parameter, that can be chosen such that $\epsilon \ll1$ with
$\ln(1/\epsilon) \ll \ln(1/v)$.
Because of rIRC safety, the latter can be ignored in computing the
observable, up to power-suppressed corrections ${\cal O}(v)$. Due to
the factorised form of the multi-gluon matrix element in
eqs.~(\ref{eq:Sigma-start-1}) and~(\ref{eq:Sigma-start}), at NLL the
contribution of unresolved emissions fully exponentiates, leading to
\begin{align}
  \label{eq:Sigma-NLL}
  \Sigma(v) &= {\cal H}(Q^2) e^{\int^{\epsilon v} [dk] M^2(k)}
  \sum_{n=0}^{\infty} \frac{1}{n!}  \int_{\epsilon v}\prod_i [dk_i]
  M^2(k_i)\,\{\Theta\left(v-\Vsc(\{\tilde
    p\},k_1,\dots,k_n)\right) \notag\notag\\
  & + \left[ \Theta\left(v-V(\{\tilde p\},k_1,\dots,k_n)\right)
    -\Theta\left(v-\Vsc(\{\tilde p\},k_1,\dots,k_n)\right)
  \right]\}\,,
\end{align}
where we have used the shorthand notations
\begin{equation}
\begin{split}
  \label{eq:R-epsilon}
  \int^{\epsilon v}\!\! [dk] \, M^2(k) = \int [dk] \, M^2(k)\,\Theta\left(\epsilon v - \Vsc(\{\tilde p\},k)\right)\,,\\
  \int_{\epsilon v}\!\! \prod_i [dk_i] \, M^2(k_i) = \prod_i\int
  [dk_i] \, M^2(k_i)\,\Theta\left( \Vsc(\{\tilde p\},k_i)-\epsilon
    v\right)\,.
\end{split}
\end{equation}
The combination of the unresolved emissions with the virtual
corrections in $\mathcal{H}(Q^2)$ gives rise to a Sudakov exponent,
representing the probability of having no emissions with $V_{\rm
sc}(\{\tilde p\},k_i)>\epsilon v$, which at NLL accuracy (\ie neglecting
corrections of relative order $\alpha_s$) reads
\begin{equation}
\label{eq:SudakovNLL}
{\cal H}(Q^2)e^{\int^{\epsilon v} [dk] M^2(k)}\simeq e^{-R(\epsilon v)}\,, 
\end{equation}
where
\begin{equation}
\label{eq:Rad-def}
R(\epsilon v)\equiv 
 \int [dk] \, M^2(k)\,\Theta\left( \Vsc(\{\tilde p\},k)-\epsilon v\right)= R(v) + \int^v_{\epsilon v} [dk] M^2(k)\,.
\end{equation}
In eqs.~(\ref{eq:Sigma-NLL}),
(\ref{eq:R-epsilon}),~\eqref{eq:SudakovNLL} and (\ref{eq:Rad-def}) any
integral over the single-emission's matrix element $[dk] M^2(k)$ has
to be interpreted as follows
\begin{align}
  \label{eq:single-emsn}
  & [dk] M^2(k) \Theta \left(\Vsc(\{\tilde{p}\},k)-\bar v\right) =
  [dk] M_{\rm sc}^2(k) \sum_{\ell=1,2} \Theta\left( d_\ell
    \,g_\ell(\phi) \left(\frac{k_t}{Q}\right)^a e^{-b_\ell
      \eta^{(\ell)}}-\bar v\right)\Theta(\eta^{(\ell)}) \notag\\&+
  \sum_{\ell=1,2}
  \frac{dk_t^2}{k_t^2}\frac{dz^{(\ell)}}{z^{(\ell)}}\left(
    z^{(\ell)}p_\ell(z^{(\ell)})-2
    C_\ell\right)\frac{\alpha_s(k_t^2)}{2\pi}\Theta\left(\frac{d_\ell
      \,g_\ell(\phi)}{(z^{(\ell)})^{b_\ell}}
    \left(\frac{k_t}{Q}\right)^{a+b_\ell}-\bar v \right)\,,
\end{align}
where in the second line we made the replacement
$e^{-\eta^{(\ell)}}=k_t/(Q z^{(\ell)})$. 
Furthermore, the two step functions in eq.~\eqref{eq:single-emsn} have
to be expanded in order to avoid power suppressed contributions and
undesired subleading logarithmic terms. At NLL, one can perform the
following approximations in computing the radiator:
\begin{align}
  \Theta\left(d_\ell \,g_\ell(\phi)
    \left(\frac{k_t}{Q}\right)^{a}e^{-b_\ell \eta^{(\ell)}}-\bar v
  \right) \simeq\, \Theta\left(\ln
    \left(\frac{k_t}{Q}\right)^{a}e^{-b_\ell
      \eta^{(\ell)}} -\ln \bar v\right)& \notag\\
  + \delta\left(\ln \left(\frac{k_t}{Q}\right)^{a}e^{-b_\ell
      \eta^{(\ell)}}-\ln \bar v\right)\ln d_\ell \,g_\ell(\phi)\,,
    \label{eq:exp-theta-soft-NLL}
\end{align}
\begin{align}
  \Theta\left(\frac{d_\ell \,g_\ell(\phi)}{z^{b_\ell}}
    \left(\frac{k_t}{Q}\right)^{a+b_\ell}-\bar v \right) \simeq\,
  \Theta\left(\ln\left(\frac{k_t}{Q}\right)^{a+b_\ell}-\ln \bar
    v\right) \,.
    \label{eq:exp-theta-coll-NLL}
\end{align}
This gives
\begin{equation}
  \label{eq:radiator}
  \begin{split}
    R(v) \simeq R_{\rm NLL}(v) & \equiv \int [dk] M_{\rm
      sc}^2(k)\sum_{\ell=1,2} \Theta\left(\ln
      \left(\frac{k_t}{Q}\right)^{a}e^{-b_\ell \eta^{(\ell)}} -\ln
      v\right) \Theta(\eta^{(\ell)}) \\& +
    \int [dk] M_{\rm sc}^2(k)\sum_{\ell=1,2} \ln\bar d_{\ell } \,\delta\left(\ln \left(\frac{k_t}{Q}\right)^{a}e^{-b_\ell \eta^{(\ell)}}-\ln v\right) \Theta(\eta^{(\ell)})\\
    & + \sum_{\ell=1,2}C_\ell B_\ell\int
    \frac{dk_t^2}{k_t^2}\frac{\alpha_s(k_t^2)}{2\pi}\Theta\left(\left(\frac{k_t}{Q}\right)^{a+b_\ell}-v
    \right)\,,
  \end{split}
\end{equation}
where
\begin{equation}
\label{eq:lndbar}
\ln\bar d_\ell = \int_0^{2\pi} \frac{d\phi}{2\pi}\ln d_\ell g_\ell(\phi)\,,
\end{equation}
and
\begin{equation}
  C_\ell B_\ell = \int_0^1 \frac{d z}{z}\left(z p_{gq}(z) - 2
    C_\ell\right)\,.
\end{equation}
In our case $C_\ell B_\ell = -3/2 C_F$. $R_{\rm NLL}(v)$ can be
parametrised as
\begin{equation}
  \label{eq:R-NLL}
  R_{\rm NLL}(v)=-L g_1(\lambda) -g_2(\lambda)\,,
\end{equation}
where $L=\ln(1/v)$, $\lambda = \alpha_s(Q) \beta_0 L$ and $\beta_0 =
(11 N_c - 4 n_f T_F)/(12\pi)$. The functions $g_1$ and $g_2$ can be
written in terms of the constants $a$, $b_\ell$, $d_\ell$ and the
functions $g_\ell(\phi)$ and are given in
Appendix~\ref{sec:radiator}.\footnote{In Appendix~\ref{sec:radiator}
  we use a modified definition of $L=\ln(x_V/v)$, and hence of
  $\lambda$, in order to estimate theoretical uncertainties from
  higher-order logarithmic corrections by varying $x_V$.}  We notice
that all integrals over real emissions in eq.~(\ref{eq:Sigma-NLL})
involve an upper and a lower bound on each $V_{\rm sc}(\{\tilde
p\},k_i)$ such that $\epsilon v < \Vsc(\{\tilde p\},k_i)\lesssim
v$. We remind that $\epsilon$ is a small parameter satisfying
$\epsilon \ll1$ and $\ln(1/\epsilon) \ll \ln(1/v)$. The upper bound
comes implicitly from the constraint that the observable is smaller
than $v$. Therefore the real-emission phase space is at most
single-logarithmic, unlike the corresponding phase space region
considered in the radiator $R(v)$, which is double logarithmic.
As a consequence, for real emissions, at NLL accuracy, one can
consider only the soft-collinear matrix element (i.e. the first line
of eq.~\eqref{eq:single-emsn}) and replace the observable with its
soft-collinear approximation, i.e.\ neglect the term in the second
line of eq.~\eqref{eq:Sigma-NLL}. This leads to
\begin{equation}
\begin{split}
\label{eq:real-rad-NLL}
\Sigma(v) = e^{-R_{\rm NLL}(v)}e^{- \int^v_{\epsilon v} [dk] M_{\rm
    sc}^2(k)} &\sum_{n=0}^{\infty} \frac{1}{n!}  \int_{\epsilon
  v}\prod_i [dk_i] M_{\rm sc}^2(k_i)\,\Theta\left(v-\Vsc(\{\tilde
  p\},k_1,\dots,k_n)\right)\,.
\end{split}
\end{equation}
Here the second exponential factor provides the {\em unresolved}
emissions that cancel the dependence on the cutoff $\epsilon$ in the
{\em resolved} real emissions, so that the result is finite and
independent of $\epsilon$.  This gives
\begin{equation}
\Sigma(v) \simeq e^{-R_{\rm NLL}(v)}\fullF(v)\,,
\end{equation}
where the function $\fullF(v)$ contains NLL corrections due to an
ensemble of soft and collinear gluons, widely separated in
rapidity~\cite{Banfi:2004yd}\footnote{The contribution from a phase
  space region where two gluons are close in rapidity is suppressed by
  one power of the logarithm, hence it contributes only to NNLL and
  will be discussed later.}, and reads
\begin{equation}
  \label{eq:F}
  \fullF(v)=e^{-\int_{\epsilon v}^{v}[dk] M^2_{\rm sc}(k)}\sum_{n=0}^{\infty}\frac{1}{n!}\int_{\epsilon v} \prod_{i=1}^n [dk_i] M^2_{\rm sc}(k_i)\, \Theta\left(v-\Vsc(\{\tilde p\},k_1,\dots,k_n)\right)\,.
\end{equation}
Although the above expression has all ingredients necessary to achieve
NLL accuracy, it contains also subleading effects. We will first
explain how to eliminate them, if one seeks a pure NLL result, and
then discuss how they can be computed at NNLL accuracy in the next
section.

We parametrise the phase space in terms of $v_i=\Vsc(\{\tilde
p\},k_i)$, \ie the value that the event shape has in the presence of
each {\em individual} emission $k_i$ (eq.~\eqref{eq:v-scaling}).
First, it is convenient to divide the phase space according to whether
an emission is collinear to $p_1$ ($\eta_i >0$) or collinear to $p_2$
($\eta_i<0$). For each emission $k_i$ we introduce the rapidity
fractions $\xi_i^{(\ell)} =\eta^{(\ell)}_i/\eta_{\rm max}^{(\ell)}$
defined as the emission's rapidity divided by the largest available
rapidity for a given value of $v_i$. $\eta_{\rm max}^{(\ell)}$ is
defined as
\begin{equation}
\label{eq:etamax}
\eta_{\rm max}^{(\ell)}=\frac{1}{a+b_\ell}\ln\frac{g_{\ell}(\phi_i) d_\ell}{v_i}\,.
\end{equation}
We introduce the two functions
\begin{equation}
  \label{eq:R'1,2}
  \begin{split}
    R'_1\left(\frac{v}{d_1 g_1(\bar \phi)}\right)&= \int [dk] M^2_{\rm sc}(k) \,(2\pi) \delta(\phi-\bar \phi)\, v\delta\left(v-\Vsc(\{\tilde p\},k)\right)\theta(\eta)  \,, \\
    R'_2\left(\frac{v}{d_2 g_2(\bar \phi)}\right)& = \int [dk]
    M^2_{\rm sc}(k)\,(2\pi) \delta(\phi-\bar \phi)\,
    v\delta\left(v-\Vsc(\{\tilde p\},k)\right)\theta(-\eta)\,.
  \end{split}
\end{equation}
Finally, we introduce $R'(v,\phi)$, defined as 
\begin{equation}
  \label{eq:R'}
  R'(v,\phi) =  R'_1\left(\frac{v}{d_1 g_1(\phi)}\right)+R'_2\left(\frac{v}{d_2 g_2(\phi)}\right)\,.
\end{equation}
Using this parametrisation, we can recast the matrix element for each
emission as follows
\begin{equation}
  \label{eq:sc-emsn-1,2}
  \begin{split}
    [dk_i] M^2_{\rm sc}(k_i) & = \frac{dv_i}{v_i}
    \frac{d\phi_i}{2\pi}\sum_{\ell_i=1,2} d\xi_i^{(\ell_i)}
    \Theta(1-\xi_i^{(\ell_i)})\Theta(\xi_i^{(\ell_i)})
    R'_{\ell_i}\left(\frac{v_i}{d_{\ell_i} g_{\ell_i}(\phi_i)}\right)
    \\ & = \frac{d\zeta_i}{\zeta_i} \frac{d\phi_i}{2\pi}
    \sum_{\ell_i=1,2} d\xi_i^{(\ell_i)}
    \Theta(1-\xi_i^{(\ell_i)})\Theta(\xi_i^{(\ell_i)})
    R'_{\ell_i}\left(\frac{\zeta_i v}{d_{\ell_i}
        g_{\ell_i}(\phi_i)}\right) \,,
  \end{split}
\end{equation}
where $\zeta_i=v_i/v$ is defined as the ratio of the observable's
value corresponding to the $i^{\rm th}$ emission to the actual
observable's value $v$.
We can now exploit a fundamental property of event shapes. Given a set
of emissions $\{k_1,\dots,k_n\}$, as long as one keeps $v_i$, $\phi_i$
and the leg $\ell_i$ to which $k_i$ is collinear fixed, the value of
an event shape does not depend on $\xi_i^{(\ell_i)}$, which can be
then integrated out analytically. This makes it possible to simplify
$\mathcal{F}(v)$ as follows
\begin{multline}
  \label{eq:F-simplified}
    \fullF(v)=e^{-\int\frac{d\phi}{2\pi}\int_{\epsilon}^{1} \frac{d\zeta}{\zeta}R'(\zeta v,\phi)}
\sum_{n=0}^{\infty}\frac{1}{n!} \prod_{i=1}^n
\int_{\epsilon}^{\infty}\frac{d\zeta_i}{\zeta_i}\int_0^{2\pi}
\frac{d\phi_i}{2\pi} \times \\ \times \sum_{\ell_i=1,2} R_{\ell_i}'\left(\frac{\zeta_i v}{d_{\ell_i} g_{\ell_i}(\phi_i)}\right)\, \Theta\left(v-\Vsc(\{\tilde p\},k_1,\dots, k_n)\right)\,,
\end{multline}
where $k_1, \dots,  k_n$ are now soft and collinear emissions
with an arbitrary rapidity fraction. As a last simplification, we can
expand each $R'_\ell$ around $v$ 
\begin{equation}
  \label{eq:R'expanded}
  R'_\ell\left(\frac{\zeta v}{d_\ell g_\ell(\phi)}\right) = R'_\ell(v)+\mathcal{O}(R''_\ell)\,\qquad R''_\ell=-v\frac{dR'_\ell(v)}{dv}\,,
\end{equation}
and neglect all contributions of order $R''_\ell$. These constitute a
NNLL leftover that will be specifically addressed in
section~\ref{sec:rap}. 
Notice that
\[
R'_\ell(v) = \int[dk] M^2_{\rm sc}(k)\,
v\delta\left(v-\frac{\Vsc(\{\tilde p\},k)}{d_\ell
  g_\ell(\phi)}\right)\theta(\eta)
\]
does not depend on $\phi$ and on $d_\ell$ any more. This function can
be further split as $R'_\ell(v) = R'_{\mathrm{NLL},\ell}(v)+\delta
R'_{\mathrm{NNLL},\ell}(v)$, where $R'_{\mathrm{NLL},\ell}$ and
$\delta R'_{\mathrm{NNLL},\ell}$ are defined in eqs.~\eqref{eq:R'ell-NLL}
and~\eqref{eq:R'ell-NNLL}, respectively. 
The NNLL term $\delta R'_{\mathrm{NNLL},\ell}$ contains running
coupling effects as well as the contribution of the cusp anomalous
dimension through the CMW scheme. This, as explained earlier, encodes
the contribution of an inclusive soft-gluon splitting. At NNLL one has
to take into account the non-inclusive nature of the observable in the
presence of the branching of a soft gluon. This non-inclusive
correction is contained in the full set of NNLL contributions (see
Section~\ref{sec:correl}), therefore the choice of the CMW scheme in
the {\it resolved} real emission becomes irrelevant (see
Section~\ref{sec:correl}).
With this simplification, ${\cal F}(v) \simeq \FNLL(\lambda)$
where
\begin{equation}
  \label{eq:F-NLL}
  \begin{split}
  \FNLL(\lambda) = \int \dZ \, \Theta\left(1-\lim_{v\to 0}\frac{\Vsc(\{\tilde p\},\{k_i\})}{v}\right)\,,
  \end{split}
\end{equation}
and subleading terms have been neglected.
In eq.~\eqref{eq:F-NLL} we have introduced the average of a function
$G(\{\tilde p\},\{k_i\})$ over the measure $d {\cal Z}$:
\begin{equation}
\label{eq:dZ}
\begin{split}
\int \dZ  G(\{\tilde p\},\{k_i\})=\epsilon^{R'_{\mathrm{NLL}}}
   \sum_{n=0}^{\infty}\frac{1}{n!} \prod_{i=1}^n
    \int_{\epsilon}^{\infty} \frac{d\zeta_i}{\zeta_i}\int_0^{2\pi}
   \frac{d\phi_i}{2\pi} \sum_{\ell_i=1,2} R'_{\mathrm{NLL}, \ell_i}G(\{\tilde p\},k_1,\dots,k_n)\,,
\end{split}
\end{equation}
where $R'_{\rm NLL}=R'_{\rm NLL, 1}+R'_{\rm NLL, 2}$.  Note that the
dependence on the regulator $\epsilon$ cancels in
eq.~\eqref{eq:dZ}. The limit $v\to 0$ in eq.~\eqref{eq:F-NLL} is
necessary to remove contributions that are power suppressed in $v$.
The existence of this limit in the step function of
eq.~\eqref{eq:F-NLL} is guaranteed by the rIRC safety property of
event shapes here considered, which implies that the quantity
$\Vsc(\{\tilde p\},k_1,\dots,k_n)/v$ is independent of $v$, with
corrections that scale as a power of $v$.  To conclude, neglecting all
terms beyond NLL accuracy, we can write $\Sigma(v)$ in the form
\begin{equation}
  \label{eq:Sigma-NLL-final}
  \Sigma(v) = e^{L g_1(\lambda)+g_2(\lambda)} \FNLL(\lambda)\,.
\end{equation}

\section{NNLL resummation}
\label{sec:nnll}
In this section we extend the above treatment to NNLL, illustrating
how the various corrections arise.
We will first discuss the general structure of the NNLL resummation
and then derive the relevant corrections.

\subsection{Logarithmic counting for the resolved real emissions}
Before extending the above treatment to NNLL, it is worth recalling
how, given rIRC safety of the observable, one can define a logarithmic
hierarchy in the resolved real emissions, and hence give a precise
definition of the multiple emissions function $\fullF(v)$ at a given
logarithmic order.
We start by considering an ensemble of $n$ soft emissions. The squared
matrix element can be expressed iteratively as a sum of products of
matrix elements with a lower number of emissions (from $1$ to $n-1$)
plus an irreducible remainder $\tilde M^2(k_1,...,k_n)$. The first few
steps of this iterative definition read
\begin{align}
 M^2(k_1)&= \tilde M^2(k_1)\,,\notag\\
M^2(k_1,k_2)&= M^2(k_1) M^2(k_2) + \tilde M^2(k_1,k_2)\,,\notag\\
M^2(k_1,k_2,k_3)&= M^2(k_1) M^2(k_2) M^2(k_3)+ (\tilde M^2(k_1,k_2)
M^2(k_3) + {\rm perm.}) + \tilde M^2(k_1,k_2,k_3)\,,\notag\\
M^2(k_1,...,k_n)&=\dots 
\label{eq:correl-decomposition}
\end{align}
The product of single-emission matrix elements clearly defines the
abelian contribution, while non-abelian colour factors are associated
with the $\tilde M^2(k_1,...,k_m)$ squared amplitudes. This makes each
single $\tilde M$ in the above decomposition invariant under gauge
transformations. The $\tilde M^2(k_1,...,k_m)$ matrix elements for
more than one emission describe the probability of emitting $m$
colour-connected soft partons, and they are therefore suppressed if
the involved emissions are very far in rapidity from each other. We
will refer to $\tilde M^2(k_1,k_2)$ as the {\it double-correlated}
contribution to the squared amplitude for multiple emissions. We will
label the correlated squared matrix elements with more than two
emissions in an analogous fashion.
We now study the logarithmic structure of each of the terms in
Eqs.~\eqref{eq:correl-decomposition}. Each resolved real emission
(i.e. an emission that contributes to the observable) is defined by
requiring that $V_{\rm sc}(\{\tilde p\},k_i)>\epsilon v$, where
$\epsilon$ is independent of $v$ because of rIRC safety. This
condition poses a lower bound on the resolved emission's phase space
which can potentially only give rise to a single logarithm of $v$ (see
for instance Eq.~\eqref{eq:sc-emsn-1,2}). When several emissions are
considered, the same argument applies, so that each emission can at
most contribute with a single logarithm. This is ensured by rIRC
safety since this condition implies that the observable will have the
same scaling independently of the number of emissions, and therefore
the condition $V_{\rm sc}(\{\tilde p\},k_i)>\epsilon v$ will still
impose a lower cutoff for all resolved emissions.
The unresolved emissions below this limit (i.e. $V_{\rm sc}(\{\tilde
p\},k_i)<\epsilon v$) can be ignored in the observable evaluation and
their role is simply to cancel the virtual IRC singularities. They
contribute exclusively to the Sudakov radiator and therefore we do not
need to consider them here.
With the above property we can immediately see that a product of $n$
independent emission matrix elements in
Eq.~\eqref{eq:correl-decomposition} gives rise at most to a
$\alpha_s^nL^n$ (i.e. a NLL) contribution, where $L=\ln 1/v$.

We now consider the double-correlated $\tilde M^2(k_1,k_2)$ term. It
involves a soft-gluon splitting into either a $q\bar q$ or $gg$ pair
and it could potentially give rise to a $\alpha_s^2 L^3$ term
($\alpha_s L$ associated with the emission of the parent gluon, and at
most two extra logarithms coming from its splitting). However, again
due to rIRC safety (see for instance Section 2.2.4 of
ref.~\cite{Banfi:2004yd} for the relevant properties), one can see
that the splitting of the parent gluon does not give rise to
additional logarithms, leaving us with a NNLL term $\alpha_s^2 L$. The
same argument can be applied to terms with more than two correlated
partons, and can be used to show that they are at most
N$^3$LL. Therefore, the sole rIRC safety property of the observable
allows one to define a logarithmic hierarchy in the multiple emissions
function and to define the relevant configurations that contribute to
a given logarithmic order.
The very same argument applies to the case of one or more emissions
emitted collinearly to the Born leg with high momentum.
Therefore, if we want to limit ourselves to, for instance, NLL
(i.e. $\alpha_s^n L^n$ terms in the multiple emissions function
$\fullF(v)$) it is sufficient to consider an ensemble of
soft-collinear independent emissions, since any configuration beyond
this one would just be at most NNLL. For a NNLL treatment, in
addition, one has to include the contribution of a single splitting of
a soft gluon (following the above argument it is easy to see that
configurations with more than one splitting are subleading), and a
single hard collinear emission. This treatment can be extended to
higher orders in a very systematic way.

In addition to the matrix element approximation, we would like to
approximate the resolved emission's phase space in order to neglect
any effects in $\fullF(v)$ which are beyond the logarithmic accuracy
that we want to achieve. We stress that this class of approximations
is not strictly necessary for the resummation, since their only
purpose is to ensure that $\fullF(v)$ is free of any contamination
from subleading effects.
For instance, at NLL, we can approximate the rapidities of all
soft-collinear emissions with the kinematic limit as done in
Section~\ref{sec:nll}, and treat the observable in the pure
soft-collinear approximation, all corrections being at most NNLL.
For a NNLL resummation, these approximations are of course not valid
anymore and one has to repeat the calculation without making
them. Alternatively, one can simply compute the NNLL corrections
associated with these approximations with respect to the NLL function
$\FNLL(\lambda)$, as it will be explained in detail in the next
section.

The last ingredient that one needs to go beyond NLL is the Sudakov
radiator. This function has the role of cancelling the infrared and
collinear singularities associated with the unresolved emissions
(i.e. $V_{\rm sc}(\{\tilde p\},k_i)<\epsilon v$) against the virtual
corrections. At NLL its structure is remarkably simple since the
unresolved real emissions fully exponentiate in the observable's space
and the cancellation of singularities is explicit. Beyond this order,
one needs to work out the exact details of real-virtual cancellations
(for instance, through renormalisation group evolution equations). We
will not present a general expression for the radiator in this
article, but we will limit ourselves to show that it only depends on
the scaling of the observable in the presence of a {\it single} soft
{\it and} collinear dressed (i.e. inclusive in its branchings)
emission. Therefore, we will show that it is universal for all
observables which have the same soft-collinear parametrisation in the
single emission case, i.e. the same $a$ and $b_\ell$ coefficients in
Eq.~\eqref{eq:v-scaling}.

\subsection{Structure of the NNLL resummation}
Using the arguments outlined in the previous section, we now derive
the general form of NNLL corrections.  We start by recalling the
procedure which lead to the NLL result. On the one hand, we
approximated the matrix element and the phase space in all emissions
appearing in the multiple emissions function of eq.~\eqref{eq:F},
neglecting subleading corrections due to the exact rapidity bound for
each resolved soft and collinear emission (see
eq.~(\ref{eq:R'expanded})), and the correct description of the
hard-collinear region (neglecting the second line of
eq.~\eqref{eq:single-emsn}). On the other hand, we replaced the
observable with its soft-collinear parametrisation $\Vsc$, neglecting
the second line of eq.~\eqref{eq:Sigma-NLL}.  We remark that, at
NNLL accuracy, these approximations have to be relaxed for a single
emission at a time, since relaxing each approximation gives rise a
correction of relative order $\alpha_s$. This implies that
configurations in which we correct more than one emission lead to
contributions beyond NNLL, that can be neglected accordingly.

A set of NNLL corrections arises from the first term of
eq.~(\ref{eq:Sigma-NLL}):
\begin{equation}
\begin{split}
\label{eq:real-rad-NNLL-1}
e^{-R_{\rm NLL}(v)}e^{- \int^v_{\epsilon v} [dk] M^2(k)} &\sum_{n=0}^{\infty} \frac{1}{n!} 
  \int_{\epsilon v}\prod_i [dk_i] M^2(k_i)\,\Theta\left(v-\Vsc(\{\tilde
    p\},k_1,\dots,k_n)\right)\,,
\end{split}
\end{equation}
where $R_{\rm NLL}$ is defined in eq.~\eqref{eq:R-NLL}.
Besides the NLL multiple emissions function $\fullF_{\rm
  NLL}(\lambda)$ of eq.~\eqref{eq:F-NLL} derived in
Sec.~\ref{sec:nll}, eq.~\eqref{eq:real-rad-NNLL-1} contains
corrections due both to the hard-collinear term of the matrix element
(given by the second line of eq.~\eqref{eq:single-emsn}), and to the
correct rapidity bounds, which at NLL are the same for all emissions
(see eq.~\eqref{eq:R'expanded}). Such corrections result in the two
NNLL contributions $\delta \mathcal{F}_{\rm hc}$
(Sec.~\ref{sec:recoil}) and $\delta \mathcal{F}_{\rm sc}$
(Sec.~\ref{sec:rap}), respectively.

Another category of NNLL corrections is contained in the remaining
term of eq.~(\ref{eq:Sigma-NLL}), namely
\begin{equation}
\begin{split}
\label{eq:real-rad-NNLL-2}
e^{-R_{\rm NLL}(v)}e^{- \int^v_{\epsilon v} [dk] M_{\rm sc}^2(k)} \sum_{n=0}^{\infty} \frac{1}{n!} 
  \int_{\epsilon v}\prod_i [dk_i] M^2(k_i) &\left[ \Theta\left(v-V(\{\tilde
    p\},k_1,\dots,k_n)\right) \right.\\ &\left.-\Theta\left(v-\Vsc(\{\tilde
    p\},k_1,\dots,k_n)\right) \right]\,,
\end{split}
\end{equation}
where we need to relax the soft-collinear approximation made for the
observable when an arbitrary emission becomes hard-collinear or is
emitted at small rapidities (large angles).  We stress that, at NNLL
accuracy, it is enough to consider an ensemble of soft and collinear
emissions, plus a single extra emission which is free to probe both
the hard-collinear and the soft-wide-angle region of the phase
space. Configurations containing more than one soft-wide-angle or
hard-collinear real emission are subleading. We can then expand
further the first step function in eq.~\eqref{eq:real-rad-NNLL-2} in
order to take into account the correct behaviour of the observable in
these limits for a single emission of the ensemble. The corresponding
NNLL corrections are: a recoil correction $\delta\mathcal{F}_{\rm
  rec}$ (computed in Sec.~\ref{sec:recoil}) which is due to the exact
kinematics of a hard-collinear emission which recoils against the
soft-collinear ensemble; a soft-wide-angle correction $\delta
\mathcal{F}_{\rm wa}$ (computed in Sec.~\ref{sec:soft-large-angle})
which is due to a soft emission that spans the whole rapidity range; a
correlated correction $\delta\mathcal{F}_{\rm correl}$ (computed in
Sec.~\ref{sec:correl}) to the inclusive treatment of the soft gluon
decay in the matrix element (encoded in the scheme of the running
coupling in the radiator $R(v)$).
An important point to stress is that the soft-collinear approximation
$\Vsc(\{\tilde p\},k_1,\dots,k_n)$ guarantees that all NNLL
corrections arising from eq.~\eqref{eq:real-rad-NNLL-2} are well
defined and finite when the corrected emission becomes unresolved.

One last NNLL contribution is due to the correction to the NLL Sudakov
radiator of eq.~\eqref{eq:radiator}. At NLL, the radiator encodes the
contribution of unresolved real emissions $k_i$ with $\Vsc(\{\tilde
p\},k_i)<\epsilon v$ and corresponding virtual corrections. Moreover,
each emission is considered to be inclusive in its two-parton
branchings.  Analogously, the NNLL Sudakov radiator has to include the
effect of the inclusive soft three-partons correlation, which can be
absorbed in a redefinition of the running coupling analogously to what
is done at NLL, together with the correct matrix element for an
inclusive double collinear emission. Furthermore, it contains exact
${\cal O}(\alpha_s)$ corrections surviving the poles cancellation
between real and virtual corrections. In formulae, we introduce a NNLL
radiator $R_{\rm NNLL}(v)$ through the replacement
\begin{equation}
{\cal H}(Q^2)e^{\int^{\epsilon v}[dk]M^2(k)} \rightarrow e^{-R_{\rm NNLL}(v) + \int_{\epsilon v}^v [dk] M^2(k)},
\end{equation}
where
\begin{align}
  \label{eq:radiator-NNLL}
  R_{\rm NNLL}(v) &= \int [dk] M_{\rm sc}^2(k) \Theta
  \left(\Vsc(\{\tilde{p}\},k)-v\right) \notag\\
&+ \sum_{\ell=1,2}\int
  \frac{dk_t^2}{k_t^2}\int_0^1\frac{dz}{z}\left(zp_\ell(z)-2C_\ell \right)\frac{\alpha_s(k_t^2)}{2\pi}\Theta\left(\frac{d_\ell \,g_\ell(\phi)}{z^{b_\ell}} \left(\frac{k_t}{Q}\right)^{a+b_\ell}-v
    \right) + \frac{\alpha_s(Q)}{\pi} h(\lambda)\,.
\end{align}
The function $\alpha_s(Q) h(\lambda)/\pi$ contains the contribution of
the triple-correlated splitting, the double hard-collinear correction
and additional ${\cal O}(\alpha_s)$ constant terms arising from
real-virtual cancellations, and corresponding running coupling
effects.  Eq.~\eqref{eq:radiator-NNLL} contains some power suppressed
terms due to the integration limits of the non-singular phase space
variables, i.e. $\phi$ in the soft limit and $\phi$, $z$ in the
hard-collinear limit. In order to neglect these terms we have relaxed
the lower bound in the $z$ integration relative to the hard-collinear
limit, and set it to zero (the physical bound being
$z>k_t/Q$). Moreover, in order to neglect power-suppressed and
subleading contributions, we can expand the two $\Theta$-functions of
eq.~\eqref{eq:radiator-NNLL} as follows:\footnote{For the NLL
  radiator, it was sufficient to consider the first two terms in the
  r.h.s. of eq.~\eqref{eq:exp-theta-soft}, and the first in the
  r.h.s. of eq.~\eqref{eq:exp-theta-coll}, respectively.}
\begin{align}
\Theta\left(d_\ell \,g_\ell(\phi)
  \left(\frac{k_t}{Q}\right)^{a}e^{-b_\ell \eta^{(\ell)}}-v
    \right) \simeq\,
    \Theta\left(\ln \left(\frac{k_t}{Q}\right)^{a}e^{-b_\ell
        \eta^{(\ell)}} -\ln v\right)& \notag\\
  + \delta\left(\ln \left(\frac{k_t}{Q}\right)^{a}e^{-b_\ell \eta^{(\ell)}}-\ln v\right)\ln
    d_\ell \,g_\ell(\phi)+\frac{1}{2}\delta'\left(\ln\left(\frac{k_t}{Q}\right)^{a}\right. &\left.e^{-b_\ell \eta^{(\ell)}}-\ln v\right)\ln^2
    d_\ell \,g_\ell(\phi)\,,
    \label{eq:exp-theta-soft}
\end{align}
\begin{align}
\Theta\left(\frac{d_\ell \,g_\ell(\phi)}{z^{b_\ell}} \left(\frac{k_t}{Q}\right)^{a+b_\ell}-v
    \right) \simeq\,
    \Theta\left(\ln\left(\frac{k_t}{Q}\right)^{a+b_\ell}-\ln v\right)
    + \delta \left(\ln \left(\frac{k_t}{Q}\right)^{a+b_\ell}-\ln v\right)\ln
    \frac{d_\ell \,g_\ell(\phi)}{z^{b_\ell}}\,.
    \label{eq:exp-theta-coll}
\end{align}
We observe that the dependence on the normalisation $d_\ell
g_\ell(\phi)$ is a local rescaling of the observable. This induces a
local shift of the logarithm $\ln 1/v$ and gives rise to subleading
contributions at each logarithmic order. This implies that, at NNLL
accuracy, the dependence on $d_\ell g_\ell (\phi)$ in the Sudakov
radiator is completely encoded in the first two integrals of
eq.~\eqref{eq:radiator-NNLL}, and it corresponds to a shift in the
logarithms of the NLL radiator (before azimuthal integration).  An
important consequence of this is that the function $h(\lambda)$
depends exclusively on the scaling in $\eta$ (or equivalently $z$) and
$k_t$ through the $a$ and $b_\ell$ coefficients. By exploiting this
property, one can conclude that the resummations of all observables
which have the same soft-collinear scaling in $k_t$ and $\eta$
(i.e. the same $a$ and $b_\ell$ coefficients) will have the same
$h(\lambda)$ function. For example, the function $h(\lambda)$ will be
the same for thrust $1-T$, $C$ parameter, and heavy jet mass $\rho_H$,
and it can be taken from~\cite{Becher:2008cf,Monni:2011gb}.
Analogously, the function $h(\lambda)$ for the jet broadenings
 $B_T$, $B_W$, thrust major $T_M$ and oblateness $O$ is identical to
 the one relative to the $k_t$ resummation (which we take from
 ref.~\cite{Banfi:2012jm} after  replacing the constant one loop
 virtual corrections with the corresponding ones in $e^+e^- \to$
 hadrons).  Practically, the function $h(\lambda)$ can be obtained by
 computing the resummation for the reference observable (e.g. the
 thrust) leaving $h(\lambda)$ unspecified, and fixing it by equating
 the resummation obtained here to the known result in the
 literature. This is similar in spirit to what has been done for the
 jet-veto in ref.~\cite{Banfi:2012yh,Banfi:2012jm}.

 We parametrise the final NNLL Sudakov radiator as
\begin{equation}
R_{\rm NNLL}(v) = - Lg_1(\lambda) - g_2(\lambda)
-\frac{\alpha_s(Q)}{\pi}g_3(\lambda)\,.
\end{equation}
The relevant expressions for the $g_1$, $g_2$, and $g_3$ functions are
reported in Appendix~\ref{sec:radiator}.
Once all these corrections have been computed, the NNLL expression for
$\Sigma(v)$ becomes
\begin{equation}
  \label{eq:Sigma-NNLL}
  \Sigma(v) = e^{L g_1(\lambda)+g_2(\lambda) + \frac{\alpha_s(Q)}{\pi}
    g_3(\lambda)}\left[\FNLL(\lambda)+\frac{\alpha_s(Q)}{\pi}\delta \FNNLL(\lambda)\right]\,.
\end{equation}
The function 
\begin{equation}
  \label{eq:deltaF}
  \delta \FNNLL = \delta \mathcal{F}_{\rm sc}
+\delta
  \mathcal{F}_{\rm hc}+\delta \mathcal{F}_{\rm rec}+\delta
  \mathcal{F}_{\rm wa}+\delta \mathcal{F}_{\rm correl}\,,
\end{equation}
represents NNLL corrections due to real radiation, and it will be
extensively discussed in the rest of this section.

Before deriving the relevant NNLL corrections to the real radiation it
is worth making an important remark. The whole resummation procedure
defined in the present section depends on a specific choice of the
variable on which the cutoff $\epsilon$ is applied. This choice is
reflected in the exponentiated part of the resummed cross section.
Our default choice is to define unresolved emissions as those for
which $\Vsc(\{\tilde p\},k) < \epsilon v$, where $\Vsc$ is defined by
eq.~\eqref{eq:v-scaling}. This choice is clearly arbitrary and one
could equally derive the same resummed results (that will be anyway
independent of the cutoff $\epsilon$) with a different definition for
the unresolved contributions. Different choices will simply lead to
different NLL terms (and beyond) in the Sudakov exponent and in the real
corrections described by the multiple emissions function, but will not
affect the final result which does not depend on such a definition.
In the present article we decide to work in the soft-collinear
prescription in which the cutoff $\epsilon$ is applied on the
soft-collinear approximation of the observable for a generic emission
$k_i$. This prescription has two advantages. On the one hand it allows
one to expand the multiple emissions function around the NLL result,
which is simply determined by the soft-collinear approximation
(meaning that the $\Vsc$ approximation of eq.~\eqref{eq:v-scaling} is
enough to account for all NLL contributions). It also ensures that all
NNLL corrections to the multiple emissions function are finite without
further regulators since the singularities of any unresolved emission
are encoded in the soft-collinear approximation.
On the other hand it allows us to define the NNLL
function $h(\lambda)$ in such a way that it is independent of
the observable's normalisation $d_\ell g_\ell(\phi)$ and it only
depends on the $a$ and $b_\ell$ coefficients. 
As stated above, this implies that the
function $ h(\lambda)$ is universal for all observables which
have the same $a$ and $b_\ell$ scaling in the soft-collinear
region.

\subsection{NNLL contributions due to resolved emissions}
\label{sec:nnllreal}

In this section we explicitly derive all corrections to the multiple
emission function $\mathcal{F}(v)$ necessary to achieve NNLL accuracy
for the cumulative distribution $\Sigma(v)$ for a generic event-shape
observable $v$. In order to do this we have to recall the basic
assumptions used to obtain eq.~(\ref{eq:Sigma-NLL-final}). They are:
\begin{itemize}
\item gluon splitting in $R(v)$ is treated inclusively;
\item each real emission $k_i$ contributing to $\mathcal{F}(v)$ is
  soft, collinear, and such that $\epsilon v < \Vsc(\{\tilde p\},k_i) <
  v$;
\item the rapidity bound of all emissions contributing to
  $\mathcal{F}(v)$ is the same.
\end{itemize}
By relaxing any of these approximations valid at NLL accuracy, we obtain
a number of NNLL corrections induced by real radiation, and introduced
in the previous chapter. We will derive them in the following order:
\begin{enumerate}
\item exact rapidity bound and running coupling corrections to the
  soft and collinear function $\mathcal{F}(v)$ ($\delta
  \mathcal{F}_{\rm sc}$);
\item one of the emissions $k_i$ is collinear but not soft, 
  generating hard-collinear ($\delta \mathcal{F}_{\rm hc}$) and 
  recoil ($\delta\mathcal{F}_{\rm rec}$) corrections;
\item one of the emissions $k_i$ is soft but at wide angle ($\delta
  \mathcal{F}_{\rm wa}$);
\item gluon decay is treated non-inclusively, giving rise to a
  correlated-emission correction ($\delta\mathcal{F}_{\rm correl}$).
\end{enumerate}

The necessary amplitudes to compute $\delta \mathcal{F}_{\rm NNLL}$
are given by the independent emission probability of
eq.~(\ref{eq:M2}), and the probability of a soft gluon branching into
either two gluons or a quark-antiquark pair (correlated emission). In
fact, for the real ensemble the observable's value $V(\{\tilde p\}, \{
k_i\})$ is bound both from above and from below. This reduces the
phase space of the real emissions to a strip which contributes with
one fewer logarithm at each order of $\alpha_s$ with respect to the
Sudakov radiator.
Therefore, to obtain the whole set of NNLL real corrections, it is
enough to use the same probability amplitudes which appear in the
definition of the Sudakov exponent at NLL, \ie the independent soft
and/or collinear emission probability, and the correlated soft-gluon
splitting.

\subsubsection{Soft-collinear NNLL contributions}
\label{sec:rap}
The first NNLL correction we consider arises from $\mathcal{F}(v)$,
when we take into account the exact rapidity bounds for a single
emission in the generated soft-collinear ensemble. At NLL, the correct
rapidity limit for the emission $k_i$,
\begin{equation}
\label{eq:rap-bound}
\eta_i^{(\ell_i)} < \frac{1}{a+b_{\ell_i}}\ln\frac{g_\ell(\phi_i) d_\ell}{\zeta_i v}\,,
\end{equation}
was effectively replaced by $1/(a+b_{\ell_i})\ln(1/v)$ through the
expansion of eq.~\eqref{eq:R'expanded}. NNLL corrections to this
approximation are obtained by considering the next term in the
expansion of $R'_\ell$, both in real and in virtual corrections, as
follows
\begin{equation}
  \label{eq:R'exp-NNLL}
  R'_\ell\left(\frac{\zeta v}{d_\ell g_\ell(\phi)}\right) \simeq R'_{\rm NLL,\ell}(v)+\delta R'_{\rm NNLL,\ell}(v)+R''_\ell(v) \ln\frac{d_\ell g_\ell(\phi)}{\zeta}\,.
\end{equation}
This gives
\begin{equation}
  \label{eq:F-simp-to-NNLL}
  \begin{split}
    \mathcal{F}(v)& \simeq\epsilon^{R'_{\rm NLL}}\left(1-\sum_\ell\left(\delta R'_{\rm NNLL,\ell}+ R''_\ell \int \frac{d\phi}{2\pi}\ln(d_\ell g_\ell(\phi))\right)\ln\frac{1}{\epsilon}-\frac 12 \sum_\ell R''_\ell
      \ln^2\frac{1}{\epsilon}\right) \times \\ & \times
    \sum_{n=0}^{\infty}\frac{1}{n!} \prod_{i=1}^n 
    \int_{\epsilon}^{\infty} \frac{d\zeta_i}{\zeta_i}\int_0^{2\pi}
    \frac{d\phi_i}{2\pi}
    \sum_{\ell_i=1,2}\left(R'_{{\rm NLL},\ell_i}+\delta R'_{{\rm NNLL},\ell_i}+R''_{\ell_i}\ln\frac{d_{\ell_i} g_{\ell_i}(\phi_i)}{\zeta_i}\right) \times
    \\ & \times \Theta\left(1-\lim_{v\to 0}\frac{\Vsc(\{\tilde p\},
        k_1,\dots, k_n)}{v}\right)\simeq \mathcal{F}_{\rm NLL}(\lambda)+
    \frac{\alpha_s(Q)}{\pi}\delta\mathcal{F}_{\rm sc}(\lambda)\,.
  \end{split}
\end{equation}
We can simplify the above equation by keeping only terms in the sum
which are linear in $ R'_{\rm NNLL,\ell}$ or 
$R''_{\ell_i}$, i.e. by correcting one emission at
a time. The latter approximation ensures that no contributions beyond
NNLL are included. Moreover, we can express the virtual correction in
eq.~\eqref{eq:F-simp-to-NNLL} as the integral over an extra dummy
emission as follows:
\begin{equation}
\label{eq:rap-virt}
\ln\frac{1}{\epsilon} = \int_\epsilon^1\frac{d\zeta}{\zeta}\,,\qquad
\frac 12\ln^2\frac{1}{\epsilon} = \int_\epsilon^1\frac{d\zeta}{\zeta}\ln\frac{1}{\zeta}\,.
\end{equation}

The final form of the soft-collinear correction then reads 
\begin{equation}
  \label{eq:F1-NNLL}
  \begin{split}
\delta\mathcal{F}_{\rm sc}(\lambda) & = 
\frac{\pi}{\alpha_s(Q)}
\int_0^\infty
    \frac{d\zeta}{\zeta}  
    \int_0^{2\pi}\frac{d\phi}{2\pi} \sum_{\ell=1,2} \left(\delta R'_{{\rm NNLL},\ell}+R''_{\ell}\ln\frac{d_{\ell} g_{\ell}(\phi)}{\zeta}\right)\int \dZ \times \\ & \times 
\left[\Theta\left(1-\lim_{v\to
        0}\frac{\Vsc(\{\tilde p\},k, \{k_i\})}{v}\right)-\Theta(1-\zeta)\Theta\left(1-\lim_{v\to
        0}\frac{\Vsc(\{\tilde p\},\{k_i\})}{v}\right)\right]\,,
  \end{split}
\end{equation}
where the average of a function over the measure $d{\cal Z}$ is
defined in eq.~\eqref{eq:dZ}.  In the first term of
eq.~\eqref{eq:F1-NNLL}, $k = k(\zeta,\phi,\ell)$ represents an
additional real emission, and the second term corresponds to virtual
corrections. In eq.~\eqref{eq:F1-NNLL} we have set the $\zeta$ lower
integration limit to zero, because singular contributions for $\zeta
\to 0$ exactly cancel between real and virtual corrections.

\subsubsection{Recoil  and hard-collinear NNLL contributions}
\label{sec:recoil}

Another source of NNLL contributions arises when one of the emissions
is collinear to any of the legs and hard, \ie it carries a sizable
fraction of emitter's longitudinal momentum. The matrix element squared
$M^2_\ell(k)$ for the emission of a gluon $k$ collinear to leg $\ell$
is given by
\begin{equation}
  \label{eq:hc-ME}
  [dk] M^2_{\ell}(k) = \frac{\alpha_s^{\rm CMW}(\tilde k_t^{(\ell)})}{4\pi}\frac{d\phi}{2\pi}  
  \frac{d(\tilde k_t^{(\ell)})^ 2}{(\tilde k_t^{(\ell)})^2} dz^{(\ell)} p_\ell(z^{(\ell)})\,,
\end{equation}
where, in our case, $p_\ell(z)=p_{gq}(z)$, given in
eq.~\eqref{eq:pgq}.  In the above equation $\tilde k_t^{(\ell)}$ is
the relative transverse momentum between the emitted gluon and the
final state parton $\tilde p_\ell$. The vectors $\tilde k_t^{(\ell)}$
satisfy
\begin{equation}
  \label{eq:ktilde-squared}
 ({\tilde k}_t^{(1)})^2 = \frac{2(\tilde p_1 k) 2(p_2 k)}{2(\tilde p_1
   p_2)}\,,\qquad ({\tilde k}_t^{(2)})^2=\frac{2(p_1 k) 2(\tilde p_2 k)}{2( p_1
   \tilde p_2)}\,.
\end{equation}
In eq.~\eqref{eq:hc-ME}, we have identified the energy fraction
relative to the splitting with the Sudakov variable $z^{(\ell)}$
defined in eq.~(\ref{eq:Sudakov}). This is justified by the fact that
all remaining emissions are soft and hence do not change the energy
fraction in an appreciable way. 

Due to recoil, the generated transverse momentum $\tilde k_t^{(\ell)}$
is different from the Sudakov transverse momentum $k_t$ of
eq.~(\ref{eq:Sudakov}), which is relative to the thrust axis. In order
to compute reliably $\Vfull(\{\tilde p\},k,k_1,\dots,k_n)$ we need to
relate $\tilde k_t^{(\ell)}$ and $k_t$.  For simplicity we consider
the case $\ell=1$ and rename $\tilde k_t^{(1)}\to \tilde k_t$.  We
start from the Sudakov parametrisation of $k$ with respect to $p_1$
and $\tilde p_1$, respectively
\begin{equation}
\label{eq:Sudakov-tilde}
k = z^{(1)}p_1 + z^{(2)}p_2 + \kappa_t = \tilde z^{(1)} \tilde p_1 + \tilde
z^{(2)} p_2 + \tilde \kappa_t\,,
\end{equation}
where $\kappa_t$ and $\tilde \kappa_t$ are spacelike vectors with
$\kappa_t^2=-k_t^2$ and $\tilde \kappa_t^2=-\tilde k_t^2$. They can be
related to the Sudakov parametrisation in the thrust axis reference
frame~\eqref{eq:Sudakov} by plugging in the parametrisation of the
recoiled momentum $\tilde p_1$ in terms of the Born momenta $p_1$ and
$p_2$
\begin{equation}
  \label{eq:Sudakov-p1}
  \tilde p_1 = z^{(1)}_p p_1 +z_p^{(2)} p_2+\pi_{t,1} \,,\qquad \pi_{t,1}^2=-p_{t,1}^2 \,,\qquad z_p^{(2)} = \frac{p_{t,1}^2}{z^{(1)}_p
    Q^2} \,,
\end{equation}
and requiring the resulting decomposition to be equal to the initial
parametrisation eq.~\eqref{eq:Sudakov}, obtaining
\begin{equation}
  \label{eq:tilde-k}
  \vec{\tilde{k}}_t = \vec{k}_t - z^{(1)}\frac{\vec{p}_{t,1}}{z^{(1)}_p} \,.
\end{equation}
From energy-momentum conservation and the fundamental property of the thrust
axis, \ie that transverse momentum is conserved separately in each
hemisphere, one has 
\begin{equation}
  \label{eq:p1-recoil}
  z^{(1)}_p \simeq 1-\!\!\!\sum_{i \in \mathcal{H}^{(1)}} \!\!z_i^{(1)}\!\! -
  z^{(1)} \simeq 1-z^{(1)} \,,\qquad \vec{p}_{t,1}  = -\!\!\! \sum_{i \in
    \mathcal{H}^{(1)}} \!\! \vec{k}_{t,i}- \vec{k}_t\,.
\end{equation}
Substituting the expressions of $z^{(1)}_p$ and $\vec{p}_{t,1}$ in
eq.~(\ref{eq:tilde-k}) we obtain
\begin{equation}
  \label{eq:ktilde}
  \vec{\tilde k}_t \simeq  \vec{k}_t - z^{(1)}\frac{\vec{p}_{t,1}}{1-z^{(1)}} =
  \vec{k}_t + \frac{z^{(1)}}{1-z^{(1)}} \left( \sum_{i \in
    \mathcal{H}^{(1)}} \!\! \vec{k}_{t,i}+\vec{k}_t\right) =
\frac{\vec{k}_t-z^{(1)} \vec{p}_{t,1}^{\,'}}{1-z^{(1)}}\,,
\end{equation}
where
\begin{equation}
  \label{eq:ptprime}
   \vec{p}_{t,1}^{\,'} = -\!\!\sum_{i \in
    \mathcal{H}^{(1)}} \!\! \vec{k}_{t,i}
\end{equation}
is the recoil due to all soft and collinear emissions. Defining also
$\vec{k}_t^{\,'} \equiv \vec{k}_t - z^{(1)}\vec{p}_{t,1}^{\,'}$ we
have that $\vec{\tilde k}_t = \vec{k}_t^{\,'}/(1-z^{(1)})$. Since
$\vec{\tilde k}_t$ and $\vec{k}_t^{\,'}$ are related by a simple
rescaling, in the collinear matrix element squared of
eq.~(\ref{eq:hc-ME}) we can replace $d\tilde k_t^2/\tilde k_t^2$ with
$d k_t^{\prime 2}/k_t^{\prime 2}$. We then obtain the relation
between the transverse momentum with respect to the thrust axis
$\vec{k}_t$ and the transverse momentum $\vec{k}_t^{\,'}$ which enters
the collinear emission phase space:
\begin{equation}
  \label{eq:kt-kt'}
  \vec{k}_t = \vec{k}_t^{\,'} + z^{(1)} \vec{p}_{t,1}^{\,'}\,.
\end{equation}
This implies that the input momentum $k$ becomes a function of
$\vec{k}_{t}^{\,'},\vec{p}_{t,1}^{\,'} ,z^{(1)}$. For the sake of
simplicity, we drop the vector superscript from now on. 

We have two NNLL contributions coming from hard-collinear
radiation. The first comes from eq.~(\ref{eq:real-rad-NNLL-2}), in which we have to take into account the exact expression of the observable when a single emission is hard and collinear:
\begin{equation}
  \label{eq:F-rec}
 \begin{split}
  &\mathcal{F}_{\rm rec}(v)=e^{-\int_{\epsilon v}^{v}[dk] M^2_{\rm sc}(k)}\sum_{n=0}^{\infty}\frac{1}{n!}\int_{\epsilon v}
   \prod_{i=1}^n [dk_i] M^2_{\rm sc}(k_i) \sum_{\ell=1,2}\int_0^1
   \!dz\, p_\ell(z) \int_0^{2\pi}\frac{d\phi}{2\pi} \int\frac{dk_t^{' 2}}{k_t^{' 2}}
   \frac{\alpha_s(k_t')}{2\pi} \times \\ & \times
  \left[ \Theta\left(v-\Vhc^{(k)}(\{\tilde
       p\},k[k_t',p'_{t,\ell} ,z],k_1,\dots,k_n)\right)
- \Theta\left(v-\Vsc(\{\tilde
       p\},k[k_t',p'_{t,\ell} ,0],k_1,\dots,k_n)\right)
\right]\,.
\end{split}
\end{equation}
In the above expression, $\Vhc^{(k)}(\{\tilde p\},k,k_1,\dots,k_n)$
denotes the expression of the observable $V$ where all emissions but
$k$ are treated in the soft-collinear approximation. In the second
term, the one containing $\Vsc(\{\tilde p\},k,k_1,\dots,k_n)$,
also emission $k$ has been treated as if it were soft and collinear, so
that its transverse momentum with respect to the emitting leg $k_t'$
is equal to $k_t$.
Notice that, in eq.~\eqref{eq:F-rec} we can replace $k_t'$ with $k_t$ in the
integration since this variable is integrated over, and use the
short-hand notation
\[
k' = k[k_t,p_{t,1}',z]\,,\qquad k = k[k_t,p_{t,1}',0]\,.
\]

To NNLL accuracy it is possible
to further simplify the phase-space for $k$. Introducing 
\begin{equation}
\label{eq:zdef}
\zeta  =\frac{1}{v} \frac{d_\ell \,g_\ell(\phi)}{z^{b_\ell}}\left(\frac{k_t}{Q}\right)^{a+b_\ell}\,,
\end{equation}
we have, at NNLL accuracy
\begin{equation}
  \label{eq:k-collinear}
  \frac{dk_t^2}{k_t^2}\frac{\alpha_s(k_t)}{2\pi}=
  \frac{\alpha_s((z^{b_\ell}\zeta
    v/(d_\ell \,g_\ell(\phi)))^{1/(a+b_\ell)}Q)}{\pi(a+b_\ell)}\frac{d\zeta}{\zeta}
\simeq
  \frac{\alpha_s(v^{1/(a+b_\ell)}Q)}{\pi(a+b_\ell)}\frac{d\zeta}{\zeta}\,.
\end{equation}
In fact, rIRC safety constrains the variable $\zeta$ to be of order
one, so that further terms arising from the expansion of the QCD
coupling around $v^{1/(a+b_\ell)} Q$ are of relative order
$\alpha_s^2$, hence at most N$^3$LL. 
 
Following what we did in sections~\ref{sec:nll} and~\ref{sec:rap}, we
eliminate all subleading contributions and obtain $\mathcal{F}_{\rm
  rec}(v)\simeq(\alpha_s(Q)/\pi) \delta \mathcal{F}_{\rm
  rec}(\lambda)$, where
\begin{equation}
  \label{eq:Frec}
  \begin{split}
  \delta\mathcal{F}_{\rm rec}(\lambda)&=\sum_{\ell=1,2}
   \frac{\alpha_s(v^{1/(a+b_\ell)}Q)}{\alpha_s(Q) (a+b_\ell)}
   \int_0^{\infty}\frac{d\zeta}{\zeta}
   \int_0^{2\pi}\frac{d\phi}{2\pi}\int \dZ
     \times \\ & \times
   \int_0^1 \!dz\,p_\ell(z)
   \left[\Theta\left(1-\lim_{v\to 0} \frac{\Vhc^{(k')}(\{\tilde
         p\},k',\{k_i\})}{v}\right)
     -\Theta\left(1-\lim_{v\to 0} \frac{\Vsc(\{\tilde
         p\},k,\{k_i\})}{v}\right)\right]\,.
  \end{split}
\end{equation}
The second NNLL contribution coming from hard collinear radiation
arises from eq.~(\ref{eq:real-rad-NNLL-1}):
\begin{equation}
  \label{eq:F-collinear}
 \begin{split}
  &\mathcal{F}_{\rm collinear}(v)=e^{-\int_{\epsilon v}^{v}[dk] M^2_{\rm sc}(k)}\sum_{n=0}^{\infty}\frac{1}{n!}\int_{\epsilon v}
   \prod_{i=1}^n [dk_i] M^2_{\rm sc}(k_i) \sum_{\ell=1,2}\int_0^1
   \!dz\, p_\ell(z) \int_0^{2\pi}\frac{d\phi}{2\pi} \int\frac{dk^2_t}{k^2_t}
   \frac{\alpha_s(k_t)}{2\pi} \times \\ & \times
  \left[ \Theta\left(v-\Vsc(\{\tilde
       p\},k,k_1,\dots,k_n)\right)
- \Theta\left(v-\Vsc(\{\tilde
       p\},k_1,\dots,k_n)\right)\Theta\left(v-\Vsc(\{\tilde p\},k)\right)
\right]\,,
  \end{split}
\end{equation}
where the second term in the square brackets represents virtual corrections.
From the above equation we see that, if $k$ is also soft, i.e. $z\to
0$, the function $ \mathcal{F}_{\rm collinear}(v)$ contains
configurations that have been already taken into account in the
function $\mathcal{F}(v)$ of eq.~(\ref{eq:F}). 
We eliminate this double counting by subtracting the NLL
contribution   
\begin{equation}
  \label{eq:dF-colsub}
 \begin{split}
\mathcal{F}_{\rm collinear}^{\rm sub.}(v) & = e^{-\int_{\epsilon v}^{v}[dk] M^2_{\rm sc}(k)}\sum_{n=0}^{\infty}\frac{1}{n!}\int_{\epsilon v}
   \prod_{i=1}^n [dk_i] M^2_{\rm sc}(k_i) \int_0^{2\pi}\frac{d\phi}{2\pi} \int\frac{dk_t^2}{k_t^2}
   \frac{\alpha_s(k_t)}{2\pi}  \sum_{\ell=1,2}2 C_\ell \int_0^1
   \!\frac{dz}{z}\\
&\times\left[\Theta\left(v-\Vsc(\{\tilde p\},k,k_1,\dots,k_n)\right) - \Theta\left(v-\Vsc(\{\tilde
       p\},k_1,\dots,k_n)\right)\Theta\left(v-\Vsc(\{\tilde p\},k)\right) \right].
  \end{split}
\end{equation}
Performing the same manipulations as for $\mathcal{F}_{\rm rec}$ we arrive at:
\begin{equation}
\mathcal{F}_{\rm collinear}(v)-\mathcal{F}_{\rm collinear}^{\rm sub.}(v)\simeq \frac{\alpha_s(Q)}{\pi}\delta\mathcal{F}_{\rm
  hc}(\lambda)\,,
\end{equation}
where
\begin{equation}
  \label{eq:Fhc}
  \begin{split}
&  \delta\mathcal{F}_{\rm  hc}(\lambda)=\sum_{\ell=1,2}
   \frac{\alpha_s(v^{1/(a+b_\ell)}Q)}{\alpha_s(Q) (a+b_\ell)}
   \int_0^{\infty}\frac{d\zeta}{\zeta}
   \int_0^{2\pi}\frac{d\phi}{2\pi}\int \dZ
   \times \\ & \times
   \int_0^1 \!\frac{dz}{z}\,(z p_\ell(z) - 2 C_\ell )
   \left[\Theta\left(1-\lim_{v\to 0} \frac{\Vsc(\{\tilde
         p\}, k, \{k_i\})}{v}\right)
     -\Theta\left(1-\lim_{v\to 0} \frac{\Vsc(\{\tilde
         p\}, \{k_i\})}{v}\right)\Theta(1-\zeta)\right]\,. 
  \end{split}
\end{equation}

\subsubsection{Soft wide-angle NNLL contributions}
\label{sec:soft-large-angle}

This contribution arises when one of the soft gluons is emitted at
wide angles.  We can parametrise the observable dependence on the
momentum of this extra gluon $k$ as
\begin{equation}
\Vwa^{(k)}(\{\tilde p\},k) = \left(\frac{k_t}{Q}\right)^a f_{\rm wa}(\eta,\phi) \,.
\label{eq:Vwa}
 \end{equation}
In general, when $\eta$ is close to zero (wide angles), the above
expression might differ from the expression of the observable after a
soft and collinear emission $k$
\begin{equation}
\Vsc(\{\tilde p\},k) = \left(\frac{k_t}{Q}\right)^a f_{\rm sc}(\eta,\phi) \,,\qquad 
f_{\rm sc}(\eta,\phi)= d_1 e^{-b_1 \eta} g_1(\phi) \Theta(\eta) + 
d_2 e^{b_2 \eta} g_2(\phi) \Theta(-\eta) \,.
\label{eq:Vsc}
\end{equation}

For a fixed value of $k_t,\eta,\phi$ for an extra emission $k$, we
denote with $\Vwa^{(k)}(\{\tilde p\},k,k_1,\dots,k_n)$ the observable
computed by keeping the full $\eta,\phi$ dependence of emission $k$,
and using the soft-collinear approximation for all other emissions.

This gives rise to the following correction
\begin{equation}
  \label{eq:dF-soft}
\begin{split}
  \mathcal{F}_{\rm wa}(v) &=
  e^{-\int_{\epsilon v}^{v}[dk] M^2_{\rm sc}(k)}\sum_{n=0}^{\infty}\frac{1}{n!}\int_{\epsilon v}
    \prod_{i=1}^n [dk_i] M^2_{\rm sc}(k_i) 2 C_F\int_0^{\infty}
    \frac{dk_t}{k_t} \frac{\alpha_s(k_t)}{\pi} \int_{-\infty}^{\infty}
    d\eta \int_0^{2\pi}\frac{d\phi}{2\pi}\times \\ & \times
\left[\Theta\left(1-\lim_{v\to 0}\frac{\Vwa^{(k)}(\{\tilde p\},k,k_1,\dots,k_n)}{v}\right)-
\Theta\left(1-\lim_{v\to 0}\frac{\Vsc(\{\tilde p\},k,k_1,\dots,k_n)}{v}\right)\right]\,.
\end{split}
\end{equation}
We can modify the phase space integration for the extra soft gluon as
follows:
\begin{equation}
\label{eq:kt-zeta}
\frac{dk_t}{k_t}\frac{\alpha_s(k_t)}{\pi} =
\frac{d\zeta}{\zeta}\frac{\alpha_s((\zeta v )^{1/a}Q)}{a \pi} \simeq
\frac{d\zeta}{\zeta}\frac{\alpha_s(v^{1/a} Q)}{a \pi}\,, 
\end{equation}
where 
\begin{equation}
\zeta  = \frac{1}{v} \left(\frac{k_t}{Q}\right)^a\,
\label{eq:zdefsoft}
\end{equation} 
is constrained to be of order one for rIRC safe observables. This
ensures that the approximation in eq.~\eqref{eq:kt-zeta} is valid, up
to corrections beyond NNLL accuracy.  This gives $\mathcal{F}_{\rm
  wa}(v)\simeq (\alpha_s(Q)/\pi)\delta\mathcal{F}_{\rm wa}(\lambda)$,
where
\begin{equation}
  \label{eq:dF-soft-final}
\begin{split}
 \delta\mathcal{F}_{\rm wa}(\lambda) &= \frac{2 C_F}{a} \frac{\alpha_s(v^{1/a}
   Q)}{\alpha_s(Q)} \int_0^{\infty} \frac{d\zeta}{\zeta}
 \int_{-\infty}^{\infty} \!\! d\eta \int_0^{2\pi}\! \frac{d\phi}{2\pi}\int \dZ
 \\ & \times
\left[\Theta\left(1-\lim_{v\to 0}\frac{\Vwa^{(k)}(\{\tilde p\}, k, \{k_i\})}{v}\right)-
\Theta\left(1-\lim_{v\to 0}\frac{\Vsc(\{\tilde p\}, k, \{k_i\})}{v}\right)\right]\,.
\end{split}
\end{equation}

\subsubsection{Soft correlated emission}
\label{sec:correl}
Unlike the hard-collinear and soft large-angle emissions, an arbitrary
amount of soft and collinear emissions contribute to $\Sigma(v)$.
Primary gluons emitted off the hard Born legs, can give rise to
subsequent branchings which need to be taken into account already
at NLL accuracy~\cite{Banfi:2004yd}. However, at this accuracy any
rIRC observable can be treated inclusively with respect to subsequent
branchings of the soft gluons. This results just in a redefinition of
the scheme for the QCD running coupling, which is now defined as the
strength of the inclusive soft radiation~\cite{Catani:1990rr}. Each
soft and collinear emission contributing to NLL accuracy is thus to be
interpreted as fully inclusive in its branchings.

A generic event-shape variable is commonly non-inclusive for such
splittings. However, for rIRC observables, non-inclusiveness only
matters starting from NNLL accuracy~\cite{Banfi:2004yd}. At NNLL, the
observable is sensitive to the details of the secondary soft
splitting, so we need to undo the inclusive branching in order to
compute the corresponding NNLL correction. Once again, in order to
achieve NNLL accuracy, only a single non-inclusive splitting can be
considered. The NNLL correlated correction has been already written in
eq.~(D.5) of ref.~\cite{Banfi:2004yd}, and reads
\begin{equation}
  \label{eq:dF-correl}
  \begin{split}
    \delta\mathcal{F_{\rm correl}}(v)&=e^{-\int_{\epsilon v}^{v}[dk] M^2_{\rm sc}(k)}\sum_{n=0}^{\infty}\frac{1}{n!}\int_{\epsilon v}
    \prod_{i=1}^n [dk_i] M^2_{\rm sc}(k_i) \frac{1}{2!}\int[dk_a]
    [dk_b] \tilde M^2(k_a,k_b) \times \\ & \times
    \left[\Theta\left(v-\Vsc(\{\tilde
        p\},k_a,k_b,k_1,\dots,k_n)\right)-\Theta\left(v-\Vsc(\{\tilde
        p\},k_a+k_b,k_1,\dots,k_n)\right)\right]\,,
  \end{split}
\end{equation}
where $\tilde M^2(k_a,k_b)$ is a two-parton correlated matrix element,
defined by 
\begin{equation}
  \label{eq:2parton-correlated}
  \tilde M^2(k_a,k_b) = M^2(k_a,k_b) - M^2(k_a)M^2(k_b)\,.
\end{equation}
To show explicitly that this contribution starts from NNLL accuracy,
we first express the two-parton correlated emission matrix element and
phase space as
\begin{equation}
  \label{eq:M2-correl}
   [dk_a] [dk_b]  \tilde M^2(k_a,k_b)= [dk_a] [dk_b] M^{2}_{\rm sc}(k_a) M^{2}_{\rm sc}(k_b) \frac{\tilde M^2(k_a,k_b)}{M^{2}_{\rm sc}(k_a) M^{2}_{\rm sc}(k_b)}\,.
\end{equation}
Neglecting terms beyond NNLL accuracy, we rewrite the $k_a$
integration as follows:
\begin{equation}
  \label{eq:M2a}
  \begin{split}
    [dk_a] M^{2}_{\rm sc}(k_a) &= \frac{dv_a}{v_a}\frac{d\phi_a}{2\pi}
    \int [dk_a] M^{2}_{\rm sc}(k_a) \sum_{\ell_a} v_a
    \delta\left(v_a-\left(\frac{k_{ta}}{Q}\right)^a e^{-b_{\ell_a}
        \eta_a^{(\ell_a)}}
    \right) \Theta\left(\eta_a^{(\ell_a)}\right) \\
    &\simeq \frac{d\zeta_a}{\zeta_a}\frac{d\phi_a}{2\pi} \int [dk_a]
    M^{2}_{\rm sc}(k_a)  \sum_{\ell_a} v
    \delta\left(v-\left(\frac{k_{ta}}{Q}\right)^a e^{-b_{\ell_a}
        \eta_a^{(\ell_a)}} \right) \Theta\left(\eta_a^{(\ell_a)}\right)\,,
  \end{split}
\end{equation}
where, in the last line, we have defined $\zeta_a=v_a/v$, and
neglected terms beyond NNLL accuracy, using the fact that rIRC safety
constrains $\zeta_a$ to be of order one.

We then parametrise the phase space of the emission $k_b$ in terms of
the variables $\kappa=k_{t,b}/k_{t,a}$, $\eta=\eta_b-\eta_a$ and
$\phi=\phi_b-\phi_a$. Notice that this is a convenient choice since
the correlated matrix element $\tilde M^2(k_a,k_b)/(M^2_{\rm sc}(k_a)
M^2_{\rm sc} (k_b))$ explicitly depends on the correlated momenta
through these variables. This leads to
\begin{equation}
    \label{eq:M2b}
    \begin{split}
      [dk_b] M^2_{\rm sc}(k_b)=  \left(\frac{2 C_F \alpha_s(k_{t,b})}{\pi}\right) \frac{d\kappa}{\kappa} \Theta(\kappa) d\eta \frac{d\phi}{2\pi} 
      \simeq \left(\frac{2 C_F \alpha_s(k_{t,a})}{\pi}\right) \frac{d\kappa}{\kappa} \Theta(\kappa) d\eta \frac{d\phi}{2\pi}\,,
    \end{split}
\end{equation}
where in the last step we have set $k_{t,b}\simeq k_{t,a}$. The latter
approximation is valid for rIRC safe observables only, with corrections
beyond NNLL accuracy.

Therefore, eq.~(\ref{eq:M2-correl}) can be rewritten as
\begin{equation}
  \label{eq:M2-correl-nnll}
   [dk_a] [dk_b]  \tilde M^2(k_a,k_b)=
   \frac{d\zeta_a}{\zeta_a}\frac{d\phi_a}{2\pi}\sum_{\ell_a=1,2}
   \left(\frac{2C_{\ell_a}\lambda}{a\pi\beta_0}R^{''}_{\ell_a}(v)\right) \frac{d\kappa}{\kappa} \Theta(\kappa) d\eta \frac{d\phi}{2\pi}C_{ab}(\kappa,\eta,\phi)\,,
\end{equation}
where 
\begin{equation}
  \label{eq:Cab}
  C_{ab}(\kappa,\eta,\phi) = \frac{\tilde M^2(k_a,k_b)}{M^{2}_{\rm sc}(k_a) M^{2}_{\rm sc}(k_b)}\,,
\end{equation}
and 
\begin{equation}
  \label{eq:Rp1l}
  \begin{split}
\int [dk] M^2_{\rm sc}(k) \Theta(\eta) \left(\frac{2 C_F
    \alpha_s(k_t)}{\pi}\right) v\delta\left(v-\left(\frac{k_t}{Q}\right)^a e^{-b_1 \eta^{(1)}}\right) =
\frac{2C_F\lambda}{a\pi\beta_0}R^{''}_{1}(v) \,, \\
\int [dk] M^2_{\rm sc}(k) \Theta(-\eta) \left(\frac{2 C_F
    \alpha_s(k_t)}{\pi}\right)
 v\delta\left(v-\left(\frac{k_t}{Q}\right)^a e^{-b_2 \eta^{(2)}}\right)
= \frac{2C_F\lambda}{a\pi\beta_0}R^{''}_{2}(v)\,. 
  \end{split}
\end{equation}
We are now in a position to write the final expression for the NNLL correlated
correction as $\delta \mathcal{F}_{\rm correl}(v)\simeq
\alpha_s(Q)/\pi\delta\mathcal{F}_{\rm correl}(\lambda)$, where 
\begin{equation}
  \label{eq:F1-correl}
  \begin{split}
    \delta\mathcal{F}_{\rm correl}(\lambda)&=\int_0^{\infty}\frac{d\zeta_a}{\zeta_a}\int_0^{2\pi}\frac{d\phi_a}{2\pi}\sum_{\ell_a=1,2}\left(\frac{2C_{\ell_a}\lambda}{a\beta_0}\frac{R^{''}_{\ell_a}(v)}{\alpha_s(Q)}\right) \int_0^{\infty} \frac{d\kappa}{\kappa} \int_{-\infty}^{\infty}\!\!\! d\eta \int_0^{2\pi}\frac{d\phi}{2\pi} \frac{1}{2!}
C_{ab}(\kappa,\eta,\phi)
\times\\& \times
\int \dZ
\left[\Theta\left(v-\Vsc(\{\tilde
        p\}, k_a,k_b,\{k_i\})\right)-\Theta\left(v-\Vsc(\{\tilde
        p\},k_a+k_b,\{k_i\})\right)\right]\,,
  \end{split}
\end{equation}
where, as usual, the observable's value does not depend on emissions'
rapidities, with the only exception of $k_b$, given by
\begin{equation}
  \label{eq:kbarb}
   k_b = \kappa\,
   k^{(\ell_a)}_{t,a}(\cosh(\eta_a+\eta),\cos(\phi_a+\phi),\sin(\phi_a+\phi),\sinh(\eta_a+\eta))\,,\qquad
   k_{t,a}^{(\ell_a)} = Q \, v_a^{\frac{1}{a}-\frac{b_{\ell_a}}{a+b_{\ell_a}} \xi_a^{(\ell_a)}}\,.
\end{equation}
Furthermore, in order to eliminate subleading effects, in the
calculation of the observable we assume that $k_b$ belongs to the same
hemisphere as $k_a$, neglecting {\em de facto} the contribution of two
emissions falling into two different hemispheres.

It is worth commenting on the connection between
Eq.~\eqref{eq:F1-correl} and the CMW scheme for the running coupling
defined in Eq.~\eqref{eq:CMW-scheme}. As already explained in
Section~\ref{sec:nll}, the term $K$ in Eq.~\eqref{eq:CMW-scheme}
encodes the contribution of the splitting of a soft gluon into either
a $q\bar q$ or a $gg$ pair. This gives rise to NLL terms in the
Sudakov radiator which are universal for all rIRC safe observables. In
the multiple emissions function $\fullF(v)$, the CMW scheme gives rise
to NNLL contributions which are contained in soft-collinear
corrections~\eqref{eq:F1-NNLL}. In the latter contribution, the
branching of a soft gluon is in fact treated inclusively.
This approximation is subsequently subtracted in the second theta
function in the correlated correction~\eqref{eq:F1-correl}, which
takes into account the correct non-inclusive nature of the observable.
Therefore, the choice of the CMW scheme in the multiple-emission
function $\fullF(v)$ is irrelevant at all logarithmic orders, since
the appropriate non-inclusive treatment of the observable is
guaranteed once one adds up all resolved real emission corrections.

\section{Validation and matched results}
\label{sec:checkandmatch}
In this section we apply the algorithm described in
Section~\ref{sec:nnll} to the following set of seven event-shape
variables: thrust $1-T$, heavy jet mass $\rho_{ H}$, total and wide
broadening $B_{ T}$, $B_{ W}$, $C$-parameter, thrust major $T_{M}$,
and oblateness $O$.  For the two observables $T_{M}$, $O$ an
NNLL resummation was not previously available. For $C$, a numerical
result was presented in~\cite{Alioli:2012fc}. On the other hand, for
the remaining four event shapes analytic results can be found in the
literature ($1-T$~\cite{Becher:2008cf}, $\rho_{H}$~\cite{Chien:2010kc},
$B_{ T}$, $B_{ W}$~\cite{Becher:2012qc}). 
As described in the previous section, we use the $h(\lambda)$ function
of thrust $1-T$ also for the resummation of both the $C$-parameter and
heavy jet mass $\rho_H$. For $1-T$, we compare our resummation
formulae to the analytic result of ref.~\cite{Monni:2011gb}, and
extract the corresponding $h(\lambda)$ function, reported in
eq.~\eqref{eq:h-thrust}.
For $\rho_H$ we then obtained the same resummed result
of~\cite{Chien:2010kc}.  Analogously, for $B_T$ and $B_W$ we have
compared our numerical expansion to the relative analytic expressions
of ref.~\cite{Becher:2012qc} and found full agreement up to (and
including) terms of order $\alpha_s^3L^2$. To check the resummation
for the observables for which we provide new results (i.e. $T_M$, $O$)
and for $C$, we subtract the numerical expansion for the differential
distributions from the predictions obtained by generating three-jet
NLO distributions with {\tt Event2}~\cite{Catani:1996vz}. In order to
get more stable distributions, we compute differences of observables,
and plot the following quantity:
\begin{equation}
\label{eq:delta}
\Delta(v_1,v_2)  = \left(\frac{1}{\sigma_0}\frac{d\sigma^{\rm
    NLO}}{d\ln\frac{1}{v_1}} - \frac{1}{\sigma_0}\frac{d\sigma^{\rm
    NNLL}|_{\rm expanded}}{d\ln\frac{1}{v_1}} \right)-
\left\{v_1\rightarrow v_2\right\}.
\end{equation}
 \begin{figure}[htp]
  \centering
  \includegraphics[width=0.90\columnwidth]{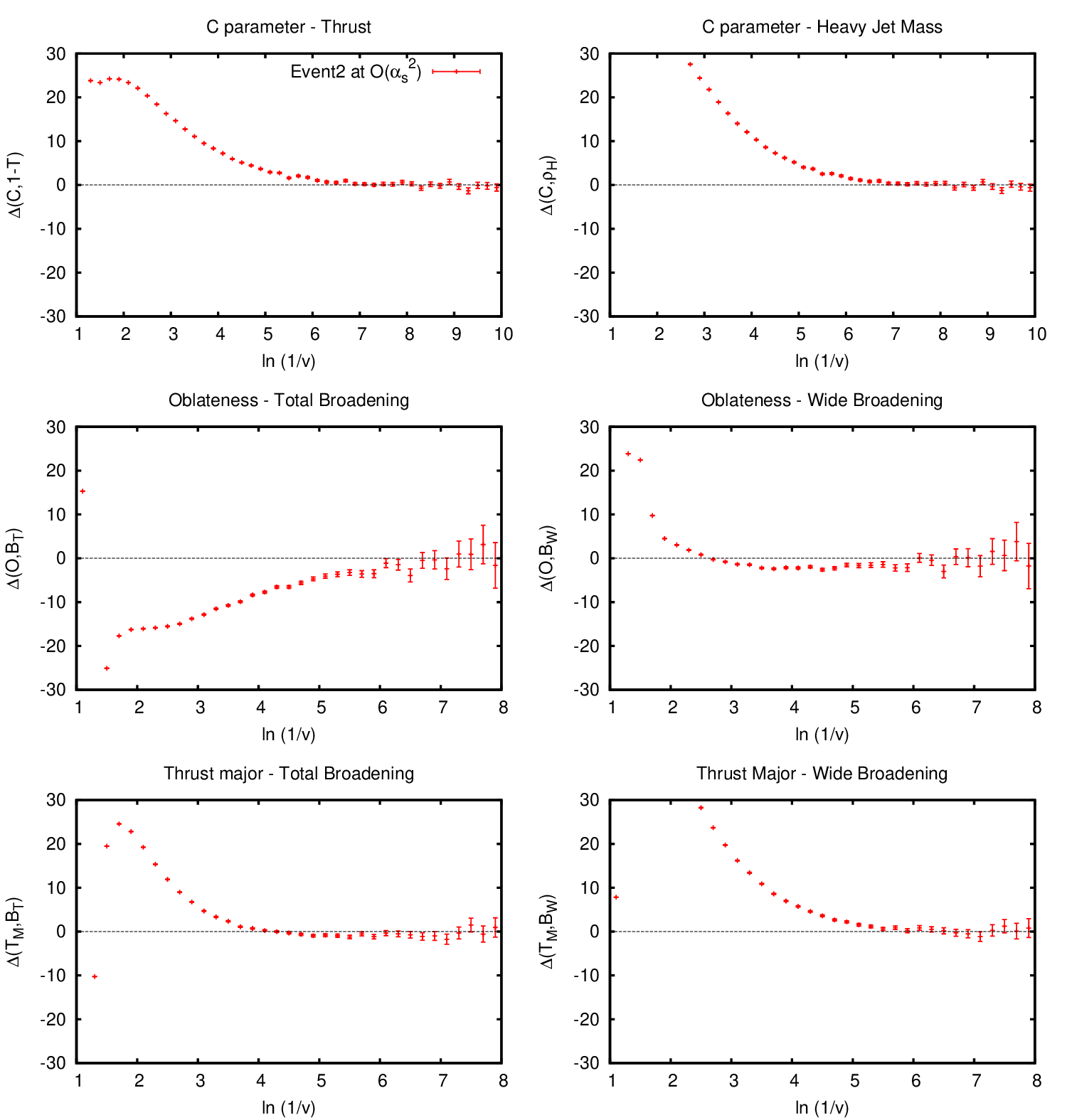}
  \caption{Difference between the NLO differential distributions of
    pairs of observables after subtracting the expansion of the NNLL
    resummation formula up to (and including) ${\mathcal
      O}(\alpha_s^2L^0)$ (see eq.~\eqref{eq:delta}).  To obtain these distributions we used about $10^{11}$ events. }
  \label{fig:event2}
\end{figure}
The results are shown in Figure~\ref{fig:event2}. There we see that
$\Delta(v_1,v_2)$ tends to zero for $v \rightarrow 0$,
providing a check of the validity of the NNLL resummation up to
${\mathcal O}(\alpha_s^2)$.  In order to check the expansion to
${\mathcal O}(\alpha_s^3)$ one would have to produce either NNLO 3-jet
or NLO 4-jet distributions which are sufficiently stable in the deep
infrared region. This can be achieved through long runs e.g. with the
generators {\tt EERAD3}~\cite{Ridder:2014wza}, however, we have not
been able to obtain distributions that were stable enough.  Alternatively, one could use {\tt
  NLOJET++}~\cite{Nagy:2003tz} to generate four-jet distributions at
NLO and consider differences of observables.  On the other hand, the
checks against the analytic results for $\rho_{H}$, $B_{T}$ and
$B_{W}$ at ${\cal O}(\alpha_s^3)$ provide us with a proof of the
validity of our NNLL resummation at this order.

As a last step, we match the resummed NNLL distributions to NNLO
fixed-order differential cross sections obtained with {\tt
  EERAD3}~\cite{Ridder:2014wza}. The matching is performed according
to the log-R scheme~\cite{CTTW,Monni:2011gb}.
As it is customary in resummed calculations, to probe the size of
subleading logarithmic terms we introduce a rescaling constant $x_V$
as \begin{equation}
\ln\frac{1}{v} = \ln\frac{x_V}{v}-\ln x_V\,,
\end{equation}
and expand the cross section around $\ln x_V/v$ neglecting subleading
terms.\footnote{For details about how the resummed formula and the
  expansion coefficients change see e.g.~ref.~\cite{Monni:2011gb}
  where one has to replace $\ln x_L\rightarrow -\ln x_V$.}
Eventually we modify the resummed logarithm $\ln x_V/v$ in order to
impose that the total cross section is reproduced at the kinematical
endpoint $v_{\rm max}$ 
\begin{equation}
\label{eq:mod-logs}
\ln\frac{x_V}{v} \rightarrow \frac{1}{p}\ln\left(1+\left(
    \frac{x_V}{v}\right)^p - \left( \frac{x_V}{v_{\rm max}}\right)^p
\right)\,. 
\end{equation}
Here, $p$ denotes a positive number which controls how quickly the
logarithms are switched off close to the endpoint. In the following we
use $p=1$.

\begin{figure}[htp]
  \centering
  \includegraphics[width=0.90\columnwidth]{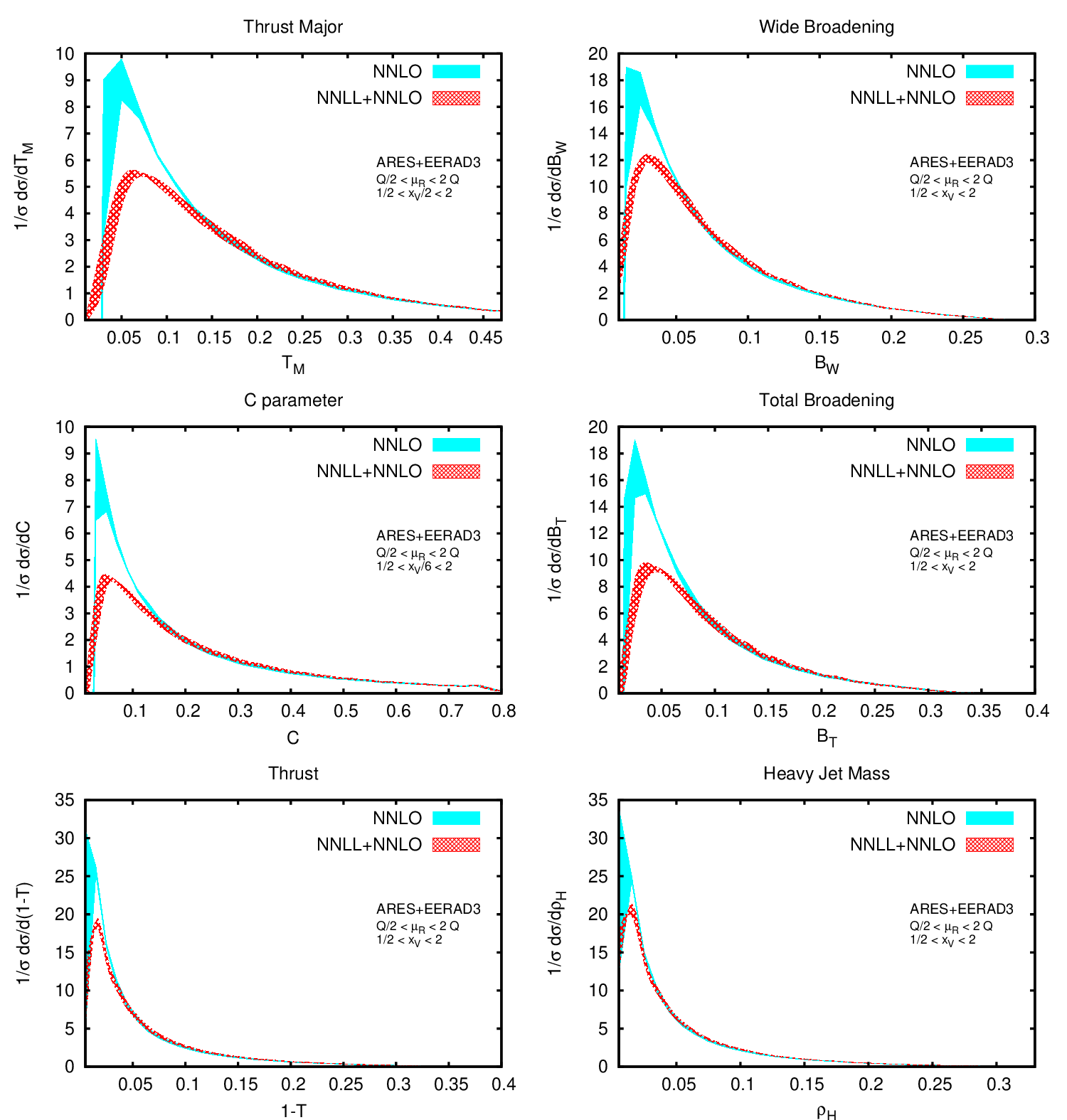}
  \caption{Differential distributions for six of the event-shape observables
    considered in the article at NNLL+NNLO (red band) and NNLO
    (light blue band).}
  \label{fig:NNLO}
\end{figure}

\begin{figure}[htp]
  \centering
  \includegraphics[width=0.90\columnwidth]{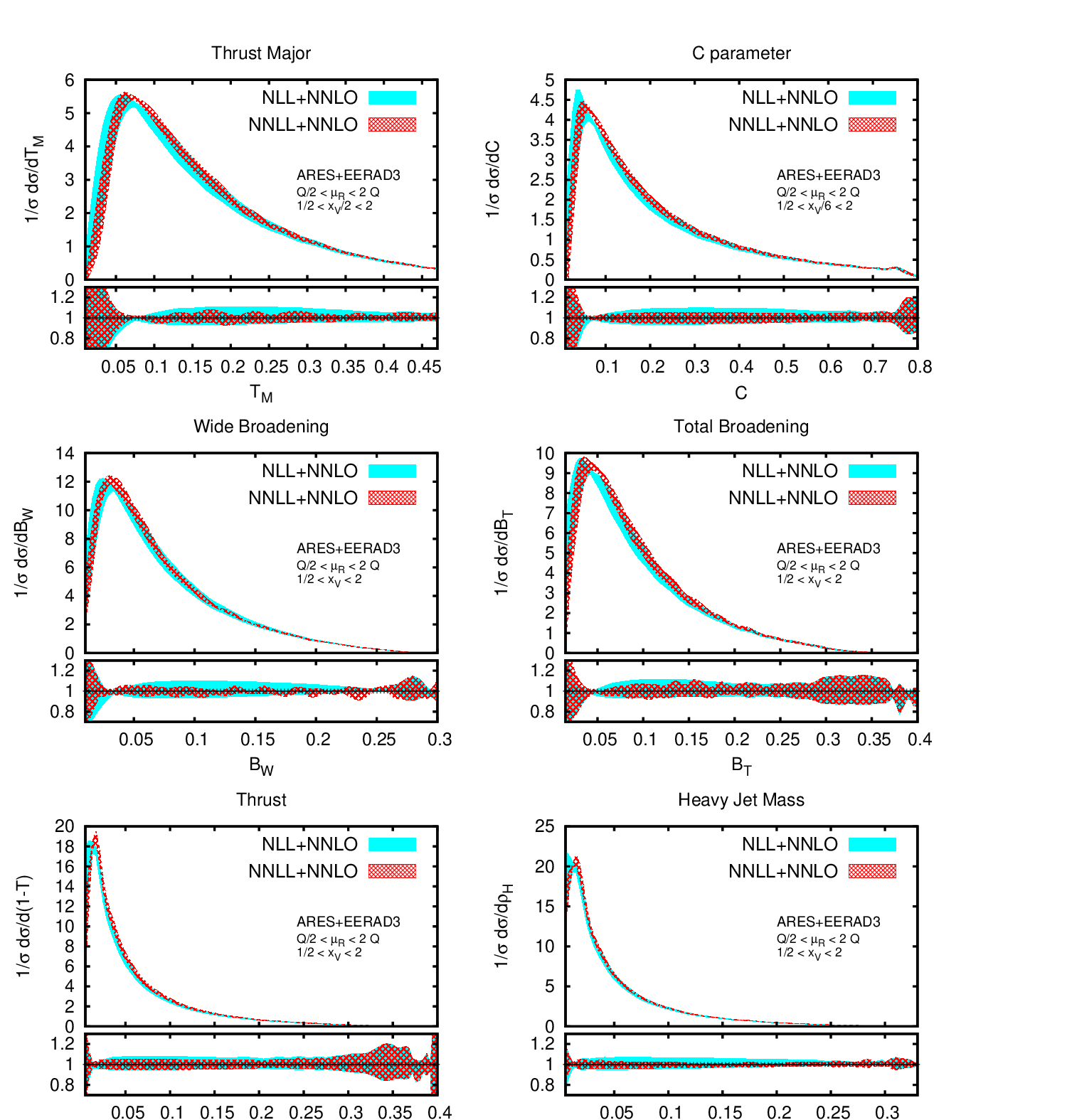}
  \caption{Matched distributions for six of the event-shape
    observables considered in the article at NNLL+NNLO (red band) and
    NLL+NNLO (light blue band). The lower panel of each plot shows the
    ratio of the NNLL+NNLO and NLL+NNLO bands to the corresponding
    central values.}
  \label{fig:NNLL+NNLO}
\end{figure}

To obtain our central predictions we set $\mu_R = Q = M_Z$,
corresponding to $\alpha_s(\mu_R) = 0.118$,
and~\cite{Dasgupta:2002dc,Banfi:2010xy}
\begin{equation}
\ln x_V = \frac{1}{2}\sum_{\ell=1,2} \left(\ln d_\ell +\int_0^{2 \pi} \frac{d\phi}{2\pi} \ln g_\ell(\phi)\right)\,.
\end{equation}
We then construct the uncertainty bands by varying $\mu_R$ and $x_V$
individually by a factor of two in either direction.
Figure~\ref{fig:NNLO} shows the comparison of the NNLL+NNLO prediction
(red bands) to the pure fixed order at NNLO accuracy (light blue
bands) for the thrust ($1-T$),
$C$-parameter, heavy-jet mass ($\rho_H$), wide- and total-broadening
($B_W$, $B_T$) and thrust major ($T_M$).
As expected, at large values of the observables the matched results
approach smoothly the fixed order distributions. On the other hand we
observe large corrections at small values of the observables, where
the NNLO distributions tend to diverge, while the NNLL+NNLO results
have a smooth Sudakov behaviour. We also notice that at
small/intermediate values of the observables the fixed order
uncertainties are artificially small, and that the matched results are
not within the fixed-order uncertainty bands.
In Figure~\ref{fig:NNLL+NNLO} we compare the NNLL+NNLO distributions
to NLL+NNLO distributions for the same set of observables. 
In the hard region, NNLL effects are small, NNLL+NNLO and NLL+NNLO
bands overlap, and the uncertainties shown are those of the NNLO
distribution. On the other hand close to the peak of the distributions
NNLL effects are important. In general NNLL corrections tend to make
the spectrum harder, and we find that uncertainties are reduced when
going from NLL+NNLO to NNLL+NNLO. This can be appreciated by looking
at the lower panel of each plot of Figure~\ref{fig:NNLL+NNLO},
representing the ratio of the NLL+NNLO and NNLL+NNLO bands to the
corresponding central values.
The oblateness distribution, not shown in Figures~\ref{fig:NNLO} and
\ref{fig:NNLL+NNLO}, has the particular feature that it is defined as
a difference of two observables.  This implies that for sufficiently
small values of $O$ the cross section is dominated by cancellations
between soft-collinear real emissions (corresponding to single
logarithmic contributions) rather than by the double logarithms
present in the Sudakov radiator.  As a consequence, the real
corrections in the multiple emission function grow faster than their
virtual counterpart, resulting in a divergence at $R'_{\rm
  NLL}=R'_c\simeq 2$~\cite{Banfi:2004yd}.  In correspondence of this
value of $R'_{\rm NLL}$, the normal hierarchy of logarithms is
reversed, and one ends up neglecting subleading logarithmic terms
which are actually numerically dominant.  A correct treatment of this
region would require a resummation of such contributions to all
logarithmic orders~\cite{Dasgupta:2001eq}.
The divergence at $R'_{\rm NLL}=R'_c$ is close to the peak of the
distribution, where the bulk of the cross section is. Therefore, one can
rely on the resummation only in the tail region, sufficiently away from the
singularity. Moreover, the singularity is pushed towards higher values
of the oblateness for higher values of $x_V$, thus one finds a large
theory  uncertainty due to the $x_V$ variation. Despite spoiling
the resummation, the above singularity does not affect the expansion
in powers of the strong coupling, so that our method can still be used
to compute correctly the coefficients of the expansion to all orders
in $\alpha_s$.\footnote{Analogously to what is observed for both $T_M$ and $C$,
  the oblateness, which has $d_\ell=2$, receives sizable NNLL
  corrections.}

\section{Conclusions}
\label{sec:conclusions}
We presented a novel method for automated resummation of event-shape
distributions to NNLL accuracy. The method is fully general and it can
be applied to any global and recursive infrared and collinear safe
event shape. Neither the factorisation of the observable into
kinematical subprocesses nor an analytic definition is required for
the method to be applied.  We implemented the algorithm in the fast
and stable numerical code {\tt ARES} which will be publicly released
soon.
For the time being, the method relies on the fact that the NNLL
Sudakov radiator is universal for all observables which have the same
scaling properties for a single soft-collinear emission. Therefore,
for the observables analysed in this article, we could extract the
relevant unknown terms of the NNLL radiator from known resummations.
Specifically, using the known result for thrust $1-T$ we can reproduce
the known resummation for the heavy jet mass $\rho_H$ and the $C$
parameter. Analogously, we extract the missing term from the $k_t$
resummations in colour singlet production in hadronic collisions,
re-derive the known results for jet broadenings $B_T$, $B_W$, and
obtain new predictions for both thrust major $T_M$ and oblateness $O$.
The calculation of the NNLL radiator for a generic observable will be
addressed in future work. 

We performed checks of our results by expanding known resummed
distributions up to ${\cal O}(\alpha_S^3)$ and comparing them to the
results in the literature.  For observables for which a NNLL
resummation was not previously known, we compare their ${\cal
  O}(\alpha_s^2)$ expansion to the fixed order generator {\tt Event2}.
We then presented NNLL+NNLO matched results, where the NNLO results
were obtained using the code {\tt EERAD3}.
We observed that NNLL corrections are in general sizable and hence
play a role in precise determinations of the strong coupling constant
using $e^+e^-$ data. A simultaneous fit using distributions of several
observables will help disentangle perturbative effects from
non-perturbative ones. In fact while perturbative effects are now
well-understood and described at NNLL+NNLO or beyond, some deeper
understanding is still required to model non-perturbative corrections, 
which are quite sizable at LEP energies.
These corrections will be much more moderate at future lepton
colliders. In these conditions, the potential of NNLL resummations can
be fully exploited and even data close or at the peak of the
distributions can be included in the fit region.

The work presented here represents a first step towards a fully
automated resummation of generic global and rIRC observables. Future
work includes the calculation of the NNLL radiator in the generic
case, the treatment of jet resolution parameters, as well as the
extension to observables at hadron colliders. Furthermore, the method
could also be applied to derive corrections beyond NNLL.

\section*{Note added}
When this work was being finalised ref.~\cite{Hoang:2014wka} appeared,
where an analytic resummation for the $C$ parameter is presented. We
compared their formulae to our analytic result for the $C$
parameter, and found full agreement at NNLL.

\section{Acknowledgements}
We would like to thank G. Salam for stimulating conversations during
the early stage of this work. We are grateful to T. Gehrmann,
A. Gehrmann-De Ridder, and G. Heinrich for useful advices concerning
the use of the generator {\tt EERAD3}, and for providing us with a
preliminary version of the code. We also thank P.~Torrielli for his
careful reading of the paper.
GZ is supported by the ERC grant 614577.  PM is supported by the
Research Executive Agency (REA) of the European Union under the Grant
Agreement number PITN-GA-2010-264564 (LHCPhenoNet), and by the Swiss
National Science Foundation (SNF) under the grant PBZHP2-147297.
The work of AB is supported by Science and Technology Facility Council 
(STFC) under grant number ST/L000504/1.
This research was partly supported by the Munich Institute for Astro-
and Particle Physics (MIAPP) of the DFG cluster of excellence "Origin
and Structure of the Universe" (PM and GZ) and by the Mainz Institute
for Theoretical Physics (MITP) of the PRISMA excellence cluster (GZ).
We would like to thank the Galileo Galilei Institute (PM and GZ) and
CERN (AB, PM) for hospitality while part of this work was carried out.

\appendix

\section{Observables definition}
\label{sec:obsdef}
In this Appendix we recall the definition of the event-shapes that we
considered in this work. 

\begin{itemize}
\item Thrust: 
\begin{equation}
\label{def-thrust}
T\equiv \max_{\vec{n}}\left.\frac{\sum_{i}\vert \vec{p_i} \cdot \vec{n}
    \vert}{Q}\right.,\qquad \tau \equiv
1-T\,, 
\end{equation}
where  $Q$  is the centre-of-mass energy and the vector $\vec{n}$ that maximizes the sum defines the
direction of the thrust axis, $\vec{n_T}$. The thrust axis divides
each event into two hemispheres, $\mathcal{H}^{(1)}$ and $\mathcal{H}^{(2)}$. 

\item Heavy-jet mass: 
\begin{equation}
\label{def-mH}
\rho_H \equiv \max_{i=1,2} \frac{M_{i}^2}{Q^2}\,, \qquad M_{i}^2 \equiv  \left( \sum_{j
  \in {\cal H}^{(i)}}   p_j \right)^2\,.
\end{equation}
\item C-parameter:
\begin{equation}
\label{def-Cpar}
C \equiv 3\left(1-\frac{1}{2}\sum\limits_{i,j}\frac{(p_i\cdot
    p_j)^2}{(p_i\cdot Q)(p_j\cdot Q)}\right), 
\end{equation}
where $Q^\mu$ is the total four-momentum. 

\item Total broadening:
\begin{equation}
B_T \equiv B_L+B_R,
\end{equation}
where
\begin{equation}
\begin{split}
\label{def-BT}
B_L \equiv &\sum_{i\in
  \mathcal{H}^{(1)}}\frac{\vert\vec{p_i}\times\vec{n_T}\vert}{2 Q}\,,
\quad 
B_R \equiv \sum_{i\in
  \mathcal{H}^{(2)}}\frac{\vert\vec{p_i}\times\vec{n_T}\vert}{2 Q}\,.
\end{split}
\end{equation}

\item Wide broadening:
\begin{equation}
\label{def-BW}
B_W \equiv \max\{B_L, B_R\}.
\end{equation}

\item Thrust-major:
\begin{equation}
\label{def-Tmaj}
T_{M} \equiv \max_{\vec{n}\cdot\vec{n_T}=0}\frac{\sum_{i}\vert
    \vec{p_i} \cdot \vec{n} \vert}{Q} ,
\end{equation}
where the vector $\vec{n}$ for which the sum is maximised  defines the
thrust-major axis.

\item Oblateness:
\begin{equation}
O \equiv T_M-T_{\text{m}},
\end{equation}
where
\begin{equation}
\label{def-Obl}
T_{\text{m}} \equiv \frac{\sum_{i}\vert p_{i,x}\vert}{Q},
\end{equation}
and where $x$ is the direction perpendicular to both the thrust and
the thrust-major axes. 
\end{itemize}

\section{Sudakov radiator}
\label{sec:radiator}
The Sudakov radiator can be parametrised as 
\begin{equation}
\label{eq:rad-gi}
R(v) = -Lg_1(\lambda) - g_2(\lambda) -\frac{\alpha_s}{\pi}g_3(\lambda)
+ ...
\end{equation}
We introduce the resummation scale $x_V$ such that
\begin{equation}
\ln \frac{1}{v} = \ln\frac{x_V}{v} - \ln x_V =
\frac{\lambda}{\alpha_s\beta_0}-\ln x_V\,,
\end{equation}
where $\lambda=\alpha_s\beta_0 \ln x_V/v$. The functions $g_1$, $g_2$ and
$g_3$ can be parametrised as
\begin{equation}
g_i(\lambda) = \sum_{\ell=1,2}g^{(\ell)}_i(\lambda)
\end{equation}
where $g_i^{(\ell)}$ can be expressed in terms of scaling parameters
$a$ and $b_{\ell}$ as follows:

\begin{equation}
\label{eq:g1-b/=0}
\begin{split}
g^{(\ell)}_1(\lambda)=\frac{A_1 \left((a+b_{\ell}-2 \lambda ) \ln \left(1-\frac{2 \lambda }{a+b_{\ell}}\right)-(a-2 \lambda ) \ln \left(1-\frac{2 \lambda }{a}\right)\right)}{4 \pi  b_{\ell} \beta_0 \lambda }\,,
\end{split}
\end{equation}

\begin{equation}
\label{eq:g2-b/=0}
\begin{split}
g^{(\ell)}_2(\lambda)&=\frac{A_2 \left(a \ln \left(1-\frac{2 \lambda }{a}\right)-(a+b_\ell) \ln \left(1-\frac{2 \lambda
   }{a+b_\ell}\right)\right)}{8 \pi ^2 b_\ell \text{$\beta_0
 $}^2}+\frac{B_1\ln \left(1-\frac{2 \lambda }{a+b_\ell}\right)}{4
\pi  \text{$\beta_0 $}}\\
&+\frac{A_1 \left(\text{$\beta_1 $} (a+b_\ell) \ln ^2\left(1-\frac{2 \lambda }{a+b_\ell}\right)+2 \text{$\beta_1 $} (a+b_\ell) \ln \left(1-\frac{2 \lambda
   }{a+b_\ell}\right)\right)}{8 \pi  b_\ell \text{$\beta_0 $}^3}\\
&-A_1\frac{\ln \left(1-\frac{2 \lambda }{a}\right) \left(a \text{$\beta_1 $} \ln \left(1-\frac{2 \lambda }{a}\right)+2 a \text{$\beta
  _1 $}\right)}{8 \pi  b_\ell \text{$\beta_0 $}^3}\\
& + \frac{A_1 \left(\ln \left(1-\frac{2 \lambda
      }{a+b_\ell}\right)-\log \left(1-\frac{2 \lambda }{a}\right)\right)}{4
  \pi  b_\ell \text{$\beta_0 $}} \ln x_V^2\\
&-\frac{A_1 \left(\ln \left(1-\frac{2 \lambda }{a+b_\ell}\right)-\ln \left(1-\frac{2 \lambda }{a}\right)\right)}{2\pi  b_\ell \text{$\beta_0 $}}\ln\bar{d_{\ell}}\,,
\end{split}
\end{equation}

\begin{equation}
\label{eq:g3-b/=0}
\begin{split}
g^{(\ell)}_3(\lambda)&=\frac{\beta_1 B_1 \left((a+b_\ell) \ln
    \left(1-\frac{2 \lambda }{a+b_\ell}\right)+2 \lambda \right)}{4
  \text{$\beta_0$}^2 (a+b_\ell-2 \lambda )}\\
&+\frac{A_2 \text{$\beta_1 $} \left(a^2 (a+b_\ell-2 \lambda ) \ln \left(1-\frac{2 \lambda }{a}\right)-(a+b_\ell)^2 (a-2 \lambda ) \ln \left(1-\frac{2 \lambda
   }{a+b_\ell}\right)+6 b_\ell \lambda ^2\right)}{8 \pi  b_\ell \text{$\beta_0 $}^3
(a-2 \lambda ) (a+b_\ell-2 \lambda )}\\
&+\frac{A_1\left(\text{$\beta_1 $}^2 (a+b_\ell)^2 (a-2 \lambda ) \ln ^2\left(1-\frac{2 \lambda }{a+b_\ell}\right)-4 b_\ell \lambda ^2 \left(\text{$\beta_0 $}
   \text{$\beta_2 $}+\text{$\beta_1 $}^2\right)\right)}{8 b_\ell
\text{$\beta_0 $}^4 (a-2 \lambda ) (a+b_\ell-2 \lambda )}\\
&-\frac{a A_1 \ln \left(1-\frac{2 \lambda }{a}\right) \left(2 \text{$\beta_0 $} \text{$\beta_2 $} (a-2 \lambda )+a \text{$\beta_1 $}^2 \ln
   \left(1-\frac{2 \lambda }{a}\right)+4 \text{$\beta_1 $}^2 \lambda
 \right)}{8 b_\ell \text{$\beta_0 $}^4 (a-2 \lambda )}\\
&+\frac{A_1 (a+b_\ell) \ln \left(1-\frac{2 \lambda }{a+b_\ell}\right) \left(\text{$\beta_0 $} \text{$\beta_2 $} (a+b_\ell-2 \lambda )+2 \text{$\beta_1 $}^2 \lambda
   \right)}{4 b_\ell \text{$\beta_0 $}^4 (a+b_\ell-2 \lambda )}\\
&- \frac{A_1}{8 (a-2 \lambda ) (a+b_\ell-2 \lambda )} \ln^2 x_V^2 +
\left[ \frac{\pi  a \text{$\beta_0 $}^2 B_1+\lambda  \left(A_2 \text{$\beta_0 $}-2 \pi  \left(A_1 \text{$\beta_1 $}+\text{$\beta_0 $}^2
   B_1\right)\right)}{4 \pi  \text{$\beta_0 $}^2 (a-2 \lambda )
(a+b_\ell-2 \lambda )} \right . \\
&\left.+\frac{A_1 \text{$\beta_1 $} \left((a+b_\ell) (a-2 \lambda )
      \ln \left(1-\frac{2 \lambda }{a+b_\ell}\right)-a (a+b_\ell-2
      \lambda ) \ln \left(1-\frac{2 \lambda
   }{a}\right)\right)}{4 b_\ell \text{$\beta_0 $}^2 (a-2 \lambda ) (a+b_\ell-2
\lambda )}\right]\ln x_V^2\\
&-\frac{A_1}{2(a-2 \lambda ) (a+b_\ell-2 \lambda )}\ln\bar{d^2_{\ell}}
+ \frac{A_1}{2(a-2 \lambda ) (a+b_\ell-2 \lambda )} \ln\bar{d_{\ell}}
\ln x_V^2\\
& -\frac{\pi  a \text{$\beta_0 $}^2 B_1+\lambda  \left(A_2 \text{$\beta_0 $}-2 \pi  \left(A_1 \text{$\beta_1 $}+\text{$\beta_0 $}^2
   B_1\right)\right)}{2\pi  \text{$\beta_0 $}^2 (a-2 \lambda )
(a+b_\ell-2 \lambda )}\ln\bar{d_{\ell}}\\
&+\frac{A_1 \text{$\beta_1 $} \left(a (a+b_\ell-2 \lambda ) \ln
    \left(1-\frac{2 \lambda }{a}\right)-(a+b_\ell) (a-2 \lambda ) \ln \left(1-\frac{2 \lambda
   }{a+b_\ell}\right)\right)}{2b_\ell \text{$\beta_0 $}^2 (a-2 \lambda
) (a+b_\ell-2 \lambda )}\ln\bar{d_{\ell}} \\
& + \frac{7 }{8 b_\ell} C_F\frac{1}{1-\frac{2}{a+b_\ell}\lambda}\Theta(b_\ell)+ h(\lambda)\,,
\end{split}
\end{equation}

where $\ln\bar{d^{n}_{\ell}}=\int_0^{2\pi}\frac{d\phi}{2\pi}\ln^n
(d_{\ell} g_{\ell}(\phi))$ and $\Theta(b_\ell) = 1 (0)$ for $b_\ell >0$
($b_\ell = 0$). The renormalisation scale dependence can be
restored using the following replacements in eq.~\eqref{eq:rad-gi}
\begin{equation}
\begin{split}
g_1(\lambda) \rightarrow&\,\, g_1(\lambda)\,,\\
g_2(\lambda) \rightarrow&\,\, g_2(\lambda) + \lambda^2 g'_1(\lambda)
\ln\frac{\mu_R^2}{Q^2}\,,\\
g_3(\lambda) \rightarrow&\,\, g_3(\lambda) + \pi\left( \beta_0\lambda
  g'_2(\lambda)+\frac{\beta_1}{\beta_0}\lambda^2g'_1(\lambda)\right)
\ln\frac{\mu_R^2}{Q^2}
 + \pi\left( \beta_0 \lambda^2 g'_1(\lambda) + \frac{\beta_0}{2}\lambda^3g''_1(\lambda)\right) \ln^2\frac{\mu_R^2}{Q^2}\,.
\end{split}
\end{equation}
The coefficients of the QCD $\beta$ function used above are defined as
\begin{align}
     &\beta_0 = \frac{11 C_A - 2 n_f}{12\pi}\,,\quad 
     \beta_1 = \frac{17 C_A^2 - 5 C_A n_f - 3 C_F n_f}{24\pi^2}\,,\\ 
     & \beta_2 = \frac{2857 C_A^3+ (54 C_F^2 -615C_F C_A -1415 C_A^2)n_f
       +(66 C_F +79 C_A) n_f^2}{3456\pi^3}\,. 
\end{align}
 The following functions are also used in the text
\begin{align}
\label{eq:R'ell-NLL}
R'_{\rm NLL,\ell}(v) &= \frac{A_1 \left(\ln \left(1-\frac{2 \lambda
      }{a+b_\ell}\right)-\ln \left(1-\frac{2 \lambda
      }{a}\right)\right)}{2\pi  b_\ell \text{$\beta_0 $}}\,,\\
\delta R'_{\rm NNLL,\ell}(v) &=\frac{\alpha_s(Q)}{\pi}\left[-\frac{A_1 \text{$\beta_1 $} \left(a (a+b_\ell-2 \lambda ) \ln \left(1-\frac{2 \lambda }{a}\right)-(a+b_\ell) (a-2 \lambda ) \ln \left(1-\frac{2 \lambda
   }{a+b_\ell}\right)+2 b_\ell \lambda \right)}{2 b_\ell \text{$\beta_0 $}^2 (a-2
\lambda ) (a+b_\ell-2 \lambda )}\right.\notag\\
&\left.+\frac{A_1 \lambda }{(a-2 \lambda )
   (a+b_\ell-2 \lambda )}\ln\frac{\mu_R^2}{Q^2}-\frac{A_1}{2 (a-2 \lambda )
   (a+b_\ell-2 \lambda )}\ln x_V^2\right.\notag\\
\label{eq:R'ell-NNLL}
&\left.+\frac{A_2 \lambda }{2 \pi  \text{$\beta_0 $} (a-2 \lambda
   ) (a+b_\ell-2 \lambda )}\right]\,,\\
\label{eq:R''ell-NNLL}
R''_\ell(v) &=\frac{\alpha_s(Q)}{\pi}\frac{A_1 }{  (a-2 \lambda ) (a+b_\ell-2 \lambda )}\,.
\end{align}
The limit $b_\ell\to 0$ (relevant for jet broadenings, thrust major and
oblateness) is finite and well defined for all the above
expressions.
The function $h(\lambda)$, implicitly defined in
eq.~\eqref{eq:radiator-NNLL}, is extracted from the resummed expression of 
\begin{equation}
\Sigma(v) = \frac{1}{\sigma} \int_0^vd v' \frac{d\sigma(v')}{dv'},
\end{equation}
(where $\sigma$ is the total cross section for $e^+e^-\to$\,hadrons)
for the two reference observables (i.e. $k_t$ and thrust $1-T$)
leading to
\begin{equation}
\begin{split}
\label{eq:h-thrust}
h^{(1-T)}(\lambda) &= -A^{(1-T)}_3\frac{\lambda ^2}{8 \pi ^2
  \text{$\beta_0 $}^2 (1-2 \lambda ) (2-2 \lambda )}
-B^{(1-T)}_2\frac{\lambda }{8 \pi \text{$\beta_0 $}(1- \lambda) }\\
& + C_F\frac{\pi^2}{24}\frac{1}{1-2\lambda} +
C_F\left(\frac{1}{4}-\frac{\pi^2}{12}\right) \frac{1}{1-\lambda} +
C_F\left(-\frac{19}{8}+\frac{7}{24}\pi^2 \right)\,,\\
\end{split}
\end{equation}
and
\begin{equation}
\begin{split}
\label{eq:h-kt}
h^{(k_t)}(\lambda) &=-A^{(k_t)}_3\frac{\lambda ^2}{8 \pi ^2
  \text{$\beta_0 $}^2 (1-2 \lambda )^2} -B^{(k_t)}_2\frac{\lambda }{4
  \pi  \text{$\beta_0$}(1-2 \lambda) } \\
&+ C_F\left(\frac{1}{4}-\frac{\pi^2}{24}\right)\frac{1}{1-2\lambda}+
C_F\left(-\frac{19}{8}+\frac{7}{24}\pi^2 \right)\,.
\end{split}
\end{equation}
Note that the above expressions imply that $g_3(0)\neq 0$. This means
that constant terms appear in the exponent, which can be expanded to
${\cal O}(\alpha_s)$ neglecting subleading terms.

\noindent The anomalous dimensions $A_i$ and $B_i$ used in the above expressions
are:
\begin{equation}
A_1= 2C_F\,,
\end{equation}
\begin{equation}
B_1= -3C_F\,,
\end{equation}
\begin{equation}
A_2= C_F\left(C_A\left(\frac{67}{9}-\frac{\pi^2}{3}\right)-\frac{10}{9}\,n_f\right)\,,
\end{equation}
the coefficient $A_2$ defines the running coupling in the CMW scheme
used in the definition of the soft emission probability.  The coefficients $B_2$ and
$A_3$ are observable dependent in our study. The expressions for the
two reference observables read
\begin{equation}
B^{(1-T)}_{2}= -2 \left(C_F^2\left(-\frac{\pi^2}{2}+\frac{3}{8}+6\zeta_3\right) + C_FC_A\left(\frac{11\pi^2}{18}+\frac{17}{24}-3\zeta_3\right)
+ C_F T_F n_f\left(-\frac{1}{6}-\frac{2}{9}\pi^2\right)\right)\,,
\end{equation}
\begin{equation}
B^{(k_t)}_{2}= B^{(1-T)}_{2} + 2\pi\beta_0\zeta_2C_F\,,
\end{equation}
\begin{equation}
\begin{split}
&A^{(1-T)}_{3}=  C_FC_A^2\left(\frac{245}{12} - \frac{67}{27} \pi^2+
  \frac{11}{3}\zeta_3 + \frac{22}{5}\zeta_2^2\right) + C_F^2T_F
n_f\left(-\frac{55}{6} + 8\zeta_3\right) - \frac{8}{27}C_F T_F^2 n_f^2 \\ &+ C_F C_AT_F
n_f\left(-\frac{209}{27}+\frac{20}{27}\pi^2  -
  \frac{28}{3}\zeta_3\right)  +\pi\beta_0C_F\left(C_A\left(\frac{808}{27}-28\zeta_3\right)-\frac{224}{27}T_F n_f\right)\,,
\end{split}
\end{equation}
\begin{equation}
A^{(k_t)}_{3}= A^{(1-T)}_{3}-8\pi^2\beta_0^2\zeta_2C_F\,.
\end{equation}

\section{Analytic NNLL results for additive observables}
\label{sec:additive}

Some event shapes have the property that they are additive, meaning
that for soft emissions 
\begin{equation}
  \label{eq:additive}
  V(\{\tilde p\}, k_1,\dots,k_n) = \sum_{i=1}^{n}V(\{\tilde p\},k_i)
  +{\cal O}(V^2)\,,
\end{equation}
while for a hard emission $k$ collinear to leg $\ell$, the
corresponding $V(\{\tilde p\},k)$ has to be replaced by
$V^{(k)}(\{\tilde p\},k[k_t', p'_{t,\ell},z^{(\ell)}])$, as defined in
section~\ref{sec:recoil}.  This is the case for instance for the
thrust, the $C$-parameter and the heavy-jet masses.
For this simpler class of observables, the NNLL corrections can be
simplified significantly. In this appendix we work out the NNLL
corrections of Sec.~\ref{sec:nnll} for these additive observables
analytically. This provides a check of the
numerical implementation of our method. We also show that for these
observables the corresponding NNLL correction $\delta \mathcal{F}_{\rm
  NNLL}$ factorises in a coefficient that multiplies the NLL function
$\mathcal{F}_{\rm NLL}(\lambda)$, that for an additive observable
reads
\begin{equation}
\label{eq:fnll-add}
\mathcal{F}_{\rm NLL}(\lambda)=\frac{e^{-\gamma_E
  R'_{\rm NLL}}}{\Gamma(1+R'_{\rm NLL})}.
\end{equation}

\subsection{Soft-collinear correction}
\label{sec:sc-correction-additive}

We consider first the soft-collinear contribution $\delta \mathcal{F}_{\rm
  sc}$ of eq.~(\ref{eq:F1-NNLL}), and use the fact that for additive
observables 
\begin{equation}
  \Vsc(\{\tilde p\}, k,\{k_i\}) = \zeta v +
\Vsc(\{\tilde p\},\{k_i\})\,. 
\label{eq:Vscadd-sc}
  \end{equation}

\begin{equation}
  \label{eq:F1-sc-add}
  \begin{split}
\delta\mathcal{F}_{\rm sc}(\lambda) & = 
\frac{\pi}{\alpha_s(Q)}
\int_0^\infty
    \frac{d\zeta}{\zeta}  
    \sum_{\ell=1,2} \left(\delta
      R'_{{\rm NNLL},\ell}+R''_{\ell}\ln\bar d_{\ell}
      +R''_{\ell}\ln\frac{1}{\zeta}\right)\int \dZ \times \\ & \times 
\left[\Theta\left(1-\zeta-\lim_{v\to
        0}\frac{\Vsc(\{\tilde p\}, \{k_i\})}{v}\right)-\Theta(1-\zeta)\Theta\left(1-\lim_{v\to
        0}\frac{\Vsc(\{\tilde p\},\{k_i\})}{v}\right)\right]\,,
  \end{split}
\end{equation}
where we used the fact that for additive observables the integral over
$\phi$ can be performed analytically.

We can define rescaled momenta $\tilde k_1,\dots,\tilde k_n$ in the
second theta function such that $\Vsc(\{\tilde p\}, \tilde k_i) =
\Vsc(\{\tilde p\}, k_i)/(1-\zeta)$. Recursive IRC safety of $V$
guarantees that
\begin{equation}
  \label{eq:sc-rescale}
  \Vsc(\{\tilde p\}, \{k_i\}) = (1-\zeta) \,\Vsc(\{\tilde p\}, \{\tilde k_i\})\,.
\end{equation}
Using the explicit expression for $d {\cal Z}$, and defining
$\tilde\zeta_i=\Vsc(\{\tilde p\},\tilde k_i)/v$, one gets 

\begin{equation}
  \begin{split}
&\delta\mathcal{F}_{\rm sc}(\lambda) = 
\frac{\pi}{\alpha_s(Q)}
\int_0^\infty
    \frac{d\zeta}{\zeta}  
    \sum_{\ell=1,2} \left(\delta
      R'_{{\rm NNLL},\ell}+R''_{\ell}\ln\bar d_{\ell}
      +R''_{\ell}\ln\frac{1}{\zeta}\right) \epsilon^{\RpNLL} \sum_{n=0}^{\infty}\frac{1}{n!}
     \prod_{i=1}^n\sum_{\ell_i=1,2}R'_{{\rm NLL},\ell_i} \\ &\times\int_0^{2\pi}\frac{d\phi_i}{2\pi} \Theta(1-\zeta)
\left[\int^{\infty}_{\frac{\epsilon}{1-\zeta}}\frac{d\tilde\zeta_i}{\tilde\zeta_i}\Theta\left(1-\lim_{v\to
        0}\frac{\Vsc(\{\tilde p\}, \{\tilde k_i\})}{v}\right)-\int_\epsilon^{\infty}\frac{d\zeta_i}{\zeta_i}\Theta\left(1-\lim_{v\to
        0}\frac{\Vsc(\{\tilde p\},\{k_i\})}{v}\right)\right]\,.
  \end{split}
\end{equation}
We can then rearrange the above equation to reconstruct the
function $\mathcal{F}_{\rm NLL}(\lambda)$. This gives 
\begin{equation}
  \begin{split}
    \label{eq:Fsc-add-final}
\delta\mathcal{F}_{\rm sc}(\lambda) & = {\cal F}_{\rm NLL}(\lambda)\frac{\pi}{\alpha_s(Q)}\int_0^1\frac{d\zeta}{\zeta}\sum_{\ell=1,2}\left(\delta
      R'_{{\rm NNLL},\ell}+R''_{\ell}\ln\bar d_{\ell}
      +R''_{\ell}\ln\frac{1}{\zeta}\right)\left(
      (1-\zeta)^{\RpNLL}-1\right)\\ &
    = - {\cal F}_{\rm NLL}(\lambda)\frac{\pi}{\alpha_s(Q)} \sum_{\ell=1,2}\left(\left(\delta
      R'_{{\rm NNLL},\ell}+R''_{\ell}\ln\bar d_{\ell}\right) \left(\psi^{(0)}(1+\RpNLL)+\gamma_E \right)\right.\\
    &\left.  +\frac{R''_{\ell}}{2} \left(\left(\psi ^{(0)}(1+\RpNLL)+\gamma_E\right)^2-
   \psi ^{(1)}(1+\RpNLL)+\frac{\pi ^2}{6}\right)\right) .
  \end{split}
\end{equation}
\subsection{Recoil correction}
\label{sec:recoil-correction-additive}

Let us now consider the recoil contribution $\delta \mathcal{F}_{\rm
  rec}$ of eq.~(\ref{eq:Frec}). 
Considering a hard emission collinear to leg $\ell$, for an additive
observable one has 
\begin{equation}
  \Vhc^{(k')}(\{\tilde p\}, k',\{k_i\}) = \left(\frac{k_t'}{Q}\right)^{a+b_\ell}
  f^{(\ell)}(z^{(\ell)},\phi)
+
\Vsc(\{\tilde p\},\{k_i\})\,,
\label{eq:Vhcadd}
   \end{equation}
and 
\begin{equation}
  \Vsc(\{\tilde p\}, k,\{k_i\}) = \left(\frac{k_t}{Q}\right)^{a+b_\ell}
  f_{\rm sc}^{(\ell)}(z^{(\ell)},\phi)
+
\Vsc(\{\tilde p\},\{k_i\})\,,
\label{eq:Vscadd}
 \end{equation}
where the presence of $k'$, rather than $k$, denotes that the full
recoil has been taken into account in the calculation of the
observable.

Using the above equations in eq.~(\ref{eq:Frec})
we get 
\begin{equation}
  \label{eq:F1-rec-additive}
  \begin{split}
  \delta\mathcal{F}_{\rm rec}(\lambda)&=\sum_{\ell=1,2}
   \frac{\alpha_s(v^{1/(a+b_\ell)}Q)}{\alpha_s(Q)(a+b_\ell)}
   \int_0^{\infty}\frac{d\zeta}{\zeta}
   \int_0^{2\pi}\frac{d\phi}{2\pi}\int \dZ
   \int_0^1 \!dz\,p_\ell(z)\times \\ & \times
   \left[\Theta\left(1-\zeta f^{(\ell)}(z,\phi)-\lim_{v\to 0}
       \frac{\Vsc(\{\tilde p\}, \{k_i \})}{v}\right)
     -\Theta\left(1-\zeta f^{(\ell)}_{\rm sc}(z,\phi)-\lim_{v\to 0}
       \frac{\Vsc(\{\tilde p\}, \{k_i \})}{v}\right)\right]\,,
  \end{split}
\end{equation}
where $\zeta v=(k_t/Q)^{a+b_\ell}$.  We can define rescaled momenta
$\tilde k_1,\dots,\tilde k_n$ in the second theta function such that
$\Vsc(\{\tilde p\}, \tilde k_i) = \Vsc(\{\tilde p\}, k_i)/(1-\zeta
f_{\rm sc}^{(\ell)}(z,\phi))$.
Recursive IRC safety of $V$ guarantees that
\begin{equation}
  \label{eq:rec-rescale}
  \Vsc(\{\tilde p\}, k_1,\dots, k_n) = (1-\zeta f_{\rm sc}^{(\ell)}(z,\phi)) \Vsc(\{\tilde p\}, \tilde k_1,\dots,
  \tilde k_n)\,.
\end{equation}
Analogously, we define soft and collinear momenta $\tilde k'_1,\dots,\tilde
k'_n$ in the theta function containing $f^{(\ell)}(z,\phi))$ such that
\begin{equation}
  \label{eq:rec-rescale2}
  \Vsc(\{\tilde p\}, k_1,\dots, k_n) = (1-\zeta f^{(\ell)}(z,\phi)) \Vsc(\{\tilde p\}, \tilde k_1',\dots,
  \tilde k_n')\,.
\end{equation}
Using the explicit expression for $d {\cal Z}$, one gets 
\begin{equation}
  \label{eq:F1-rec-additive-rescaled}
  \begin{split}
  \delta\mathcal{F}_{\rm rec}(\lambda)&=\sum_{\ell=1,2}
   \frac{\alpha_s(v^{1/(a+b_\ell)}Q)}{\alpha_s(Q)(a+b_\ell)}
   \int_0^{\infty}\frac{d\zeta}{\zeta}
   \int_0^{2\pi}\frac{d\phi}{2\pi}
 \int_0^1 \!dz\,p_\ell(z) \epsilon^{\RpNLL} \sum_{n=0}^{\infty}\frac{1}{n!}
     \prod_{i=1}^n\sum_{\ell_i=1,2}R'_{{\rm NLL},\ell_i} \int_0^{2\pi}\frac{d\phi_i}{2\pi}\times \\ & \times
   \left[\Theta(1\!-\! \zeta f^{(\ell)}(z))
     \int_{0}^{\infty}\frac{d\tilde\zeta_i'}{\tilde\zeta_i'}
    \Theta\left(\tilde \zeta'_i -
       \frac{\epsilon}{1\!-\! \zeta
         f^{(\ell)}(z)}\right)\Theta\left(1-\lim_{v\to 0}
       \frac{\Vsc(\{\tilde p\}, \tilde k_1', \ldots, \tilde k_n')}{v}\right)
\right.\\&\left.
-\Theta(1\!-\!\zeta f^{(\ell)}_{\rm sc} (z))
     \int_{0}^{\infty}\frac{d\tilde\zeta_i}{\tilde\zeta_i}
\Theta\left(\tilde \zeta_i -
       \frac{\epsilon}{1\!-\!\zeta f^{(\ell)}_{\rm sc} (z)}\right)
\Theta\left(1-\lim_{v\to 0} \frac{\Vsc(\{\tilde p\}, \tilde k_1, \ldots,
    \tilde k_n )}{v}\right)
\right]\,.
  \end{split}
\end{equation}
We can then rearrange the above equation to reconstruct the
function $\mathcal{F}_{\rm NLL}(\lambda)$. This gives 
\begin{equation}
  \label{eq:F1-rec-additive-final}
  \begin{split}
 \delta\mathcal{F}_{\rm rec}(\lambda)&=\mathcal{F}_{\rm NLL}(\lambda)\sum_{\ell=1,2}
   \frac{\alpha_s(v^{1/(a+b_\ell)}Q)}{\alpha_s(Q)(a+b_\ell)}
   \int_0^{2\pi}\frac{d\phi}{2\pi}
  \int_0^1 \!dz\,p_\ell(z) \times \\ & \times
   \int_0^{\infty} \frac{d\zeta}{\zeta} \left[
(1-\zeta f^{(\ell)}(z,\phi))^{\RpNLL}\Theta (1-\zeta
f^{(\ell)}(z,\phi))-(1-\zeta f^{(\ell)}_{\rm sc}(z,\phi))^{\RpNLL}\Theta
(1-\zeta f^{(\ell)}_{\rm sc}(z,\phi))\right]\\
& = \mathcal{F}_{\rm NLL}(\lambda)\sum_{\ell=1,2}
   \frac{\alpha_s(v^{1/(a+b_\ell)}Q)}{\alpha_s(Q)(a+b_\ell)} \int_0^{2\pi}\frac{d\phi}{2\pi}
   \int_0^1 \!dz\,p_\ell(z) \ln\frac{f^{(\ell)}_{\rm sc}(z,\phi)}{f^{(\ell)}(z,\phi)}\,.
  \end{split}
\end{equation}
As an example, we consider the thrust. One can show that its
expression in terms of Sudakov variables is
\begin{equation}
  \label{eq:tau-def}
  1-T = \sum_{i=1}^{n} \frac{k_{ti}}{Q} e^{-|\eta_i|} +
\frac{1}{Q^2} \sum_{\ell=1,2}
\frac{\left(\sum_{i\in \mathcal{H}^{(\ell)}}\vec
        k^{(\ell)}_{ti}\right)^2}{1-\sum_{i\in \mathcal{H}^{(\ell)}}
        z_i^{(\ell)}} \,.
\end{equation}
Suppose $k$ is collinear to leg $\tilde p_1$. Using the Sudakov
parametrisation of eq.~(\ref{eq:Sudakov-tilde})  we then have
\begin{equation}
  \label{eq:tau-def2}
  1-T \simeq  \sum_{i=1}^{n} \frac{k_{ti}}{Q} e^{-|\eta_i|} + \frac{k_t^2}{z^{(1)}Q^2}+
  \frac{k_t^2}{(1-z^{(1)})Q^2} = \sum_{i=1}^{n} \frac{k_{ti}}{Q}
  e^{-|\eta_i|}  + \frac{k_t^2}{z^{(1)}(1-z^{(1)})Q^2}\,,
\end{equation}
where we have used the fact that the hard-collinear $k_t$ is larger than all
soft-collinear $k_{ti}$, and therefore $k_t\simeq k_t'$. A hard collinear emission gives an additive contribution to
the observable, so that we can apply
eq.~(\ref{eq:F1-rec-additive-final})  with 
\[
f^{(\ell)} (z^{(\ell)},\phi) =
\frac{1}{z^{(\ell)}(1-z^{(\ell)})}\,,\qquad 
f^{(\ell)}_{\rm sc} (z^{(\ell)},\phi) =
\frac{1}{z^{(\ell)}}\,.
\]
This gives
\begin{equation}
  \begin{split}
  \label{eq:F1-rec-thrust}
  \delta\mathcal{F}_{\rm rec}(\lambda) & = 
  \mathcal{F}_{\rm NLL}(\lambda) 2 C_F\frac{\alpha_s(\sqrt{\tau} Q)}{2\alpha_s(Q)} 
  \int_0^1 \!\!dz\, \frac{(1+(1-z)^2)}{z} \ln(1-z)\\ &=
\mathcal{F}_{\rm NLL}(\lambda)\frac{C_F \alpha_s(\sqrt{\tau} Q)}{\alpha_s(Q)} \left(\frac{5}{4}-\frac{\pi^2}{3}\right)
\,.
  \end{split}
\end{equation}
This result holds also for the $C$-parameter and the heavy-jet mass,
which behave as $1\!-\!T$ in the collinear region. 

\subsection{Hard-collinear correction}
\label{sec:hard-coll-corr}
In a similar way, we compute here the hard-collinear function
$\delta\mathcal{F}_{\rm hc}(\lambda)$ of eq.~(\ref{eq:Fhc}). 
Using eq.~\eqref{eq:Vscadd} we obtain
\begin{equation}
  \label{eq:F-hc-additive}
\begin{split}
 \delta \mathcal{F}_{\rm hc}(\lambda) &= \sum_{\ell=1,2}
   \frac{\alpha_s(v^{1/(a+b_\ell)}Q)}{\alpha_s(Q)(a+b_\ell)}\int_{0}^{\infty}
  \frac{d\zeta}{\zeta}\int_0^{\pi}\frac{d\phi}{2\pi}\int_0^1
  \frac{dz}{z}\left(zp_\ell(z)-2C_\ell\right)\int \dZ
 \times \\ & \times
    \left[\Theta\left(1-\zeta-
        \lim_{v\to 0} \frac{\Vsc(\{\tilde p\}, \{k_i \} )}{v}\right)-\Theta(1-\zeta)
      \Theta\left(1-\lim_{v\to
      0}\frac{\Vsc(\{\tilde p\}, \{k_i \}) } {v}\right)\right]\,.
\end{split}
\end{equation}
Rescaling the momenta in a similar way as we have done in the previous
section we get
\begin{equation}
  \label{eq:F-hc-rescaled-final}
\begin{split}
  \delta\mathcal{F}_{\rm hc}(\lambda) &= \mathcal{F}_{\rm
    NLL}(\lambda) \sum_{\ell=1,2}
   \frac{\alpha_s(v^{1/(a+b_\ell)}Q)}{\alpha_s(Q)(a+b_\ell)}\int_{0}^{\infty}
  \frac{d\zeta}{\zeta}\int_0^{2\pi}\frac{d\phi}{2\pi} \int_0^1
  \frac{dz}{z}\left(zp_\ell(z)-2C_\ell\right) \times \\ & \times
    \left[\left(1-\zeta\right)^{\RpNLL}\Theta\left(1-\zeta \right)   
    -\Theta(1-\zeta)
      \right] = \mathcal{F}_{\rm NLL}(\lambda)\sum_{\ell=1,2}
   \frac{\alpha_s(v^{1/(a+b_\ell)}Q)}{\alpha_s(Q)(a+b_\ell)}\times\\
&\times C_\ell B_\ell\int_{0}^{1}
  \frac{d\zeta}{\zeta} \left[(1-\zeta)^{\RpNLL}-1\right]\,.
\end{split}
\end{equation}
For thrust, using the explicit expression for $\mathcal{F}_{\rm NLL}$~\eqref{eq:fnll-add} we obtain
\begin{equation}
  \label{eq:F-hc-thrust}
\begin{split}
  \delta\mathcal{F}_{\rm hc} (\lambda) = \frac{\alpha_s(\sqrt{\tau}
    Q)}{\alpha_s(Q)} C_F\frac{3}{2}\left(\psi^{(0)}(1+R'_{\rm
        NLL})+\gamma_E\right) \mathcal{F}_{\rm NLL}.
\end{split}
\end{equation}

\subsection{Soft large-angle correction}
\label{sec:soft-large-angle-additive}

We consider now the case of a NNLL correction induced by a soft
large-angle emission, eq.~(\ref{eq:dF-soft-final}) . Then we have

\begin{equation}
  \Vwa^{(k)}(\{\tilde p\}, k,\{k_i\}) = \left(\frac{k_t}{Q}\right)^{a}
  f_{\rm wa}(\eta,\phi)
+
\Vsc(\{\tilde p\},\{k_i\})\,,
\label{eq:Vwaadd}
  \end{equation}
\begin{equation}
  \Vsc(\{\tilde p\}, k,\{k_i\}) = \left(\frac{k_t}{Q}\right)^{a}
  f_{\rm sc}(\eta,\phi)
+
\Vsc(\{\tilde p\},\{k_i\})\,,
\label{eq:Vscadd-wa}
  \end{equation}
where $f_{\rm sc}(\eta,\phi)$ and $f_{\rm wa}(\eta,\phi)$ are defined
in eqs.~\eqref{eq:Vwa} and~\eqref{eq:Vsc}.  
Performing a similar rescaling as for the recoil correction one finds
\begin{equation}
  \label{eq:F1-soft-additive-final}
  \begin{split}
 \delta \mathcal{F}_{\rm wa}(\lambda)&= \mathcal{F}_{\rm NLL}(\lambda) \frac{2 C_F}{a}
  \frac{\alpha_s(v^{\frac{1}{a}}Q)}{\alpha_s(Q)}\int_0^{2\pi}\frac{d\phi}{2\pi}
   \int_{-\infty}^\infty \!\!d\eta \ln\frac{f_{\rm
       sc}(\eta,\phi)}{f_{\rm wa}(\eta,\phi)}\,.
  \end{split}
\end{equation}
For the
thrust and the heavy-jet mass:
\begin{equation}
f_{\rm wa} (\eta,\phi) = f_{\rm sc} (\eta,\phi) = e^{-|\eta|}\,,
\end{equation}
so that $\delta \mathcal{F}_{\rm wa}(\lambda)=0$. In the case of the
$C$-parameter instead we have
\begin{equation}
f_{\rm wa}(\eta,\phi) =\frac{3}{\cosh\eta}\, \quad {\rm and}\quad f_{\rm sc} (\eta,\phi) = 6\, e^{-|\eta|}\,.
\end{equation}
This gives
\begin{equation}
  \label{eq:F1-soft-Cparameter}
  \begin{split}
\delta  \mathcal{F}_{\rm wa}(\lambda)&= \mathcal{F}_{\rm NLL}(\lambda) 2C_F
  \frac{\alpha_s(CQ)}{\alpha_s(Q)}
\int_{-\infty}^\infty \!\!\! d\eta \,\ln(2 \cosh \eta e^{-|\eta|}) =
\mathcal{F}_{\rm NLL}(\lambda) C_F
  \frac{\alpha_s(CQ)}{\alpha_s(Q)} \frac{\pi^2}{6}\,, 
 \end{split}
\end{equation}
where $C$ is the value of the $C$-parameter. 

\subsection{Soft correlated correction}
\label{sec:soft-correlated-additive}

The correlated correction presented in eq.~\eqref{eq:F1-correl} depends on the
difference 
\begin{equation}
  \label{eq:theta-diff}
    \Theta\left(v-\Vsc(\{\tilde
        p\},k_a,k_b,k_1,\dots,k_n)\right)-\Theta\left(v-\Vsc(\{\tilde
        p\},k_a+k_b,k_1,\dots,k_n)\right)\,,
\end{equation}
which is in general non-zero for additive observables.  However, the
above correction vanishes if the observable $\Vsc$ is inclusive,
i.e. $\Vsc(k_a,k_b)=\Vsc(k_a+k_b)$. There are observables which are
inclusive in particular regions of the phase space. As an example, the
thrust $T$ happens to be inclusive only for emissions that propagate
into the same hemisphere (defined by the thrust axis itself). In this
case, the difference~\eqref{eq:theta-diff} is non-zero if the two
correlated soft partons $k_a$, $k_b$ move into opposite
hemispheres. However, this configuration requires the parent gluon to
be emitted at small rapidities, which in the limit $T\rightarrow 1$
gives rise to a correction which is at most N$^3$LL, and can be
neglected accordingly. The other additive observables treated in this
article are also inclusive in the relevant phase space regions, so we
can conclude that for $T$, $C$, and $\rho_H$, at NNLL
\begin{equation}
  \label{eq:F1-correl-T}
  \begin{split}
  \delta\mathcal{F}_{\rm correl}(\lambda) =  0\,. 
 \end{split}
\end{equation}

\section{Monte Carlo determination of real emission corrections}
\label{sec:monte-carlo-determ}
Both the NLL function $\mathcal{F}_{\mathrm{NLL}}(\lambda)$ and the NNLL correction
$\delta \mathcal{F}_{\mathrm{NNLL}}(\lambda)$ can be computed efficiently with a Monte Carlo
procedure. In this appendix we recall the procedure devised in
ref.~\cite{Banfi:2001bz}, simplifying the notation so that it can be
easily adapted to the NNLL case. We then discuss the MC determination
of all NNLL corrections. 

\subsection{The function $\mathcal{F}_{\rm NLL}$}
\label{sec:MC-F0}
We now recall the procedure of ref.~\cite{Banfi:2004yd} to efficiently
compute the function $\mathcal{F}_{\mathrm{NLL}}(\lambda)$ via a Monte Carlo
procedure. The first observation is that in the sum in
eqs.~(\ref{eq:F-NLL}) and~(\ref{eq:dZ}) 
the term with zero emissions is negligibly small
due to the factor $\epsilon^{R'_{\rm NLL}}$. Second, in all other terms we can
pick up the hardest emission $k_1$ (the one for which $V_{\rm sc}(\{p\},k_1)$ is the largest of all $V_{\rm sc}(\{p\},k_i)$) and neglect all emissions $\bar
k_i$ with $v_i<\epsilon v_1$, with corrections suppressed by powers of
$v_1\sim v$. This gives 
\begin{equation}
  \label{eq:F-MC-NLL}
  \begin{split}
    \mathcal{F}_{\mathrm{NLL}}(\lambda)&=\epsilon^{R'_{\rm NLL}}
    \sum_{\ell_1=1,2} R'_{\rm NLL,\ell_1}\int_{0}^{\infty} \frac{d\zeta_1}{\zeta_1} \zeta_1^{R'_{\rm NLL}}\int_0^{2\pi} \frac{d\phi_1}{2\pi} \times \\ & \times
    \sum_{n=0}^{\infty}\frac{1}{n!} \prod_{i=2}^{n+1} \sum_{\ell_i=1,2} R'_{\rm NLL,\ell_i}\int_{\epsilon \zeta_1}^{\zeta_1} \frac{d\zeta_i}{\zeta_i}\int_0^{2\pi} \frac{d\phi_i}{2\pi} \, \Theta\left(1-\lim_{v\to 0}\frac{V_{\rm sc}(\{\tilde p\},k_1,\dots,k_{n+1})}{v}\right)\,.
  \end{split}
\end{equation}
We now introduce $\tilde\zeta_i=\zeta_i/\zeta_1$, with corresponding
momenta $\tilde k_i$ such that $V_{\rm sc}(\{\tilde p\},\tilde k_i) =
v_i/\zeta_1$. Since $V$ is rIRC safe we have
\begin{equation}
  \label{eq:rescaling}
  V_{\rm sc}(\{\tilde p\},k_1,\dots,k_{n+1}) = \zeta_1V_{\rm sc}( \{\tilde p\},\tilde k_1,\dots,\tilde k_{n+1})\,.
\end{equation}
Substituting into eq.~(\ref{eq:F-MC-NLL}) we have 
\begin{equation}
  \label{eq:F-MC-NLL-rescaled}
  \begin{split}
    \mathcal{F}_{\mathrm{NLL}}(\lambda)&=\epsilon^{R'_{\rm NLL}}
    \sum_{n=0}^{\infty}\frac{1}{n!} \prod_{i=2}^{n+1} \sum_{\ell_i=1,2} R'_{\rm NLL,\ell_i}\int_{\epsilon}^{1} \frac{d\tilde\zeta_i}{\tilde\zeta_i}\int_0^{2\pi} \frac{d\phi_i}{2\pi} \, 
    \times \\ & \times
    \sum_{\ell_1=1,2} R'_{\rm NLL,\ell_1}\int_{0}^{\infty} \frac{d\zeta_1}{\zeta_1} \zeta_1^{R'_{\rm NLL}}\int_0^{2\pi} \frac{d\phi_1}{2\pi} 
\Theta\left(1-\zeta_1 \lim_{v\to 0}\frac{V_{\rm sc}(\{\tilde p\},\tilde k_1,\dots,\tilde k_{n+1})}{v}\right)\,.
  \end{split}
\end{equation}
The $\zeta_1$ integration can be trivially performed to get
\begin{equation}
  \label{eq:F-MC-NLL-final}
  \begin{split}
    \mathcal{F}_{\mathrm{NLL}}(\lambda)&=
    \sum_{\ell_1=1,2} \frac{R'_{\rm NLL,\ell_1}}{R'_{\rm NLL}}\int_0^{2\pi} \frac{d\phi_1}{2\pi} 
   \times \\ & \times
  \epsilon^{R'_{\rm NLL}} \sum_{n=0}^{\infty}\frac{1}{n!} \prod_{i=2}^{n+1} \sum_{\ell_i=1,2} R'_{\rm NLL,\ell_i}\int_{\epsilon}^{1} \frac{d\tilde\zeta_i}{\tilde\zeta_i}\int_0^{2\pi} \frac{d\phi_i}{2\pi} \, 
\exp\left(-R'_{\rm NLL}\ln\lim_{v\to 0}\frac{V_{\rm sc}(\{\tilde p\},\tilde k_1,\dots,\tilde k_{n+1})}{v}\right)\,.
  \end{split}
\end{equation}

\subsection{The function $\delta\mathcal{F}_{\rm sc}$}
\label{sec:MC-F1}
We now extend the procedure devised in section \ref{sec:MC-F0} so as
to be able to efficiently compute $\delta\mathcal{F}_{\rm sc}$ with a Monte
Carlo procedure. First we observe that without any secondary emission
there is no contribution to $\delta\mathcal{F}_{\rm sc}$. We isolate the
hardest emission $k_1$ among $k_1,\dots,k_n$. 

We first consider the case in which the special emission is not the hardest of all, \ie $\zeta<\zeta_1$. This gives

\begin{equation}
\label{eq:F11-NNLL-MC}
  \begin{split}
    \delta \mathcal{F}^{<}_{\rm sc} &= 
    \frac{\pi}{\as(Q)} \sum_{\ell_1=1,2}  R'_{\rm NLL,\ell_1}    \int_0^{\infty}
    \frac{d\zeta_1}{\zeta_1} \zeta_1^{R'_{\rm NLL}}   \int_0^{2\pi}\frac{d\phi_1}{2\pi}
    \int_0^{\zeta_1}
    \frac{d\zeta}{\zeta}  \int_0^{2\pi}\frac{d\phi}{2\pi}\sum_{\ell=1,2}\left(\delta R'_{\rm NNLL,\ell}+R''_{\ell}\ln\left(\frac{d_{\ell}g_{\ell}(\phi)}{\zeta} \right)\right)
   \times \\& \times \left[ \epsilon^{R'_{\rm NLL}}
    \sum_{n=0}^{\infty}\frac{1}{n!} \prod_{i=2}^{n+1} \sum_{\ell_i=1,2}  R'_{\rm NLL,\ell_i}
    \int_{\epsilon \zeta_1}^{\zeta_1} \frac{d\zeta_i}{\zeta_i}\int_0^{2\pi}
    \frac{d\phi_i}{2\pi} \right]
\times \\ & \times
\left[ 
    \Theta\left(1-\lim_{v\to
        0}\frac{V_{\rm sc}(\{\tilde p\},k, k_1,\dots,
        k_{n+1})}{v}\right)
    -\Theta(1-\zeta)\Theta\left(1-\lim_{v\to
        0}\frac{V_{\rm sc}(\{\tilde p\}, k_1,\dots,
        k_{n+1})}{v}\right)
\right]\,.
  \end{split}
\end{equation}
We now rescale all momenta as in section~\ref{sec:MC-F0}, and obtain

\begin{equation}
  \label{eq:F11-NNLL-MC-rescale}
  \begin{split}
     \delta \mathcal{F}^{<}_{\rm sc} &= 
    \frac{\pi}{\as(Q)}\sum_{\ell_1=1,2}  R'_{\rm NLL,\ell_1}    \int_{\epsilon}^{\infty}
    \frac{d\zeta_1}{\zeta_1} \zeta_1^{R'_{\rm NLL}}   \int_0^{2\pi}\frac{d\phi_1}{2\pi} 
\int_0^{1}
    \frac{d\tilde\zeta}{\tilde\zeta}  \int_0^{2\pi}\frac{d\phi}{2\pi}\sum_{\ell=1,2}\left(\delta R'_{\rm NNLL,\ell}+R''_{\ell}\ln\left(\frac{d_{\ell}g_{\ell}(\phi)}{\tilde{\zeta}\zeta_1} \right)\right)
   \times \\& \times \left[ \epsilon^{R'_{\rm NLL}}
    \sum_{n=0}^{\infty}\frac{1}{n!} \prod_{i=2}^{n+1} \sum_{\ell_i=1,2}  R'_{\rm NLL,\ell_i}
    \int_{\epsilon}^{1} \frac{d\tilde\zeta_i}{\tilde\zeta_i}\int_0^{2\pi}
    \frac{d\phi_i}{2\pi} \right]
  \times \\ & \times
  \left[
        \Theta\left(1-\zeta_1\lim_{v\to
        0}\frac{V_{\rm sc}(\{\tilde p\},\tilde k, \tilde k_1,\dots,\tilde
        k_{n+1})}{v}\right)
-\Theta(1-\zeta_1\tilde\zeta)
    \Theta\left(1-\zeta_1\lim_{v\to
        0}\frac{V_{\rm sc}(\{\tilde p\}, \tilde k_1,\dots,\tilde
        k_{n+1})}{v}\right)
   \right]\,.
  \end{split}
\end{equation}
Performing the $\zeta_1$ integration we get

\begin{equation}
  \label{eq:F11-NNLL-MC-rescale2}
  \begin{split}
      \delta \mathcal{F}^{<}_{\rm sc}&= 
   \frac{\pi}{\as(Q)}\sum_{\ell_1=1,2} \frac{R'_{\rm NLL,\ell_1}}{R'_{\rm NLL}}\int_0^{2\pi}\frac{d\phi_1}{2\pi}
    \int_0^1 \frac{d\tilde\zeta}{\tilde\zeta} \int_0^{2\pi}\frac{d\phi}{2\pi}
    \times \left[ \epsilon^{R'_{\rm NLL}} \sum_{n=0}^{\infty}\frac{1}{n!}
      \prod_{i=2}^{n+1} \sum_{\ell_i=1,2}R'_{\rm NLL,\ell_i}
      \int_{\epsilon}^{1}
      \frac{d\tilde\zeta_i}{\tilde\zeta_i}\int_0^{2\pi}
      \frac{d\phi_i}{2\pi} \right] \times \\ & \times \left[
      \sum_{\ell=1,2}
      \left(\left(\delta R'_{\rm NNLL,\ell}+R''_{\ell}\left(\frac{1}{R'_{\rm NLL}}+\ln\left(\frac{d_{\ell}g_{\ell}(\phi)}{\tilde{\zeta}}\right)+\ln\lim_{v\to0}\frac{V_{\rm sc}(\{\tilde p\},\tilde k, \tilde k_1,\dots,\tilde
          k_{n+1})}{v}\right)\right)\times  \right. \right. \\&  \left. \left. \times \exp\left(-R'_{\rm NLL}\ln\lim_{v\to
          0}\frac{V_{\rm sc}(\{\tilde p\},\tilde k, \tilde k_1,\dots,\tilde
          k_{n+1})}{v}\right) \right. \right. \\
    & \left. \left.  -\left(\delta R'_{\rm NNLL,\ell}+R''_{\ell}\left(\frac{1}{R'_{\rm NLL}}+\ln\left(\frac{d_{\ell}g_{\ell}(\phi)}{\tilde{\zeta}}\right)+\ln\max\left[\tilde\zeta,\lim_{v\to
            0}\frac{V_{\rm sc}(\{\tilde p\},\tilde k_1,\dots,\tilde
            k_{n+1})}{v}\right]\right)\right)\times  \right. \right. \\&  \left. \left. \times \exp\left(-R'_{\rm NLL}\ln\max\left[\tilde\zeta,\lim_{v\to
            0}\frac{V_{\rm sc}(\{\tilde p\},\tilde k_1,\dots,\tilde
            k_{n+1})}{v}\right]\right)\right) \right].
  \end{split}
\end{equation}

The second contribution arises when the special emission is the
hardest of all, i.e. $\zeta>\zeta_1$. This gives

\begin{equation}
  \label{eq:F12-NNLL-MC}
  \begin{split}
     \delta \mathcal{F}^{>}_{\rm sc}&= 
    \frac{\pi}{\as(Q)}\int_{0}^{\infty}
    \frac{d\zeta}{\zeta} \zeta^{R'_{\rm NLL}}   \int_0^{2\pi}\frac{d\phi}{2\pi}\sum_{\ell=1,2}\left(\delta R'_{\rm NNLL,\ell}+R''_{\ell}\ln\left(\frac{d_{\ell}g_{\ell}(\phi)}{\zeta} \right)\right)
     \times \\&
    \times \left[ \epsilon^{R'_{\rm NLL}}
    \sum_{n=0}^{\infty}\frac{1}{n!} \prod_{i=1}^{n} \sum_{\ell_i=1,2}  R'_{\rm NLL,\ell_i}
    \int_{\epsilon \zeta}^{\zeta} \frac{d\zeta_i}{\zeta_i}\int_0^{2\pi}
    \frac{d\phi_i}{2\pi} \right]
\times \\ & \times
\left[
    \Theta\left(1-\lim_{v\to
      0}\frac{V_{\rm sc}(\{\tilde p\},k, k_1,\dots,
        k_{n})}{v}\right)
    -\Theta(1-\zeta)
  \Theta\left(1-\lim_{v\to
      0}\frac{V_{\rm sc}(\{\tilde p\},k_1,\dots,
        k_{n})}{v}\right)
  \right]\,.
  \end{split}
\end{equation}

Defining now $\tilde\zeta_i=\zeta_i/\zeta$ we obtain

\begin{equation}
  \label{eq:F12-NNLL-MC-rescaled}
  \begin{split}
     \delta \mathcal{F}^{>}_{\rm sc}&= 
    \frac{\pi}{\as(Q)}\int_{0}^{\infty} \frac{d\zeta}{\zeta} \zeta^{R'_{\rm NLL}}\int_0^{2\pi}\frac{d\phi}{2\pi}\sum_{\ell=1,2}\left(\delta R'_{\rm NNLL,\ell}+R''_{\ell}\ln\left(\frac{d_{\ell}g_{\ell}(\phi)}{\zeta} \right)\right)
    \times \\ & \times \left[ \epsilon^{R'_{\rm NLL}} \sum_{n=0}^{\infty}\frac{1}{n!}
      \prod_{i=1}^{n} \sum_{\ell_i=1,2} R'_{\rm NLL,\ell_i} \int_{\epsilon
      }^{1} \frac{d\tilde\zeta_i}{\tilde\zeta_i}\int_0^{2\pi}
      \frac{d\phi_i}{2\pi} \right]
    \times \\ & \times
    \left[    \Theta\left(1-\zeta\lim_{v\to 0}\frac{V_{\rm sc}(\{\tilde p\},\tilde k, \tilde
        k_1,\dots,\tilde k_{n})}{v}\right)
    -\Theta(1-\zeta)
  \Theta\left(1-\zeta\lim_{v\to
      0}\frac{V_{\rm sc}(\{\tilde p\},\tilde k_1,\dots,\tilde
        k_{n})}{v}\right)
    \right]\,.
  \end{split}
\end{equation}

We can now perform the integration over $\zeta$ to obtain

\begin{equation}
  \label{eq:F12-NNLL-MC-final}
  \begin{split}
      \delta \mathcal{F}^{>}_{\rm sc}&=
    \frac{\pi}{\as(Q)} \frac{1}{R'_{\rm NLL}}\int_0^{2\pi}\frac{d\phi}{2\pi}\times \left[ \epsilon^{R'_{\rm NLL}}
      \sum_{n=0}^{\infty}\frac{1}{n!}  \prod_{i=1}^{n}
      \sum_{\ell_i=1,2} R'_{\rm NLL,\ell_i} \int_{\epsilon }^{1}
      \frac{d\tilde\zeta_i}{\tilde\zeta_i}\int_0^{2\pi}
      \frac{d\phi_i}{2\pi} \right]\times \\ & \times \left[
      \sum_{\ell=1,2}
      \left(\left(\delta R'_{\rm NNLL,\ell}+R''_{\ell}\left(\frac{1}{R'_{\rm NLL}}+\ln d_{\ell}g_{\ell}(\phi)+\ln\lim_{v\to0}\frac{V_{\rm sc}(\{\tilde p\},\tilde k, \tilde k_1,\dots,\tilde
        k_{n+1})}{v}\right)\right)\times \right. \right. \\& \left. \left. \times \exp\left(-R'_{\rm NLL}\ln\lim_{v\to
          0}\frac{V_{\rm sc}(\{\tilde p\},\tilde k, \tilde k_1,\dots,\tilde
          k_{n+1})}{v}\right) \right. \right. \\
    & \left. \left.  -\left(\delta R'_{\rm NNLL,\ell}+R''_{\ell}\left(\frac{1}{R'_{\rm NLL}}+\ln d_{\ell}g_{\ell}(\phi)+\ln\max\left[1,\lim_{v\to
            0}\frac{V_{\rm sc}(\{\tilde p\},\tilde k_1,\dots,\tilde
            k_{n+1})}{v}\right]\right)\right)\times \right. \right. \\&  \left. \left. \times \exp\left(-R'_{\rm NLL}\ln\max\left[1,\lim_{v\to
            0}\frac{V_{\rm sc}(\{\tilde p\},\tilde k_1,\dots,\tilde
            k_{n+1})}{v}\right]\right)\right) \right].
  \end{split}
\end{equation}

Of course, the NNLL correction $\delta \mathcal{F}_{\rm sc}=\delta \mathcal{F}^{<}_{\rm sc}+\delta \mathcal{F}^{>}_{\rm sc}$.

\subsection{The function $\delta\mathcal{F}_{\rm hc}$}
\label{sec:MC-F1-hc}

We start from eq.~(\ref{eq:Fhc}), and select $k_1$, the emission with the largest value among the $\zeta_i$. 

We consider first the case $\zeta<\zeta_1$. This gives
\begin{equation}
  \label{eq:Fhc<-NNLL-MC}
  \begin{split}
    \delta\mathcal{F}_{\mathrm{hc}}^{<} &= 
    \sum_{\ell_1=1,2}  R'_{\rm NLL,\ell_1}    \int_0^{\infty}
    \frac{d\zeta_1}{\zeta_1} \zeta_1^{R'_{\rm NLL}}   \int_0^{2\pi}\frac{d\phi_1}{2\pi}
    \int_0^{\zeta_1}
    \frac{d\zeta}{\zeta}  \int_0^{2\pi}\frac{d\phi}{2\pi}\sum_{\ell=1,2}\frac{\as(v^{\frac{1}{a+b_{\ell}}}Q)}{\as(Q)(a+b_{\ell})}\int_0^1 \frac{dz}{z}(zp_{\ell}(z)-2C_{\ell})
  \times \\& \times 
   \left[ \epsilon^{R'_{\rm NLL}}
    \sum_{n=0}^{\infty}\frac{1}{n!} \prod_{i=2}^{n+1}
    \sum_{\ell_i=1,2}  R'_{\rm NLL,\ell_i}
    \int_{\epsilon \zeta_1}^{\zeta_1} \frac{d\zeta_i}{\zeta_i}\int_0^{2\pi}
    \frac{d\phi_i}{2\pi} \right]
\times \\ & \times
\left[
  \Theta\left(1-\lim_{v\to
        0}\frac{V_{\rm sc}(\{\tilde p\},k, k_1,\dots,
        k_{n+1})}{v}\right)
    -\Theta(1-\zeta)\Theta\left(1-\lim_{v\to
        0}\frac{V_{\rm sc}(\{\tilde p\}, k_1,\dots,
        k_{n+1})}{v}\right)
\right]\,.
  \end{split}
\end{equation}

We now define $\tilde\zeta = \zeta/\zeta_1$ and $\tilde\zeta_i = \zeta_i/\zeta_1$. Using the rIRC safety properties of the observable we get
\begin{equation}
  \label{eq:Fhc<-NNLL-MC-rescale}
  \begin{split}
    \delta\mathcal{F}_{\mathrm{hc}}^{<} &= \sum_{\ell_1=1,2}
    R'_{\rm NLL,\ell_1} \int_0^{\infty} \frac{d\zeta_1}{\zeta_1}
    \zeta_1^{R'_{\rm NLL}}
    \int_0^{2\pi}\frac{d\phi_1}{2\pi} \int_0^1
    \frac{d\tilde\zeta}{\tilde\zeta} \int_0^{2\pi}\frac{d\phi}{2\pi}  \sum_{\ell=1,2}\frac{\as(v^{\frac{1}{a+b_{\ell}}}Q)}{\as(Q)(a+b_{\ell})}\int_0^1 \frac{dz}{z}(zp_{\ell}(z)-2C_{\ell})
  \times \\& \times 
   \left[ \epsilon^{R'_{\rm NLL}}
    \sum_{n=0}^{\infty}\frac{1}{n!} \prod_{i=2}^{n+1}
    \sum_{\ell_i=1,2}  R'_{\rm NLL,\ell_i}
    \int_{\epsilon}^{1} \frac{d\tilde\zeta_i}{\tilde\zeta_i}\int_0^{2\pi}
    \frac{d\phi_i}{2\pi} \right]
\times \\ & \times
\left[ 
    \Theta\left(1-\zeta_1\lim_{v\to
        0}\frac{V_{\rm sc}(\{\tilde p\},\tilde k, \tilde k_1,\dots,\tilde
        k_{n+1})}{v}\right)
    -\Theta(1-\zeta_1\tilde\zeta)\Theta\left(1-\zeta_1\lim_{v\to
        0}\frac{V_{\rm sc}(\{\tilde p\}, \tilde k_1,\dots,\tilde
      k_{n+1})}{v}\right)
\right]\,.
\end{split}
\end{equation}

This allows us to perform the integration with respect to $\zeta_1$ and obtain
\begin{equation}
  \label{eq:Fhc<-NNLL-MC-rescale2}
  \begin{split}
    \delta\mathcal{F}_{\mathrm{hc}}^{<} &= \sum_{\ell_1=1,2}
    \frac{R'_{\rm NLL,\ell_1}}{R'_{\rm NLL}} 
    \int_0^{2\pi}\frac{d\phi_1}{2\pi} \int_0^1
    \frac{d\tilde\zeta}{\tilde\zeta} \int_0^{2\pi}\frac{d\phi}{2\pi}  \sum_{\ell=1,2}\frac{\as(v^{\frac{1}{a+b_{\ell}}}Q)}{\as(Q)(a+b_{\ell})}\int_0^1 \frac{dz}{z}(zp_{\ell}(z)-2C_{\ell})
    \times \\& \times 
   \left[ \epsilon^{R'_{\rm NLL}}
    \sum_{n=0}^{\infty}\frac{1}{n!} \prod_{i=2}^{n+1}
    \sum_{\ell_i=1,2}  R'_{\rm NLL,\ell_i}
    \int_{\epsilon}^{1} \frac{d\tilde\zeta_i}{\tilde\zeta_i}\int_0^{2\pi}
    \frac{d\phi_i}{2\pi} \right]
\times \\ & \times
\left[
    \exp\left(-R'_{\rm NLL}\ln\lim_{v\to
        0}\frac{V_{\rm sc}(\{\tilde p\},\tilde k, \tilde k_1,\dots,\tilde
        k_{n+1})}{v}\right)  \right. \\ & \left.
    -\exp\left(-R'_{\rm NLL}\ln\max\left[\tilde\zeta,\lim_{v\to
        0}\frac{V_{\rm sc}(\{\tilde p\}, \tilde k_1,\dots,\tilde
        k_{n+1})}{v}\right]\right)
\right]\,.
\end{split}
\end{equation}

We next consider the case $\zeta>\zeta_1$. This gives
 \begin{equation}
  \label{eq:Fhc>-NNLL-MC}
  \begin{split}
    \delta\mathcal{F}_{\mathrm{hc}}^{>} &= 
    \int_{0}^{\infty}
    \frac{d\zeta}{\zeta} \zeta^{R'_{\rm NLL}}  \int_0^{2\pi}\frac{d\phi}{2\pi}\sum_{\ell=1,2}\frac{\as(v^{\frac{1}{a+b_{\ell}}}Q)}{\as(Q)(a+b_{\ell})}\int_0^1 \frac{dz}{z}(zp_{\ell}(z)-2C_{\ell})   
    \times \\& \times
     \left[ \epsilon^{R'_{\rm NLL}}
    \sum_{n=0}^{\infty}\frac{1}{n!} \prod_{i=1}^{n} \sum_{\ell_i=1,2}
    R'_{\rm NLL,\ell_i}
    \int_{\epsilon \zeta}^{\zeta} \frac{d\zeta_i}{\zeta_i}\int_0^{2\pi}
    \frac{d\phi_i}{2\pi} \right]
\times \\ & \times
\left[
  \Theta\left(1-\lim_{v\to
      0}\frac{V_{\rm sc}(\{\tilde p\},k, k_1,\dots,
        k_{n})}{v}\right)
    -\Theta(1-\zeta)
  \Theta\left(1-\lim_{v\to
      0}\frac{V_{\rm sc}(\{\tilde p\},k_1,\dots,
        k_{n})}{v}\right)
  \right]\,.
  \end{split}
\end{equation}

Defining $\tilde \zeta_i =\zeta_i/\zeta$ and exploiting the rIRC safety properties of the observable, we find
\begin{equation}
  \label{eq:Fhc>-NNLL-MC-rescale}
  \begin{split}
    \delta\mathcal{F}_{\mathrm{hc}}^{>} &= 
    \int_{0}^{\infty}
    \frac{d\zeta}{\zeta} \zeta^{R'_{\rm NLL}}  \int_0^{2\pi}\frac{d\phi}{2\pi}\sum_{\ell=1,2}\frac{\as(v^{\frac{1}{a+b_{\ell}}}Q)}{\as(Q)(a+b_{\ell})}\int_0^1 \frac{dz}{z}(zp_{\ell}(z)-2C_{\ell})
     \times \\& \times
    \left[ \epsilon^{R'_{\rm NLL}} \sum_{n=0}^{\infty}\frac{1}{n!}
      \prod_{i=1}^{n} \sum_{\ell_i=1,2} R'_{\rm NLL,\ell_i}
      \int_{\epsilon}^{1}
      \frac{d\tilde\zeta_i}{\tilde\zeta_i}\int_0^{2\pi}
      \frac{d\phi_i}{2\pi} \right] \times \\ & \times \left[
      \Theta\left(1-\zeta\lim_{v\to 0}\frac{V_{\rm sc}(\{\tilde p\},\tilde k, \tilde
          k_1,\dots,\tilde k_{n})}{v}\right) -\Theta(1-\zeta)
      \Theta\left(1-\zeta\lim_{v\to 0}\frac{V_{\rm sc}(\{\tilde p\},\tilde
          k_1,\dots,\tilde k_{n})}{v}\right) \right]\,.
  \end{split}
\end{equation}

This allows us to perform the integration with respect to $\zeta$, to obtain
\begin{equation}
  \label{eq:Fhc>-NNLL-MC-final}
  \begin{split}
    \delta\mathcal{F}_{\mathrm{hc}}^{>} &= 
   \frac{1}{R'_{\rm NLL}}  \int_0^{2\pi}\frac{d\phi}{2\pi}\sum_{\ell=1,2}\frac{\as(v^{\frac{1}{a+b_{\ell}}}Q)}{\as(Q)(a+b_{\ell})}\int_0^1 \frac{dz}{z}(zp_{\ell}(z)-2C_{\ell})     \times \\&
    \times \left[ \epsilon^{R'_{\rm NLL}} \sum_{n=0}^{\infty}\frac{1}{n!}
      \prod_{i=1}^{n} \sum_{\ell_i=1,2} R'_{\rm NLL,\ell_i}
      \int_{\epsilon}^{1}
      \frac{d\tilde\zeta_i}{\tilde\zeta_i}\int_0^{2\pi}
      \frac{d\phi_i}{2\pi} \right] \times \\ & \times \left[
      \exp\left(-R'_{\rm NLL}\ln\lim_{v\to 0}\frac{V_{\rm sc}(\{\tilde p\},\tilde k, \tilde
          k_1,\dots,\tilde k_{n})}{v}\right) \right. \\ & \left. 
      -\exp\left(-R'_{\rm NLL}\ln\max\left[1,\lim_{v\to 0}\frac{V_{\rm sc}(\{\tilde p\},\tilde
          k_1,\dots,\tilde k_{n})}{v}\right]\right) \right]\,.
  \end{split}
\end{equation}

\subsection{The function $\delta\mathcal{F}_{\rm rec}$}
\label{sec:MC-F1-rec}

We start from eq.~(\ref{eq:Frec}), and again pick up $k_1$, the emission with the largest value among the $\zeta_i$. For $\zeta < \zeta_1$ we have

\begin{equation}
  \label{eq:dF-rec-a}
 \begin{split}
   \delta\mathcal{F}_{\rm rec}^{<}&=
   \sum_{\ell_1=1,2} R'_{\rm NLL,\ell_1} \int_0^{\infty}\frac{d\zeta_1}{\zeta_1}\zeta_1^{R'_{\rm NLL}}
   \int_0^{2\pi}\frac{d\phi_1}{2\pi}
   \sum_{\ell=1,2}
   \frac{\alpha_s(v^{1/(a+b_\ell)}Q)}{\as(Q)(a+b_\ell)} \int_0^1
   \!dz\,
   p_\ell(z)\int_0^{\zeta_1}\frac{d\zeta}{\zeta}
   \int_0^{2\pi}\frac{d\phi}{2\pi}
   \times \\ & \times   
   \left[\epsilon^{R'_{\rm NLL}}\sum_{n=0}^{\infty}\frac{1}{n!}
     \prod_{i=2}^{n+1}\sum_{\ell_i=1,2}R'_{\rm NLL,\ell_i}
     \int_{\epsilon\zeta_1}^{\zeta_1}\frac{d\zeta_i}{\zeta_i}
     \int_0^{2\pi}\frac{d\phi_i}{2\pi}\right] \times \\ & \times
   \left[\Theta\left(1-\lim_{v\to 0} \frac{V^{(k')}_{\rm hc}(\{\tilde p'\},k',
         k_1,\dots,k_{n+1})}{v}\right)-\Theta\left(1-\lim_{v\to 0}
       \frac{V_{\rm sc}(\{\tilde p\},k,k_1,\dots,
         k_{n+1})}{v}\right)\right]\,.
  \end{split}
\end{equation}

As usual, defining $\tilde \zeta_i=\zeta_i/\zeta_1$, and integrating over $\zeta_1$, we get

\begin{equation}
  \label{eq:dF-rec-a-fine}
 \begin{split}
   \delta\mathcal{F}_{\rm rec}^{<}&=
   \sum_{\ell_1=1,2} \frac{R'_{\rm NLL,\ell_1}}{R'_{\rm NLL}} \int_0^{2\pi}\frac{d\phi_1}{2\pi}
   \sum_{\ell=1,2}
   \frac{\alpha_s(v^{1/(a+b_\ell)}Q)}{\as(Q)(a+b_\ell)} \int_0^1
   \!dz\,
   p_\ell(z)\int_0^1\frac{d\tilde\zeta}{\tilde\zeta}
   \int_0^{2\pi}\frac{d\phi}{2\pi}
   \times \\ & \times   
   \left[\epsilon^{R'_{\rm NLL}}\sum_{n=0}^{\infty}\frac{1}{n!}
     \prod_{i=2}^{n+1}\sum_{\ell_i=1,2}R'_{\rm NLL,\ell_i}
     \int_{\epsilon}^{1}\frac{d\tilde\zeta_i}{\tilde\zeta_i}
     \int_0^{2\pi}\frac{d\phi_i}{2\pi}\right] \times \\ & \times
   \left[\exp\left(-R'_{\rm NLL}\ln\lim_{v\to 0} \frac{V_{\rm hc}^{(k')}(\{\tilde p'\},\tilde k',\tilde
         k_1,\dots,\tilde k_{n+1})}{v}\right) \right. \\& \left.
     -
     \exp\left(-R'_{\rm NLL}\ln\lim_{v\to 0} \frac{V_{\rm sc}(\{\tilde p\},\tilde k,\tilde
         k_1,\dots,\tilde k_{n+1})}{v}\right)\right]\,.
  \end{split}
\end{equation}

Similarly, for $\zeta>\zeta_1$ we define $\tilde\zeta_i=\zeta_i/\zeta$ and integrate over $\zeta$, thus obtaining

\begin{equation}
  \label{eq:dF-rec-a-fine2}
 \begin{split}
   \delta\mathcal{F}_{\rm rec}^{>}&= \frac{1}{R'_{\rm NLL}}\sum_{\ell=1,2}
   \frac{\alpha_s(v^{1/(a+b_\ell)}Q)}{\as(Q)(a+b_\ell)} \int_0^1
   \!dz\, p_\ell(z) \int_0^{2\pi}\frac{d\phi}{2\pi}
   \times \\ & \times
   \left[\epsilon^{R'_{\rm NLL}}\sum_{n=0}^{\infty}\frac{1}{n!}
     \prod_{i=1}^n\sum_{\ell_i=1,2}R'_{\rm NLL,\ell_i}
     \int_{\epsilon}^{1}\frac{d\tilde\zeta_i}{\tilde\zeta_i}
     \int_0^{2\pi}\frac{d\phi_i}{2\pi}\right] \times \\ & \times
   \left[\exp\left(-R'_{\rm NLL}\ln\lim_{v\to 0} \frac{V_{\rm hc}^{(k')}(\{\tilde
         p'\},\tilde k',\tilde k_1,\dots,\tilde k_n)}{v}\right)
   \right. \\& \left. -
     \exp\left(-R'_{\rm NLL}\ln\lim_{v\to 0} \frac{V_{\rm sc}(\{\tilde p\},\tilde
         k,\tilde k_1,\dots,\tilde k_n)}{v}\right)\right]\,.
  \end{split}
\end{equation}

\subsection{The function $\delta\mathcal{F}_{\rm wa}$}
\label{sec:MC-F1-wal}

We start from eq.~(\ref{eq:dF-soft-final}), and repeat the above
procedure obtaining the two contributions
\begin{equation}
  \label{eq:dF-soft-<}
\begin{split}
 \delta\mathcal{F}_{\rm wa}^{<} &= \sum_{\ell_1=1,2}
    \frac{R'_{\rm NLL,\ell_1}}{R'_{\rm NLL}} 
    \int_0^{2\pi}\frac{d\phi_1}{2\pi} \int_0^1
    \frac{d\tilde\zeta}{\tilde\zeta} \int_0^{2\pi}\frac{d\phi}{2\pi}\frac{2 C_F}{a} \frac{\alpha_s(v^{1/a}
   Q)}{\alpha_s(Q)} 
 \int_{-\infty}^{\infty} \!\! d\eta\\&\times
\left[ \epsilon^{R'_{\rm NLL}}
    \sum_{n=0}^{\infty}\frac{1}{n!} \prod_{i=2}^{n+1}
    \sum_{\ell_i=1,2}  R'_{\rm NLL,\ell_i}
    \int_{\epsilon}^{1} \frac{d\tilde\zeta_i}{\tilde\zeta_i}\int_0^{2\pi}
    \frac{d\phi_i}{2\pi} \right]
 \\ & \times
\left[
    \exp\left(-R'_{\rm NLL}\ln\lim_{v\to
        0}\frac{\Vwa^{(k)}(\{\tilde p\}, \tilde k, \tilde
          k_1,\dots,\tilde k_{n+1})}{v}\right)  \right. \\ & \left.
    -\exp\left(-R'_{\rm NLL}\ln\lim_{v\to
        0}\frac{\Vsc(\{\tilde p\}, \tilde k, \tilde
          k_1,\dots,\tilde k_{n+1})}{v}\right)
\right]\,,
\end{split}
\end{equation}
and
\begin{equation}
  \label{eq:dF-soft->}
\begin{split}
 \delta\mathcal{F}_{\rm wa}^{>} &= 
    \frac{1}{R'_{\rm NLL}} 
   \int_0^{2\pi}\frac{d\phi}{2\pi}\frac{2 C_F}{a} \frac{\alpha_s(v^{1/a}
   Q)}{\alpha_s(Q)} 
 \int_{-\infty}^{\infty} \!\! d\eta\\&\times
\left[ \epsilon^{R'_{\rm NLL}}
    \sum_{n=0}^{\infty}\frac{1}{n!} \prod_{i=1}^{n}
    \sum_{\ell_i=1,2}  R'_{\rm NLL,\ell_i}
    \int_{\epsilon}^{1} \frac{d\tilde\zeta_i}{\tilde\zeta_i}\int_0^{2\pi}
    \frac{d\phi_i}{2\pi} \right]
 \\ & \times
\left[
    \exp\left(-R'_{\rm NLL}\ln\lim_{v\to
        0}\frac{\Vwa^{(k)}(\{\tilde p\}, \tilde k, \tilde
          k_1,\dots,\tilde k_{n})}{v}\right) \right. \\ & \left.
    -\exp\left(-R'_{\rm NLL}\ln\lim_{v\to
        0}\frac{\Vsc(\{\tilde p\}, \tilde k, \tilde
          k_1,\dots,\tilde k_{n})}{v}\right)
\right]\,.
\end{split}
\end{equation}

\subsection{The function $\delta\mathcal{F}_{\rm correl}$}
\label{sec:MC-F1-correl}

We start from eq.~(\ref{eq:F1-correl}), and pick up $k_1$, the
emission with the largest value among the $\zeta_i$. We also restrict
$\kappa$ to be less than one, getting rid of the factor $1/2!$ in
front of $C_{ab}(\kappa,\eta,\phi)$.

We consider first the case $\zeta_a < \zeta_1$. This gives
\begin{equation}
  \label{eq:F1-correl-<}
  \begin{split}
    \delta\mathcal{F}_{\rm correl}^{<}&=
    \int_0^{\infty}\frac{d\zeta_1}{\zeta_1}\zeta_1^{R'_{\rm NLL}}\int_0^{2\pi}\frac{d\phi_1}{2\pi}\sum_{\ell_1=1,2}
    R'_{\rm NLL,\ell_1}
    \int_0^{\zeta_1}\frac{d\zeta_a}{\zeta_a}\int_0^{2\pi}\frac{d\phi_a}{2\pi}\sum_{\ell_a=1,2}
    \frac{2C_{\ell_a}\lambda}{a\beta_0}\frac{R''_{\ell_a}(v)}{\as(Q)} \times \\& \times\int_0^1 \frac{d\kappa}{\kappa} \int_{-\infty}^{\infty}\!\!\! d\eta \int_0^{2\pi}\frac{d\phi}{2\pi} C_{ab}(\kappa,\eta,\phi) \times\\
    &\times\left[\epsilon^{R'_{\rm NLL}}\sum_{n=0}^{\infty}\frac{1}{n!}
      \prod_{i=2}^{n+1}\sum_{\ell_i=1,2}R'_{\rm NLL,\ell_i}
      \int_{\epsilon\zeta_1}^{\zeta_1}\frac{d\zeta_i}{\zeta_i}
      \int_0^{2\pi}\frac{d\phi_i}{2\pi}\right] \times \\ & \times
    \left[\Theta\left(v-V_{\rm sc}(\{\tilde p\},k_a,k_b,
        k_1,\dots,k_{n+1})\right)-\left(v-V_{\rm sc}(\{\tilde
        p\},k_a+k_b,k_1,\dots,
        k_{n+1})\right)\right]\,.
  \end{split}
\end{equation}
We now define $\tilde\zeta_a=\zeta_a/\zeta_1$, and
$\tilde\zeta_i=\zeta_i/\zeta$, and correspondingly we define the
rescaled momenta $\tilde k_a, \tilde k_b$ and $\tilde k_i$. Notice
that $\kappa,\eta$ and $\phi$ stay unchanged in the rescaling
process. Integrating over $\zeta_1$ we get
\begin{equation}
  \label{eq:F1-correl-<-fine}
  \begin{split}
    \delta\mathcal{F}_{\rm correl}^{<}&=
\int_0^{2\pi}\frac{d\phi_1}{2\pi}\sum_{\ell_1=1,2} \frac{R'_{\rm
    NLL,\ell_1}}{R'_{\rm NLL}}
\int_0^{1}\frac{d\tilde\zeta_a}{\tilde\zeta_a}\int_0^{2\pi}\frac{d\phi_a}{2\pi}\sum_{\ell_a=1,2} \frac{2C_{\ell_a}\lambda}{a\beta_0}\frac{R''_{\ell_a}(v)}{\as(Q)}\times \\& \times\int_0^1 \frac{d\kappa}{\kappa} \int_{-\infty}^{\infty}\!\!\! d\eta \int_0^{2\pi}\frac{d\phi}{2\pi} C_{ab}(\zeta,\eta,\phi) \times\\
    &\times\left[\epsilon^{R'_{\rm NLL}}\sum_{n=0}^{\infty}\frac{1}{n!}
     \prod_{i=2}^{n+1}\sum_{\ell_i=1,2}R'_{\ell_i}
     \int_{\epsilon}^{1}\frac{d\tilde\zeta_i}{\tilde\zeta_i}
     \int_0^{2\pi}\frac{d\phi_i}{2\pi}\right] \times \\ & \times
\left[\exp\left(-R'_{\rm NLL}\ln\lim_{v\to 0}\frac{V_{\rm sc}(\{\tilde
        p\},\tilde k_a,\tilde k_b,\tilde k_1,\dots,\tilde
        k_{n+1})}{v}\right)\right.\\&\left. -
    \exp\left(-R'_{\rm NLL}\ln\lim_{v\to 0} \frac{V_{\rm sc}(\{\tilde
        p\},\tilde k_a+\tilde k_b,\tilde k_1,\dots,\tilde k_{n+1})}{v}\right)\right]\,.
  \end{split}
\end{equation}
Similarly, for $\zeta_a>\zeta_1$ we define $\tilde \zeta_i = \zeta_i/\zeta_a$, and integrate over $\zeta_a$ to obtain
\begin{equation}
  \label{eq:F1-correl->-fine}
  \begin{split}
    \delta\mathcal{F}_{\rm correl}^{>}&=
   \frac{1}{R'_{\rm NLL}} \sum_{\ell_a=1,2} \frac{2C_{\ell_a}\lambda}{a\beta_0}\frac{R''_{\ell_a}}{\as(Q)}\int_0^{2\pi}\frac{d\phi_a}{2\pi}
    \int_0^1 \frac{d\kappa}{\kappa} \int_{-\infty}^{\infty}\!\!\! d\eta \int_0^{2\pi}\frac{d\phi}{2\pi} C_{ab}(\zeta,\eta,\phi) \times\\
    &\times\left[\epsilon^{R'}\sum_{n=0}^{\infty}\frac{1}{n!}
     \prod_{i=1}^{n}\sum_{\ell_i=1,2}R'_{\rm NLL,\ell_i}
     \int_{\epsilon}^{1}\frac{d\tilde\zeta_i}{\tilde\zeta_i}
     \int_0^{2\pi}\frac{d\phi_i}{2\pi}\right] \times \\ & \times
\left[\exp\left(-R'_{\rm NLL}\ln\lim_{v\to
      0}\frac{V_{\rm sc}(\{\tilde
        p\},\tilde k_a,\tilde k_b,\tilde k_1,\dots,\tilde
        k_{n})}{v}\right) \right. \\ & \left. -
    \exp\left(-R'_{\rm NLL}\ln\lim_{v\to 0} \frac{V_{\rm sc}(\{\tilde
        p\},\tilde k_a+\tilde k_b,\tilde k_1,\dots,\tilde k_{n})}{v}\right)\right]\,.
  \end{split}
\end{equation}

\section{Expansion to ${\cal O}(\alpha_s^3)$}
\label{sec:expansion_emsn}
To conclude, we give the numerical expansion of the multiple
emissions function for the observables analysed in the article.
We recall the form of the resummed cross section
\begin{equation}
  \Sigma(v) = e^{L g_1(\lambda)+g_2(\lambda) + \frac{\alpha_s(Q)}{\pi}
    g_3(\lambda)}\left[\FNLL(\lambda)+\frac{\alpha_s(Q)}{\pi}\delta \FNNLL(\lambda)\right]\,,
\end{equation}
where we expand the multiple emissions contribution as
\begin{equation}
\FNLL(\lambda)+\frac{\alpha_s(Q)}{\pi}\delta \FNNLL(\lambda) =
\sum_{i,j} {\cal F}_{ij} \left(\frac{\alpha_s}{2\pi}\right)^i L^j\,.
\end{equation}
In order to perform the matching to NNLO, we need the ${\cal
  F}_{ij}$ coefficients up to ${\cal O}(\alpha_s^3)$. The results
are summarized in Table~\ref{tab:expansionFF}.
\begin{table}[htp!]
\centering
\begin{tabular}{|c|c|c|c|c|c|c|c|}
\hline
 & $T$ & $C$& $\rho_H$ &$B_T$ &$B_W$ &$T_M$ &$O$ \\
\hline
\hline
${\cal F}_{22}$ & -23.394(6) & -23.394(6) & -11.697(4) & -74.121(6)
&-27.332(7) & -53.287(7) & 42.975(9) \\
${\cal F}_{33}$ & -208.252(3) & -208.252(3) & -119.324(2) & -724.49(2)& -371.76(2) & -563.24(7) &513.96(8) \\
\hline
\hline
${\cal F}_{10}$ & -5.4396 &  -1.0532 & -5.4396 & 0 & 0 & 0 &0 \\
${\cal F}_{21}$ & -19.951(7) & -70.157(1) & -20.401(9) & 61.45(2) &
59.65(2) & -10.080(9) & 80.79(5) \\
${\cal F}_{32}$ & -463.51(6) & -1427.72(5) & -247.79(4) & -717.1(1) &
335.8(9) & -1287.0(8)  &-79.(5) \\
\hline
\end{tabular}
\caption{Expansion coefficients for the multiple emissions
  function at NLL (${\cal F}_{22}$, ${\cal
    F}_{33}$), and NNLL (${\cal F}_{10}$, ${\cal
    F}_{21}$, and ${\cal F}_{32}$) up to ${\cal O}(\alpha_s^3)$.
  The error is meant to be on the digit in brackets. The
  numbers shown are just indicative, and the numerical precision can
  be increased.}
\label{tab:expansionFF}
\end{table}

\end{document}